\documentclass[twocolumn,preprintnumbers,superscriptaddress,amsmath,amssymb,nobalancelastpage,10pt,noeprint]{revtex4-2}

\usepackage{physics}
\usepackage{braket}
\usepackage{miller}
\usepackage{changes}
\usepackage{multirow}
\usepackage{float}
\usepackage{bm}
\usepackage{verbatim}
\usepackage{xfrac}
\usepackage{array}
\usepackage{siunitx}
\usepackage{ulem}
\usepackage{amssymb}
\usepackage{amsmath}
\usepackage{svg}
\usepackage{hhline}
\usepackage[hidelinks]{hyperref}
\usepackage{lineno}
\usepackage[french,english]{babel}
\usepackage{lineno}
\usepackage[hidelinks]{hyperref}

\newcommand{\kk}{\mathbf{k}}
\newcommand{\rr}{\mathbf{r}}

\begin{document}
        \title{Quantized Hall drift in a frequency-encoded photonic Chern insulator}
        
        \author{A. Chénier}
        \affiliation{Département de Physique, Université de Montréal, C.P. 6128, Succursale center-Ville, Montréal, Québec, Canada H3C 3J7}
        
        \author{B. d'Aligny}
        \affiliation{Pitaevskii BEC Center, INO-CNR and Dipartimento di Fisica, Università di Trento, via Sommarive 14, I-38123 Trento, Italy}
        \affiliation{Ecole Polytechnique, Institut Polytechnique de Paris, 91128, Palaiseau, France}
        
        \author{F. Pellerin}
        \affiliation{Département de Physique, Université de Montréal, C.P. 6128, Succursale center-Ville, Montréal, Québec, Canada H3C 3J7}

        \author{P.-É. Blanchard}
        \affiliation{Département de Physique, Université de Montréal, C.P. 6128, Succursale center-Ville, Montréal, Québec, Canada H3C 3J7}
        
        \author{T. Ozawa}
        \affiliation{Advanced Institute for Materials Research (WPI-AIMR), Tohoku University, Sendai 980-8577, Japan}
        
        \author{I. Carusotto}
        \affiliation{Pitaevskii BEC Center, INO-CNR and Dipartimento di Fisica, Università di Trento, via Sommarive 14, I-38123 Trento, Italy}
        
        \author{P. St-Jean}
        \affiliation{Département de Physique, Université de Montréal, C.P. 6128, Succursale center-Ville, Montréal, Québec, Canada H3C 3J7}
        \affiliation{Institut Courtois, Université de Montréal, Montréal, Québec, Canada}

        \begin{abstract}
       The quantization of transport and its resilience to backscattering are key features for leveraging topological matter in applications that demand stringent noise mitigation, such as metrology and quantum information processing. Due to the bosonic nature of light, engineering such robust, ``one-way'' channels in synthetic photonic systems imposes the implementation of topological models with broken time-reversal symmetry; this is challenging since photons possess neither an electric charge nor a magnetic moment. Here, we propose and demonstrate a novel approach to realizing photonic Chern insulators -- topological insulators with broken time-reversal symmetry -- by encoding a Haldane-like model in the synthetic frequency dimension of an optical fiber loop platform. The bands' topology is assessed by reconstructing the Bloch states geometry across the Brillouin zone. We further highlight its consequences by measuring a driven-dissipative analogue of the quantized transverse Hall conductivity. Our results open new avenues for harnessing topologically protected light propagation in frequency-multiplexed photonic systems, with applications ranging from precision metrology to photonic quantum processors.
        \end{abstract}
        \maketitle
        
        \setcounter{topnumber}{2}
        \setcounter{bottomnumber}{2}
        \setcounter{totalnumber}{4}
        \renewcommand{\topfraction}{0.85}
        \renewcommand{\bottomfraction}{0.85}
        \renewcommand{\textfraction}{0.15}
        \renewcommand{\floatpagefraction}{0.7}	


\section{Introduction}

Chern insulators are a topological phase of matter with broken time-reversal symmetry ($\mathcal{T}$)~\cite{hasan_colloquium_2010, qi_topological_2011}. A prominent example is provided by the quantum Hall effect, where the defining feature of topology is the emergence of backscattering-immune channels along the edges giving rise to quantized plateaus in the transverse conductivity~\cite{thouless_quantized_1982, prange_quantum_1990}. These robust, quantized channels hold significant value for applications requiring extensive resilience against environmental noise, e.g. in metrology~\cite{von_klitzing_essay_2019}, quantum computing~\cite{nayak_non-abelian_2008} and information processing~\cite{zangeneh-nejad_analogue_2021}.

Extending quantum Hall physics to photonics could enable the design of devices -- classical or quantum -- where the flow of light is similarly quantized and unidirectional, leading to enhanced performance and reduced footprint~\cite{raghu_analogs_2008, ozawa_topological_2019-1, price_roadmap_2022}. However, this endeavor currently faces a two-fold challenge. On the one-hand, breaking $\mathcal{T}$ through magneto-optical effects is technically challenging, particularly at optical or near-infrared frequencies where the Verdet constant of most materials is very small. Therefore, implementing $\mathcal{T}$-broken topological phases in these frequency bands has so far required either operating in a lasing regime~\cite{bahari_nonreciprocal_2017, klembt_exciton-polariton_2018} or mapping time onto spatial coordinates~\cite{kraus_topological_2012, rechtsman_photonic_2013}. On the other hand, $\mathcal{T}$-invariant photonic topological insulators, inspired by the spin-~\cite{hafezi_imaging_2013, wu_scheme_2015} and valley Hall effects~\cite{ma_all-si_2016, noh_observation_2018}, are easier to realize but are inherently prone to backscattering~\cite{rosiek_observation_2023}. Indeed, Kramers' degeneracy theorem which precludes the hybridization of counter-propagating electrons in $\mathcal{T}$-invariant topological matter doesn't apply to bosonic fields~\cite{lu_topological_2016}.

Moreover, transport quantization in driven-dissipative photonic systems arises from fundamentally different physical mechanisms compared to condensed matter, cold atoms or unitary quantum walks~\cite{derrico_two-dimensional_2020}. In these latter systems, the particle number is conserved and the quantization is typically associated with a well-defined band occupation. In contrast, driven-dissipative systems are inherently out-of-equilibrium and a steady-state can emerge from a complex interplay between coherent drive, dissipation, and the underlying band geometry. As a non-equilibrium steady-state necessarily involves non-hermitian effects, the observation of a quantized transport of light typically requires distinct theoretical and experimental frameworks to account for this driven-dissipative balance~\cite{ozawa_anomalous_2014}.

In this work, we introduce and demonstrate a novel approach to realizing photonic Chern insulators that do not rely on strong external magnetic fields, and evidence the onset of a quantized Hall drift of light. We focus on the Haldane model~\cite{haldane_model_1988} where $\mathcal{T}$ is broken by adding complex next-nearest-neighbor couplings to a honeycomb lattice. This model was first realized with cold atoms in optical lattices~\cite{jotzu_experimental_2014, flaschner_observation_2018} and, more recently, has also been reported in the solid-state in temporally modulated monolayers~\cite{mitra_light-wave-controlled_2024} and in twisted Moiré bilayers~\cite{zhao_realization_2024, cai_signatures_2023}. The key challenge in implementing Haldane phases, especially in the photonic context~\cite{he_floquet_2019, liu_gain-induced_2021, sridhar_quantized_2024}, lies in the strict requirements for inter-site connectivity within the lattice.

\begin{figure*}[t]
		\includegraphics[trim=0cm 0cm 0cm 0cm,  width=\textwidth]{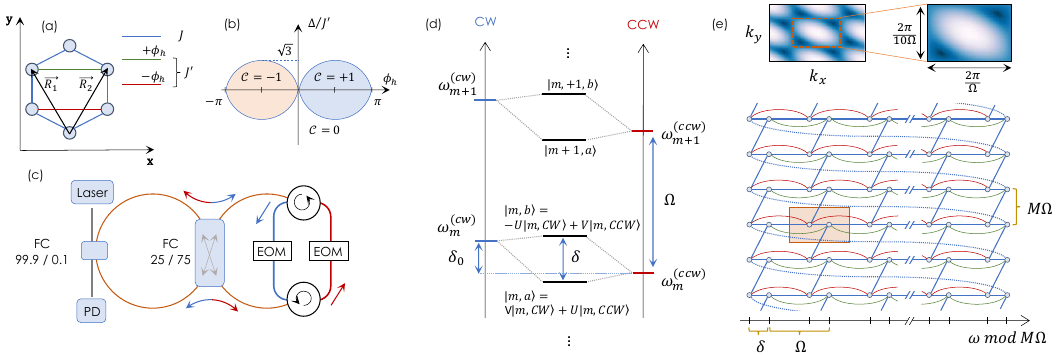}
    \caption{\textbf{Implementation of the Haldane-like model in the synthetic frequency dimension.} (a) Schematic representation of our simplified Haldane model with NN couplings (blue lines) and a single pair of NNN couplings (red and green lines) of opposite phase ($\pm\phi_h$) for each sub-lattice. (b) Topological phase diagram, as a function of $\phi_{h}$ and $\Delta/J'$, exhibiting three distinct phases with Chern number $\mathcal{C}=0,\pm1$. (c) Schematic representation of the experimental setup depicting a single loop whose CCW (red arrows) and CW (blue arrows) circulating modes are coupled with a 25/75 fiber coupler (FC). On the right side, a pair of optical circulators allows modulating CW and CCW modes individually. The loop is probed with a photodiode (PD), using a transmission line coupled to the main loop with a $99.9\%$ transmission FC. (d) 
    Optical mode structure for two pairs of supermodes $\ket{i,a}$, $\ket{i,b}$ with $i=m, m+1$, formed from the coupling of CW and CCW modes split by a frequency $\delta_0$. (e) Representation of the brick-wall Haldane-like Hamiltonian with a folding period of $M\Omega$ - each site represents a supermode of the fiber loop. A unit cell is depicted in orange. Blue lines indicate NN couplings with dotted lines indicating the twisted boundary conditions. NNN couplings are indicated with red and green lines for the $\ket{a}$ and $\ket{b}$ sublattices, respectively (for the sake of clarity, boundary conditions are not depicted for NNN couplings). The insets above depicts the associated reciprocal space (left) and a zoom on the first BZ.}
    \label{fig:haldane_model}
\end{figure*}

To overcome this challenge, we capitalize on the concept of synthetic dimensions which roots in the use of some internal degrees of freedom of a particle to mimic spatial coordinates~\cite{ozawa_topological_2019, yuan_synthetic_2018-1}. Here, we make use of an optical fiber loop platform inspired by electro-optics frequency combs~\cite{parriaux_electro-optic_2020} to encode a honeycomb lattice in frequency space. Independent tuning of the inter-site hopping amplitude and phase is achieved via suitable electro-optical modulation of the loop's refractive index~\cite{ozawa_synthetic_2016, dutt_experimental_2019, sriram_quantized_2025}. Although our platform uses magneto-optical components (i.e. optical isolators and circulators) to improve the cavity lifetime, the breaking of time-reversal symmetry in frequency space, necessary to open a topological gap, is achieved
by tuning the phase of the refractive index temporal modulation. This allows implementing a Haldane-like model across its topological phase diagram while maintaining clearly resolved bandgaps. Taking profit of the driven-dissipative nature of our system, we realize a comprehensive study of the bulk topology of this Haldane-like model. First, we perform a detailed tomography of the Bloch modes to extract the Berry curvature over the Brillouin zone (BZ) and the resulting Chern number of each band. Then, we evidence an anomalous transverse drift of light in frequency space induced by a synthetic electric field. As a hallmark of quantum Hall physics, we show that the frequency integral of this transverse drift is quantized thus providing a driven-dissipative analogue of the quantum Hall conductivity. 

The structure of the article is the following. In Sec.\ref{sec:Haldane}, we give a short review of the Haldane model. Its implementation in the synthetic frequency set-up is presented in Sec.\ref{sec:implementation}. In Sec.\ref{sec:experiments}
we summarize our experimental results for the band structure, the tomography of the Bloch bands and the extraction of the Berry curvature, and for the anomalous displacement and the quantized Hall drift. Conclusions and perspectives are finally given in Sec.\ref{sec:conclu}. Details on the experimental system and the experimental protocols are given in the Appendices.

\section{The Haldane model} 
\label{sec:Haldane}
The Haldane model describes a honeycomb lattice with complex next-nearest neighbor (NNN) couplings. In the tight-binding approximation, it is described by a 2-band Hamiltonian:
\begin{equation}
\label{eq:hamiltonian}
    \mathcal{H}(\textbf{k}) = \textbf{h}(\textbf{k})\cdot\Vec{\sigma},
\end{equation}
with $\textbf{k}$ spanning the BZ and $\Vec{\sigma}=(\sigma_x, \sigma_y, \sigma_z)$ the Pauli matrices. Eigenstates can be expressed as spinors:
\begin{equation}
\label{eq:wavefunction}
    \ket{\psi^{(\pm)}_{\textbf{k}}}=\mathrm{cos}(\theta_{\textbf{k}}^{(\pm)}/2)\ket{a} + \mathrm{sin}(\theta_{\textbf{k}}^{(\pm)}/2)e^{i\phi_{\textbf{k}}^{(\pm)}}\ket{b}.
\end{equation}
with $\{\ket{a}, \ket{b}\}$ the sub-lattice basis, $\theta_{\textbf{k}}^{(\pm)}$ and $\phi_{\textbf{k}}^{(\pm)}$ the polar and azimuthal angles on the Bloch sphere, and superscripts $\pm$ identifying the lower/upper band.

\begin{figure*}[t]
		\includegraphics[trim=0cm 0cm 0cm 0cm,  width=\textwidth]{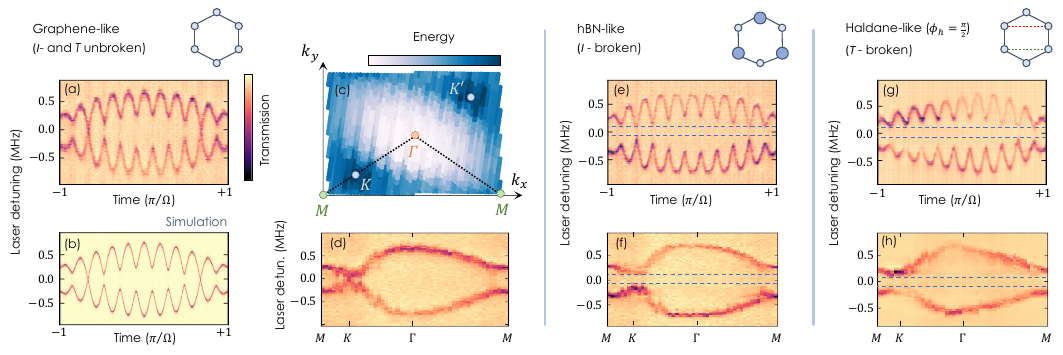}
    \caption{\textbf{Band structure measurements for graphene-like, hBN-like and Haldane-like models.} (a),(e),(g) Band structure measurements for configurations emulating graphene (a), hexagonal boron nitride (e) and the Haldane model with $\phi_h=\pi/2$ (g). The band dispersion is obtained by measuring the time-resolved transmission of the loop as a function of time and laser detuning. (b) The simulation of the graphene-like configuration agrees with the experimental results in (a). (c) Lower band dispersion as a function of $(k_x,k_y)$; a fine coverage of the BZ is obtained by scanning the phase of one driving component. (d),(f),(h) Band dispersion along a trajectory through high-symmetry points $M-K-\Gamma-M$ indicated in (c) for graphene-like (d), hBN-like (f) and the Haldane-like model (h). The latter two clearly show a bandgap, indicated by dashed horizontal lines.}
    \label{fig:band_structure}
\end{figure*}

In a honeycomb lattice with only nearest-neighbor (NN) couplings, as for graphene, inversion ($\mathcal{I}$) and $\mathcal{T}$ impose $h_z(\textbf{k})=0$. This leads to the emergence of band touching points in reciprocal space, labeled $K$ and $K'$ with $K'=-K$, around which the band dispersion is linear, similarly to Dirac fermions. These Dirac points carry remarkable topological signatures~\cite{xiao_berry_2010}, evidenced by singularities in the Berry curvature $\Omega^{(\pm)}_{\textbf{k}} = \nabla \times\bra{\psi^{(\pm)}_{\textbf{k}}}i\,\partial_{\textbf{k}}\ket{\psi^{(\pm)}_{\textbf{k}}}$.

The degeneracies at Dirac points can be lifted either by breaking $\mathcal{I}$ or $\mathcal{T}$. In the former case, e.g. with inequivalent atoms in the unit cell as in hexagonal boron nitride (hBN), a constant mass term $h_{z}(\textbf{k}) = \Delta$ emerges. The bands are however topologically trivial with a vanishing Chern number $\mathcal{C}^{(\pm)}=\frac{1}{2\pi}\int_{BZ} \Omega^{(\pm)}_{\textbf{k}}\cdot\,\mathrm{d}^2\textbf{k}$, as $\mathcal{T}$ imposes $\Omega_{\textbf{k}}$ to be anti-symmetric in $\textbf{k}$. In the latter case, the Berry curvature no longer necessarily integrates to zero,
giving rise to topological bands with non-zero Chern number.

In the Haldane model, $\mathcal{T}$ is broken by adding complex NNN couplings with amplitude $J'$ and opposite phase $\pm\phi_h$ for each sub-lattice. Differently from the original Haldane model, we only consider one pair of NNN coupling terms along $\hat{x}$ (Fig.~\ref{fig:haldane_model} (a)). This simplification does not qualitatively modify the topological phase diagram shown in Fig.~\ref{fig:haldane_model} (b). The NNN coupling terms induce a $\textbf{k}$-dependent mass term $h_z(\textbf{k})=J'\mathrm{sin}(\textbf{k}\cdot(\textbf{R}_{2}-\textbf{R}_{1}))\sin{(\phi_h})$ which has the same amplitude but opposite sign at $K$ and $K'$, leading to topologically non-trivial phases for $|\Delta|<|\sqrt{3}J'\mathrm{sin}(\phi_h)|$.

\section{Implementation in the synthetic frequency dimension}
\label{sec:implementation}

We realize our Haldane-like model using the frequency-periodic cavity modes of an optical fiber loop (Fig.~\ref{fig:haldane_model} (c)) to form a honeycomb lattice in the brickwall geometry (Fig.~\ref{fig:haldane_model} (e)). Unit cells are separated by the loop's free spectral range $\Omega$ with the two sublattices formed from the hybridized supermodes, split by frequency $\delta$, that emerge from coupling the quasi-resonant clockwise (CW) and counter-clockwise (CCW) modes (see Fig.~\ref{fig:haldane_model} (d) and Appendix A).

Inter-site hoppings are tailored by independently modulating the CW and CCW fields with electro-optic phase modulators (EOM) driven at frequencies corresponding to the frequency spacing between pairs of supermodes~\cite{dutt_experimental_2019, dutt_single_2020}. NN hoppings are realized by driving both EOMs with Fourier components at frequencies $\delta$, $\Omega-\delta$ and $M\Omega+\delta$ (see Appendix C). The first two terms which describe coupling between adjacent frequency modes provide a first synthetic dimension. The second synthetic dimension is provided by the third, higher-frequency terms that couple modes separated by $2M+1$ lattice sites~\cite{yuan_synthetic_2018, senanian_programmable_2023, cheng_multi-dimensional_2023}. The resulting synthetic lattice shown in Fig.~\ref{fig:haldane_model} (e) presents a reorganization of the modes in frequency space with a $M\Omega$ folding period. An effective brickwall lattice with twisted boundary conditions (dotted lines) emerges. The associated reciprocal space and a zoom-in on the first BZ are depicted in the top inset.

Finally, the NNN couplings of the Haldane-like model are implemented by including an additional component to the EOMs' driving field at frequency $\Omega$ with opposite phase $\pm\phi_{mod}$ for each EOM. For this model specifically, we work in a regime where the bare CW and CCW modes are not degenerate, so that each supermode has a dominant CW or CCW character, i.e. $\abs{U}^{2} \gg \abs{V}^{2}$ in Fig.~\ref{fig:haldane_model} (d). This is critically important to obtain opposite phases for the  NNN couplings of the $\ket{a,b}$ sub-lattices, even though their splittings are equal $\omega_{m}^{(a/b)}-\omega_{m+1}^{(a/b)}=\Omega$. The value of $\phi_h$ is determined by the ratio U/V: if $U/V>1$, $\phi_h = \phi_{mod}$, and if $U/V<1$, $\phi_h = -\phi_{mod}$ (see Eq. (S.27) in Supplementary Materials \cite{SM}). 

\begin{figure*}[t]
		\includegraphics[trim=0cm 0cm 0cm 0cm,  width=\textwidth]{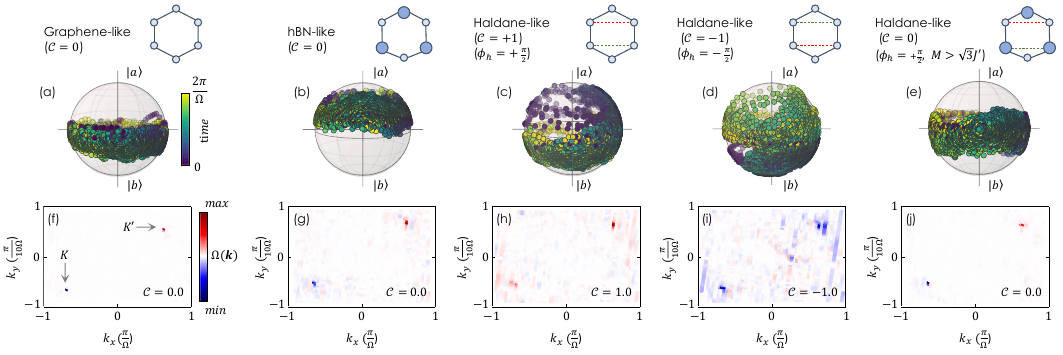}
    \caption{\textbf{Extraction of the Berry curvature and Chern number.} (a)-(e) Tomographic measurement of the lower band's eigenstates $\ket{\psi^{(+)}_{\textbf{k}}}$ across the full BZ for the cases of graphene-like (a) and hBN-like (b) lattices, for topologically non-trivial Haldane-like model with $\phi_h=+\pi/2$ (c) and $\phi_h=-\pi/2$ (d), and topologically trivial Haldane-like model with $\phi_h=+\pi/2$ and $\Delta>\sqrt{3}J'$(e). The north and south poles of the Bloch spheres correspond to the sub-lattice basis states $\ket{a}$ and $\ket{b}$ respectively. The color code refers to the time bin of each $k$-point. (f)-(j) The Berry curvature, extracted using the Fukui method from the momentum-resolved eigenstates shown in Panels (a)-(e), exhibit clear peaks at the positions of the Dirac point $K$ and $K'$. The value of the Chern number obtained by integrating the Berry curvature over the BZ is indicated in the corner of each panel.}
    \label{fig:berry_phase}
\end{figure*}

\section{Results}
\label{sec:experiments}
\subsection{Band structure measurement}

While edge states are the natural observable for topological photonics systems in real-space~\cite{ozawa_topological_2019, mittal_photonic_2019}, synthetic frequency dimensions offer direct access to 
the bulk energy bands through time-resolved transmission measurement using a continuous-wave laser~\cite{dutt_experimental_2019}. The principle underlying this measurement (see Appendix C and Supplementary Materials \cite{SM}) is that the periodicity of the eigenmodes in frequency gives rise to time-periodic pulse trains with superposed long ($T=\frac{2\pi}{\Omega}$) and short periods ($T'=\frac{2\pi}{M\Omega}$) components. Each time bin $t_i$ probes a single point $\textbf{k}=(t_i\,\mathrm{mod}\,T,t_i\,\mathrm{mod}\,T')$ within a BZ whose dimensions are inversely proportional to $\Omega$ and $M\Omega$, respectively associated to the short- and long-range periodicity in frequency space. The transmitted signal exhibits intensity dips whenever the laser frequency is resonant with an
energy band of the lattice.

This periodic signal is monitored on a photodiode while scanning the frequency of the laser. We plot in Fig.~\ref{fig:band_structure} the result averaged over 150 periods of length $T$ with $M=10$. For the case with only NN coupling terms with equal amplitude, which emulates the physics of graphene (Fig.~\ref{fig:band_structure} (a)), we observe two bands with one slow (period $T$) and 10 fast (period $T'$) oscillations, respectively associated to scans of the BZ along $k_x$ and $k_y$. The band touching points associated to each Dirac point ($K$ and $K'$) are clearly visible, in very good agreement with theoretical simulations (Fig.~\ref{fig:band_structure} (b)).

This measurement is then reproduced by varying the phase of the Fourier component with frequency $\Omega-\delta$ to follow slightly shifted trajectories in reciprocal space, thus producing a finer coverage of the BZ. In Fig.~\ref{fig:band_structure} (c), we plot the extracted energy of the lower band, showing very clearly the position of the Dirac points in accordance with the theory (inset in Fig.~\ref{fig:haldane_model} (e)).
Then, by selecting data at time bins along the high-symmetry $M-K-\Gamma-M$ trajectory (dashed line), it is possible to generate the band structure in Fig.~\ref{fig:band_structure} (d).

Breaking $\mathcal{I}$ in this lattice, thus emulating the physics of hBN, is realized by slightly shifting the EOM driving frequencies associated to NN coupling terms to $\delta+\Delta$, $\Omega-\delta-\Delta$ and $M\Omega+\delta+\Delta$ with $\Delta/2\pi=\SI{200}{\kilo\hertz}$. In the synthetic lattice, this detuning realizes a staggered on-site potential giving rise to a mass term $h_z(\textbf{k})=\Delta/2$. Accordingly, we observe at each Dirac point the opening of a bandgap of $E_g \sim \SI{175}{\kilo\hertz}$ (Fig.~\ref{fig:band_structure} (e)-(f)), significantly larger than the optical linewidth. 

Breaking $\mathcal{T}$ and realizing the Haldane-like Hamiltonian is achieved by adding the NNN component at frequency $\Omega$. Using a phase of $\phi_h=\pi/2$, setting $\Delta=0$ and working in a regime where $\abs{U}^2\gg\abs{V}^2$, we again open a significant bandgap (Fig.~\ref{fig:band_structure} (g)-(h)). 
Interestingly, the intensity of each band changes as a function of $\textbf{k}$, with the lower (upper) band reaching a vanishing intensity at $K$ ($K'$); this is a direct consequence of the mass term $h_z(\textbf{k})$ since the transmission measurement is performed along the CCW direction (see Appendix C).

\subsection{Extraction of the Berry curvature}

These transmission measurements further allow performing a full tomography of the eigenmodes across the BZ~\cite{pellerin_wave-function_2024}. As the eigenmodes are linear combinations of CW and CCW modes but we measure the transmission along the CCW direction, the transmitted intensity is modulated in time at frequency $\delta$ and, for each time bin in the BZ, the phase of this beating exactly follows the azimuthal angle $\phi_{\textbf{k}}$ in Eq.~\ref{eq:wavefunction}. The polar angle $\theta_{\textbf{k}}$ indicating the eigenmodes' relative weight on the $\ket{a,b}$ sublattices is extracted from the intensity ratio between the two bands (see Appendix C).


The experimentally extracted values of $\phi_{\textbf{k}},\theta_{\textbf{k}}$ across the BZ can be plotted on a Bloch sphere (Fig.~\ref{fig:berry_phase} (a)-(e)). For the case of graphene (a), the lower band's eigenstates $\ket{\psi_{\textbf{k}}^{(+)}}$ stays confined to the vicinity of the equator, as expected for $h_z=0$. For hBN (b), the evolution is shifted to the northern hemisphere evidencing the effect of the constant mass term $h_z=\Delta/2$. For Haldane-like lattices with $\phi_h=\pm\pi/2$ (c)-(d), the situation is drastically different: the opposite sign of the mass term $h_z(\textbf{k})$ between $K$ and $K'$ makes $\ket{\psi_{\textbf{k}}^{(+)}}$ explore the full Bloch sphere, indicating a non-trivial topology. The trajectory on the Bloch sphere is reversed when switching the sign of $\phi_h$. When we add a constant mass term $\Delta>\sqrt{3}J'$ (e), the Bloch sphere is no longer entirely covered.

In order to quantitatively assess the bands' topology, we take profit of this tomography to extract the value of the Berry curvature across the entire BZ~\cite{gianfrate_measurement_2020}. To do this, we use the technique developed in Ref.~\cite{fukui_chern_2005} for discretized BZs. The result is plotted below each Bloch sphere and the associated Chern number, extracted by summing $\Omega_{\textbf{k}}$ over the BZ, is indicated in the corner.

For graphene-like lattices, the Berry curvature is negligible over the entire BZ except in the close vicinity of $K$ and $K’$. Sharply peaked Berry curvatures with opposite signs are observed close to $K$ and $K’$ points, leading to a vanishing Chern number; this is consistent with $\mathcal{T}$-symmetry of the graphene-like lattice. For hBN-like lattices, the situation is very similar as the Hamiltonian is still $\mathcal{T}$-symmetric but the peaks are broader in reciprocal space, as the Berry curvature spreads in $\textbf{k}$ when a bandgap opens. For the Haldane-like model, we clearly see that the two peaks now have the same sign, either positive for $\mathcal{C}=+1$ or negative for $\mathcal{C}=-1$, as a direct consequence of the broken $\mathcal{T}$. The extracted Chern numbers are consistent with the expected values for topologically non-trivial phases. In the presence of a constant mass term $\Delta>\sqrt{3}J'$, the peaks recover opposite signs: this gives a vanishing Chern number as expected for a topologically trivial phase.

\subsection{Anomalous displacement and quantized Hall drift}

One of the most representative consequences of topology in quantum Hall~\cite{thouless_quantized_1982, prange_quantum_1990} and magnetic~\cite{chang_colloquium_2023} systems is the emergence of plateaus in the transverse conductivity at values determined by the bands' Chern number. In such conservative systems, the topological properties of transport can be probed by tuning the band filling through the chemical potential. In contrast, driven-dissipative photonic systems lack particle-number conservation, and transport manifests as the displacement of an out-of-equilibrium steady-state resulting from the interplay between the drive properties (frequency, phase, amplitude and spatial profile), losses and the underlying band geometry. It is characterized by probing the spatial distribution of the steady-state field intensity induced by a continuous-wave coherent pump. In this context, the topological nature of a band manifests through the system’s out-of-equilibrium response integrated over the entire bandwidth as the driving frequency is swept ~\cite{ozawa_anomalous_2014}. Here, we perform this task and evidence a quantized Hall drift of light in frequency space.

\begin{figure*}[t]
		\includegraphics[trim=0cm 0cm 0cm 0cm,  width=\textwidth]{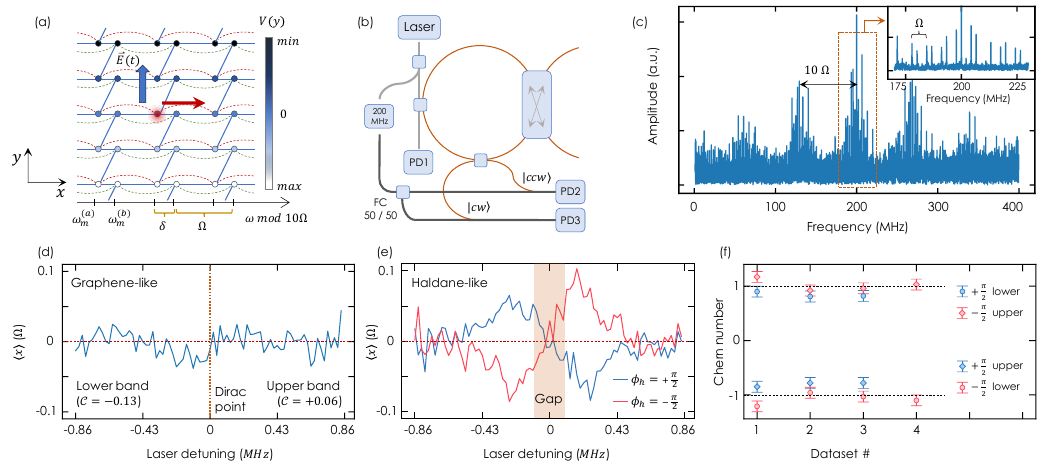}
    \caption{\textbf{Photonic analogue of the transverse Hall conductivity in frequency space.} (a) Schematic representation of the experimental protocol for measuring the transverse displacement. The effective scalar potential induced by the detuning along $\hat{y}$ is depicted by the blue colour gradient. The driven site is indicated by the red spot and the displacement by the red arrow. (b) Experimental setup for the heterodyne measurement. The local oscillator consists of the laser signal shifted by $\SI{200}{\mega\hertz}$ and split in two to beat with both the CCW and CW modes, respectively measured with photodiodes PD2 and PD3. (c) Example of a heterodyne spectrum exhibiting broad regions separated by $10\Omega$ with a high density of narrowly-spaced peaks. The inset shows a zoom-in on the central region exhibiting pairs of peaks separated by $\Omega$. (d)-(e) Anomalous transverse displacement in units of the unit cell ($\Omega$) as a function of the laser detuning  for a graphene-like lattice (d) and two topologically distinct phases of the Haldane-like model (e) . For the graphene-like lattice, we indicate the Chern number extracted by integrating the displacement over the laser detuning bandwidth corresponding to the lower and upper band. (f) Summary of the extracted Chern numbers extracted for the Haldane-like model where 3 measurements are taken with $\phi_h=+\frac{\pi}{2}$ (blue) and 4 measurements taken with $\phi_h=-\frac{\pi}{2}$ (red). Circles (diamonds) correspond to integrating over a bandwidth corresponding to the lower (upper) band. Taking the mean of the absolute value of all these points, we obtain $\abs{\mathcal{C}}_{mean}=0.95$ with a standard deviation of $\sigma=0.14$. The error bars reflect an experimental uncertainty of $\pm 0.10$ on the extracted Chern numbers, obtained by propagating the variance of the graphene-like measurement through Eq. (\ref{eq:Chern_extraction_main}) (see Appendix C).}
    \label{fig:transv_displacement}
\end{figure*}

We implement a synthetic electric field by adding a detuning $\lambda$ to the Fourier component associated to the NN coupling term along $\hat{y}$. In the rotating frame, this induces a linearly varying detuning with respect to the loop's eigenmodes. Within the synthetic lattice picture (Fig.~\ref{fig:transv_displacement} (a)), this can be interpreted as a scalar potential $V(y)=-\lambda y$ giving rise to a synthetic electric field $\textbf{E}=\lambda\hat{y}$~\cite{li_dynamic_2021, oliver_bloch_2023}. The transverse displacement~\cite{xiao_berry_2010} induced by this synthetic electric field is measured  according to the theoretical proposals in~\cite{ozawa_anomalous_2014,ozawa_steady-state_2018} by injecting monochromatic light at a single lattice site and then detecting the center-of-mass displacement of the steady-state distribution of light in frequency space. In our synthetic-dimension framework this is extracted by means of a heterodyne measurement (Fig.~\ref{fig:transv_displacement} (b)) where the signal radiating from the CCW and CW modes of the loop is mixed with the laser shifted by $\SI{200}{\mega\hertz}$. A typical heterodyne spectrum obtained by adding the CW and CCW signals' Fourier transform is presented in Fig.~\ref{fig:transv_displacement} (c). Regions with closely spaced peaks are visible with a spacing of $10\Omega$, which correspond to distinct rows of the lattice; the inset gives a magnified view over the central region, where the spacing $\Omega$ between the pairs of peaks corresponds to the lattice periodicity along $\hat{x}$. From this spectrum, we extract the wavepacket's center-of-mass position $\langle x \rangle = \sum \left[\frac{\omega}{\Omega}\,\textrm{mod}\,M\Omega\right]\, I_{\omega} / \sum I_{\omega}$, where the modulo $M\Omega$ serves to isolate the displacement along $\hat{x}$ (see Appendix C).

The expected anomalous displacement is limited to only a fraction of a unit cell, notably because of the relatively strong dissipation of our system. In order to accurately extract such a small quantity, we make use of frequency modulation to create a time-varying synthetic electric field $\lambda(t)= \lambda_0 \mathrm{sin}(\Omega_{\lambda}t)$. The modulation frequency $\frac{\Omega_{\lambda}}{2\pi}=\SI{100}{\kilo\hertz}$ is slow enough to allow the system to follow its instantaneous out-of-equilibrium steady-state. We then perform a quadrature demodulation of the heterodyne signal with $\lambda(t)$ as the reference signal to isolate the Fourier component of $\langle x \rangle$ that oscillates exactly at $\Omega_{\lambda}$. Since the displacement is obtained by driving the system at one specific frequency at a time, it is sensitive to several contributions that are not linked to the Berry curvature, including the quantum metric (see Supplementary Materials \cite{SM}). The transverse displacement associated solely to the Berry curvature -- the so-called anomalous displacement -- is then isolated by combining measurements obtained by driving different sub-lattice sites, which allow canceling out all non-Berry contributions, according to the specific procedure discussed in Appendix C.

The results are presented in Fig.~\ref{fig:transv_displacement} (d)-(e) as a function of the laser detuning.  As the linewidth of each state in the bands is smaller than the overall width of the bands, the observed frequency-dependent anomalous displacement probes the Berry curvature averaged over states spectrally overlapping with the laser and provides evidence of an anomalous Hall effect~\cite{ozawa_anomalous_2014,wimmer_experimental_2017, gianfrate_measurement_2020, rosen_synthetic_2024}. For the Haldane-like model in (e), we observe significant displacements when the laser is resonant with eigenmodes with a strong Berry curvature, i.e. at the top of the lower band and the bottom of the upper band. This displacement has opposite signs for the upper and lower bands as expected, and gets reversed when switching the topological phase. In contrast, no displacement is observed for the case of a graphene-like lattice (d). 

While this anomalous displacement is of geometrical origin, a topologically quantized Chern number results from integrating the Berry curvature over the BZ. Here, it can be obtained by integrating the transverse displacement over the laser detuning:
\begin{equation}
\label{eq:Chern_extraction_main}
    \mathcal{C}=-\frac{\gamma^2}{4\pi^2\,\lambda_0}\,\int_{\rm band}\!d\omega_L\,\braket{\Delta x}_{\omega_{L}}\,\int \,\frac{\!d^2\kk}{(\epsilon_{\kk}-\omega_L)^2+(\gamma/2)^2}\, 
\end{equation}
with $\epsilon_{\kk}$ the band dispersion, $\braket{\Delta x}_{\omega_{L}}$  the displacement of the center-of-mass position along $\hat{x}$ measured with a laser detuning $\omega_L$ and $\lambda_0$ the amplitude of the oscillating synthetic electric field (see Supplementary Materials \cite{SM}).

The result of such a procedure is presented in Fig.~4 (d) for a measurement on a graphene-like lattice and in (f) for seven different measurements on Haldane-like lattices. Within the experimental uncertainty and the approximations made, values compatible with $0$ and $\pm1$ are obtained, in accordance with the theory and with the geometry of the band states previously measured in Fig.~\ref{fig:berry_phase}. This quantization of the photonic transverse Hall drift is a clear signature of its robustness and universality, endowed by the bands' underlying topology.

\section{Conclusion}
\label{sec:conclu}

Our work demonstrates a novel approach to realizing photonic Chern insulators. In particular, the remarkable tunability and the intrinsic driven-dissipative nature of our platform has allowed us to perform a thorough characterization of the bulk topological properties, including the measurement of the band structures and Berry curvature. Importantly, we have shown the onset of a driven-dissipative analogue of the quantized transverse Hall conductivity, thus paving the way to exploring the consequences of quantization in open quantum systems.

We envision that our platform will allow investigating driven-dissipative topological models that extend beyond the three physical dimensions~\cite{ozawa_synthetic_2016} or involve non-Abelian gauge fields~\cite{cheng_non-abelian_2025}. It also offers new avenues for further exploring the consequences of topologically non-trivial bands, including directional light-matter interaction phenomena in chiral environments~\cite{bello_unconventional_2019, owens_chiral_2022, de_bernardis_chiral_2023}, the emergence of one-way topological edge modes when translational symmetry in frequency space is broken~\cite{ozawa_synthetic_2016, dutt_creating_2022, reid_topological_2025} and their peculiar coupling to bulk states~\cite{zhu_reversal_2024}. On a longer run, extending our platform to the quantum non-linear regime opens a promising route to emulating topological quantum matter with fluids of light~\cite{chiu_classification_2016, carusotto_photonic_2020, wang_observation_2009}. 

It is important to point out that, although our system makes use of magneto-optical elements (i.e. optical isolators and circulators), these are not inherently essential as their main purpose is to facilitate the control of the electromagnetic field and reduce spurious crosstalks. By replacing the CW and CCW propagating modes of a single fiber loop with  a recently proposed~\cite{villa_mean-chiral_2024} and experimentally demonstrated~\cite{sridhar_quantized_2024,sridhar_measuring_2025} configuration consisting of a pair of coupled fiber ring resonators, a similar model could be implemented without the need for those components. This will eventually allow integrating on a chip a frequency-encoded photonic Chern insulator, e.g. using modulated microring resonators~\cite{dinh_reconfigurable_2024}

From a technological perspective, our work opens new pathways, inspired by topological band theory, to robustly engineer the flow of light in frequency space. This advancement holds significant potential for a wide range of applications involving multi-frequency dynamics, including pulsed lasers~\cite{leefmans_topological_2024, yang_mode-locked_2020}, frequency combs~\cite{mittal_topological_2021, hu_realization_2020}, signal processing~\cite{zhang_broadband_2019, buddhiraju_arbitrary_2021}, sensing~\cite{budich_non-hermitian_2020, mcdonald_exponentially-enhanced_2020}, time crystals~\cite{lustig_topological_2018, zheludev_time_2024}, neural networks~\cite{fan_experimental_2023, wanjura_fully_2024} and analog quantum simulation~\cite{javid_chip-scale_2023, bartlett_programmable_2024, arguello-luengo_synthetic_2024}.\\

\section{Acknowledgements}

We acknowledge insightful discussions with W. A. Coish. PSJ acknowledges financial support from Québec's Fonds de Recherche--Nature et Technologies (FRQNT), Canada's Natural Sciences and Engineering Research Council (NSERC), the Alliance Quantum Program grant
funded by NSERC for the project entitled “A new generation of hardware efficient superconducting qubits” and Québec's Minstère de l'Économie, de l'Innovation et de l'Énergie. IC acknowledges financial support from the Provincia Autonoma di Trento, from the Q@TN Initiative, and from the National Quantum Science and Technology Institute through the PNRR MUR Project under Grant PE0000023-NQSTI, co-funded by the European Union - NextGeneration EU. TO acknowledges financial support from JSPS KAKENHI Grant Number JP24K00548, JST PRESTO Grant No. JPMJPR2353, and JST CREST Grant Number JPMJCR19T1.

\section{Data availability}

The data that support the findings of this article are openly available~\cite{DA}.

\vspace{20pt}

\section{Appendix A: Experimental setup}

\subsection{Optical setup}

\begin{figure*}
    \centering
    \includegraphics[width=\textwidth]{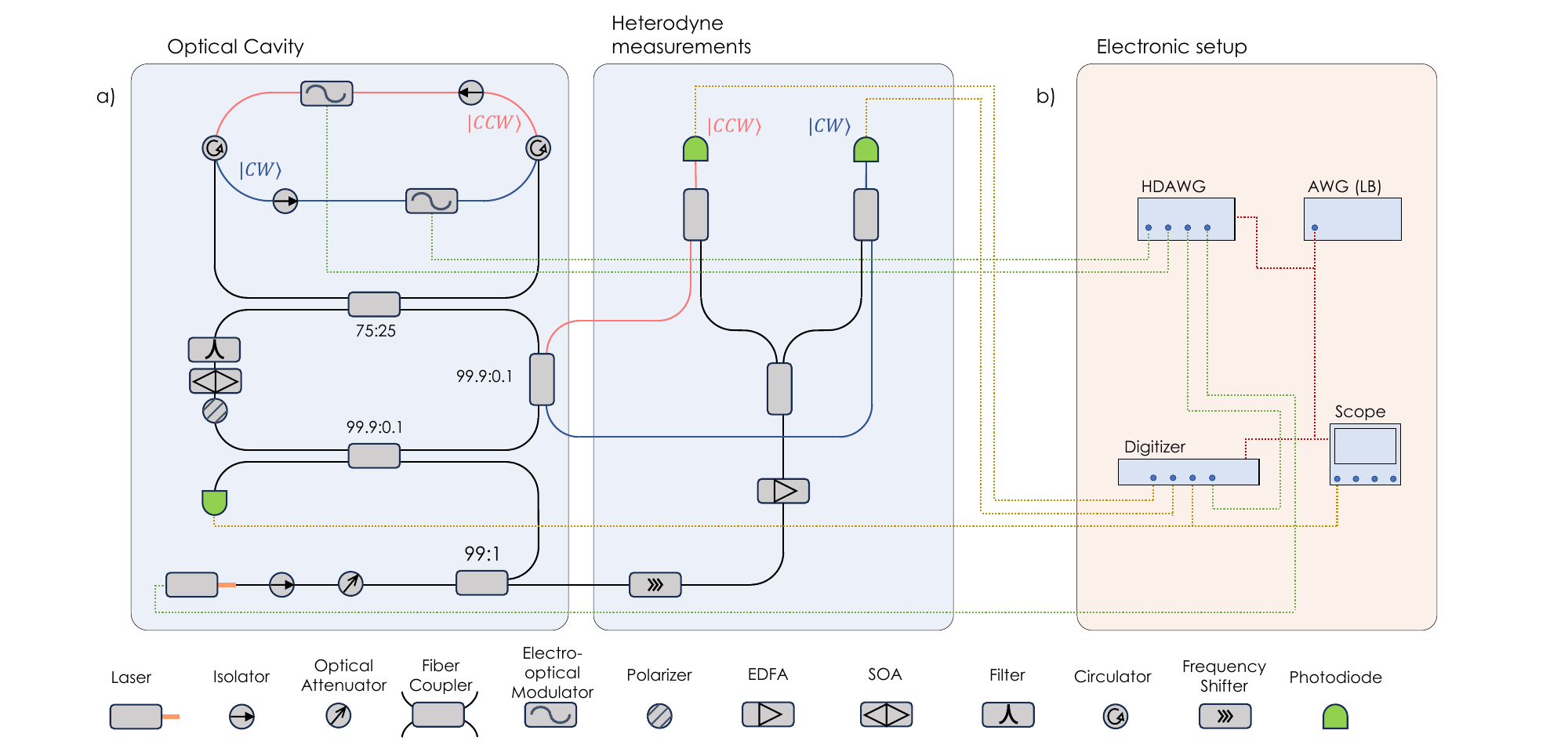}
    \caption{Schematic representation of the experimental setup. (a) Optical setup, including the cavity and the heterodyne measurement and (b) electronic setup, including electrical controllers and digitizers.}
    \label{fig:setup}
\end{figure*}

A schematic of the optical setup is presented in Fig.~\ref{fig:setup} (a). We use a Grade 3 Rio Orion laser emitting with a central wavelength of $1542.27$~nm and with a linewidth of 3.1 kHz. A tunable optical attenuator is used to limit the optical power in the cavity and mitigate non-linear effects. The output of the laser is split by a fiber coupler with 1 \% sent toward the main loop and the rest continuing to a $200$~MHz frequency shifter to be used as a local oscillator (LO) in the heterodyne measurement. A 99.9:0.1 fiber coupler is used as the entry port to the main loop. The transmitted intensity is measured with a $250$~MHz bandwidth InGaAs photodiode connected to a digitizer. This signal is used for measuring band structures and performing wave-function tomography.

All optical components in the setup are polarization maintaining to ensure perfect alignment with the optical axes of the electro-optic modulators; a polarizer is used to further ensure proper polarization filtering. A semiconductor optical amplifier (SOA) is used to amplify the optical signal inside the cavity in order to compensate for the losses caused by the different components inside the loop and ensure a high quality factor ($Q\sim4\times10^{9}$). The resulting finesse of the cavity is $\mathcal{F}\sim120$. A $100$~GHz bandwidth filter is used to suppress the amplification of undesired modes. 

Inside the loop, a 75:25 fiber coupler is used to hybridize the two directions of propagation inside the cavity. As described in the main text, the resulting supermodes form the lattice sites in the synthetic frequency dimension. They are given by:
\begin{align}
\label{eq:supermodes}
\begin{split}
    &\ket{m,a}=U\ket{m, \mathrm{CCW}}+V\ket{m, \mathrm{CW}}\\
    &\ket{m,b}=V\ket{m, \mathrm{CCW}}-U\ket{m, \mathrm{CW}}
\end{split}
\end{align}
\noindent
split by $\delta = 2\sqrt{g^2 + (\delta_0/2)^2}$, with $g\sim \SI{1.1}{\mega\hertz}$ the strength of the effective coupling induced by the FC and $\delta_0=\omega_{\mathrm{m}}^{(cw)} - \omega_{\mathrm{m}}^{(ccw)}$ the splitting between the CW and CCW uncoupled modes. The precise value of this splitting changes depending on the measurement (see Appendix B: Calibration procedure). On the other hand, the lattice periodicity $\omega_{m}^{(a/b)}-\omega_{m+1}^{(a/b)}$ is constant and is given by the free spectral range: $\Omega/2\pi \sim \SI{6.6}{\mega\hertz}$.

In order to efficiently implement nearest-neighbor (NN) and next-nearest-neighbor coupling terms (NNN) -- the latter with opposite phases for different sub-lattices --, we separate the CW and CCW propagating modes with a pair of circulators. This allows time-modulating independently CW and CCW modes with iXBlue lithium niobate electro-optic phase modulators. Each branch also includes an isolator to ensure unidirectional propagation in-between the circulators (see Appendix C: Experimental Protocol, Section A. Band structure Measurements for details on the signal sent to the EOMs). It is important to note here that while circulators and isolators break time-reversal symmetry due to magneto-optics effects, they do so in real space and not in frequency space. As such, they play no role in our implementation of a $\mathcal{T}$-broken topological phase. The experiment could be adapted to make no use of these components as their only role is to provide a better signal-to-noise ratio 

A second 99.9:0.1 fiber coupler is used as an output port to realize the heterodyne measurement. The use of a second port allows probing simultaneously the CW and CCW modes. The LO for the heterodyne measurement is provided by the laser shifted by $200$~MHz and amplified with an erbium doped fiber amplifier~(EDFA). This LO is then split in half with a 50:50 fiber coupler and then mixed (using 99:1 fiber couplers) with the signal radiating from the CW and CCW modes output ports. Two InGaAs photodiodes with a bandwidth of $600$~MHz are used for collecting these mixed signals.

\subsection{Electronic setup}

A schematic of the full electronic setup, used to control and monitor the platform's state, can be seen in figure \ref{fig:setup} (b). The electronic signal from the photodiode measuring the transmission signal is split in two toward a $2$~GHz-bandwidth Rigol oscilloscope and a Tektronix 6 Series Low Profile digitizer (with a bandwidth of $2.5$~GHz and a sampling rate of $25$~GSs$^{-1}$). This allows probing the signal over two different time windows, each with high temporal resolution; the window measured on the oscilloscope is used as a calibration probe and the window measured on the digitizer is used to register the actual experiment data (see Appendix B: Calibration Procedure).

The electrical signals driving the EOMs and tuning the laser frequency are provided by a Zurich Instruments High Definition Arbitrary Waveform Generator (HDAWG). A low bandwidth arbitrary waveform generators, generating a low frequency square wave, is also used to trigger all our electrical instruments simultaneously.

\section{Appendix B: Calibration procedure}

The calibration consists in determining the values of $\Omega$ (the FSR) and $\delta=2\sqrt{g^2+(\delta_{0}/2)^{2}}$ (the splitting between the two sub-lattices) with $\delta_{0}$ the CW-CCW frequency splitting. Accurate determination of these two frequencies is critical for driving the EOMs with the appropriate signal in order to not induce spurious on-site energies and/or synthetic electric fields. Hereafter, we detail the procedure for calibrating each frequency.\\

\subsection{Calibration of the free spectral range $\Omega$}
\begin{figure}
    \includegraphics[width=0.4\textwidth]{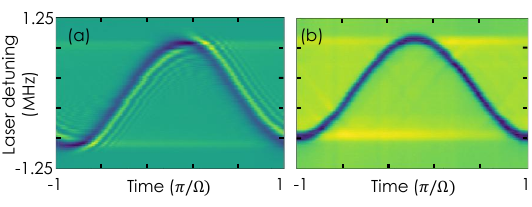}
    \caption{Band dispersions obtained when calibrating the FSR with $|\Omega'-\Omega|\approx40$~kHz in (a) and $|\Omega'-\Omega| <2$~kHz in (b).}
    \label{fig:FSR}
\end{figure}

The FSR is the easiest frequency splitting to calibrate as it does not fluctuate significantly in time. Hence, it can be measured a single time prior to the experiment and be kept identical for all measurements. While a good first approximation can be made using the fact that $\Omega =\frac{2\pi c}{nL}$, this doesn't yield a precise enough value. To increase the precision, we send the following signals to the EOMs: 
\begin{equation*}
     V^{(cw/ccw)}(t) = \frac{V_{nnn}}{2}e^{-i\Omega' t}+\mathbf{c.c}
\end{equation*}
This results in the coupling of next-nearest-neighbors which are exactly split by a frequency $\Omega$. Since each sub-lattice is coupled independently by this driving field and that no coupling between sub-lattices is introduced, the effective bands emerging from each frequency mode are identical and correspond to that of a lattice of identical sites with a unique hopping strength. This leads to band dispersions of the form: 
\begin{equation*}
     \omega(t) = -2J'\mathrm{cos}(\Omega' t).
\end{equation*}

A slight detuning between $\Omega'$ and $\Omega$ results, in the rotating frame, in a linearly varying frequency detuning between the drive and the frequency eigenmodes, and thus to an effective synthetic electric field. This effective field leads to asymmetric features with respect to $k=0$ in the measurement of the dispersion relation. Hence, we can tune $\Omega'$ until we reach symmetric bands (Fig.\ref{fig:FSR}). This procedure can be further optimized by changing the modulation frequency to $N\Omega$, thus enhancing by a factor $N$ the accuracy of the calibration. This led to a value of $\Omega = 2\pi\times6.628$~MHz with an uncertainty of less than $5$~kHz when modulating up to $N=10$.\\

\subsection{Calibration of the sub-lattice spacing $\delta$}

Calibration of the sub-lattice spacing $\delta$ is much more challenging as thermal fluctuations continuously modify the relative length of the CW and CCW branches between the circulators, hence changing the frequency splitting between the hybridized modes on a timescale of the order of several hundreds ms. This relatively slow drift was achieved by passively stabilizing the setup thermally; this drift characteristic time is much larger than typical measurement times ($\sim 10-100 \mu s$). Moreover, for the Haldane-like model measurements, we need to work in a regime with a very specific detuning between CW and CCW modes such that the frequency eigenmodes in Eq.(\ref{eq:supermodes}) have $U\neq V$. This allows modulating the two sub-lattices independently -- using the modulators in the CW and CCW branches -- which is a necessary requirement for engineering NNN couplings with opposite phase for the two sub-lattices. When implementing the Haldane-like model, we typically operate in a regime where $|U|^{2}/|V|^{2}\sim20$ such that we can still have significant NN couplings (which scale as $UV$) with moderate EOM drive amplitude. 

Hence, the challenge consists in fixing a specific and exact value of $\delta$, such that the the EOM drives do not induce undesirable variations of the effective on-site energies, while thermal fluctuations continuously modify the relative length of the CW and CCW branches hence continuously changing $\delta_{0}$.

To do this, we developed a two-step protocol that is implemented for every measurement realized in this work. This protocol consists in separating the acquisition process in two distinct time frames separated by $100$~ms (which is typically faster than the typical thermal fluctuations rate); this dual-time-frame sequence repeats continuously with a period imposed by the low-bandwidth AWG which triggers all other electrical instruments. The first time frame is used to calibrate the value of $\delta$ while the second is dedicated to the measurement itself. Each time frame has a duration of $50$~ms during which the laser frequency scans a bandwidth comparable to the FSR in order to measure at least two band structures. During the first time frame used for calibration, the EOMs are driven with the following signal:
\begin{equation}
\label{eq:mod_signals}
    V^{(cw/ccw)}_{nn}(t) = \pm \frac{1}{2} V^{(0)} \left[e^{-i\delta' t} -  e^{-i(\Omega-\delta')t} \right] + \mathbf{c.c}
\end{equation}
with $+$ and $-$ corresponding to the modulation of the CW and CCW modes respectively, and $\Omega$ obtained following the procedure described above. This driving field induces a coupling between NN lattice sites with identical strength $V^{(0)}$. Upon such driving, the effective band structures follow the following dispersion:
\begin{equation}
    \omega_{\pm}(t) = \mp\sqrt{2J(1+\mathrm{cos}(ka))+\left( \Bar{\Delta}/2 \right)^{2}}
\end{equation}
where $J$ is a coupling coefficient proportional to $V^{(0)}$ and $\Bar{\Delta}=\delta'-\delta$ is the detuning between the driving frequency and the sub-lattice spacing (whose effect can be viewed as a staggered on-site potential). 
\begin{figure}
    \includegraphics[width=0.45\textwidth]{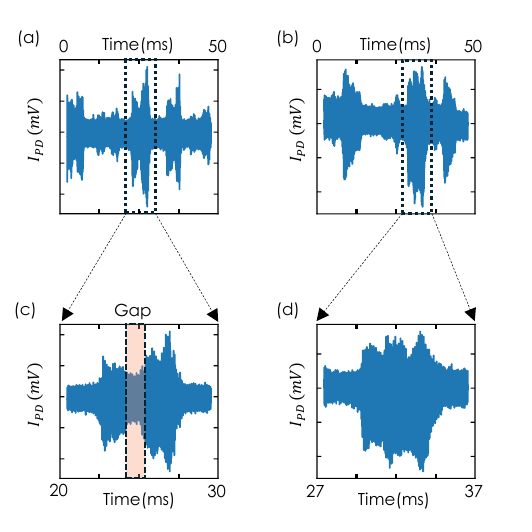}
    \caption{Density of states can be observed visually from the transmission signal on both the digitizer and the oscilloscope  when averaging the signal over a time-window much bigger than the period of a BZ. (a) shows densities of states that exhibit a gap and (c) is a zoom on one of them. This describes a regime where $\delta'\neq\delta$. Meanwhile, (b) depicts densities of states that are gapless, which can be more easily seen when zooming in (d). In this case, we have $\delta'\approx\delta$ and we can acquire the data from the digitizer.}
    \label{fig:Density of states}
\end{figure}
This dispersion relation is gapped whenever $\Bar{\Delta}\neq0$; hence, probing the density of states (DOS) during this time frame (to identify moments where the DOS is not gapped) allows identifying when thermal fluctuations lead to $\delta=\delta'$ (Fig.~\ref{fig:Density of states}). Therefore, it is possible to define a value of $\delta'$ with which we want to work and then wait until thermal fluctuations lead to a gapless DOS indicating perfect matching between the driving frequencies and the loop's eigenmode splitting, i.e. $\Bar{\Delta}=0$. Then, in the second time frame, the EOMs are driven with the fields described in the main text for each experimental part, using $\delta'$ as the definition of the sub-lattice spacing. For graphene-like and hBN-like measurements, $\delta'$ is chosen such that the CW and CCW modes are degenerate, i.e $\delta_{0}=0$. This is because these models don't require NNN couplings. However, to implement the Haldane-like model, $\delta'$ is chosen such that $\delta_{0}\approx 1.1$~MHz. This corresponds to values of $|U|=0.82$ and $|V|=0.18$; a regime where $|U|^{2}/|V|^{2}\sim20$. This regime allows for a good balance between the amplitudes of the NN and NNN couplings, and ensures that the NNN hopping phase, $\phi_{h}$, is exactly equal to the phase of the modulated signal sent to the EOMs, $\phi_{mod}$ (see Supplementary Materials \cite{SM}). As $\delta$ drifts over a much longer timescale than that of individual measurements, this regime does not fluctuate throughout an entire experiment and the value of $\phi_h$ remains stable.

More in specific, the procedure for acquiring data exactly when $\Bar{\Delta}=0$ is the following. The electrical response from the photodiode measuring the transmitted signal is split in half with one part going to the oscilloscope and the other to the digitizer (see above section on the electronic setup). We ensure that the time delays for triggering the scope and the digitizer are such that the former acquires signal during the first time frame while the latter acquires
during the second time frame; separating the signal in two distinct time windows ensures a high temporal resolution for each measurement. As soon as we observe a gapless density of states on the oscilloscope, we stop the acquisition which automatically transfers the memory buffer of the digitizer (containing the experimental signal measured upon good experimental conditions) to a computer where it can be analysed as described below. 

\section{Appendix C: Experimental protocol}
\subsection{Band Structure Measurements}
\label{sec:Band_Measurements}

The band structures shown in Figure 2 of the main text are obtained by modulating the laser frequency with a staircase triangular waveform with frequency $10$~Hz and amplitude $0.075$~V. Upon this modulation, we probe $50$~ms time-windows (as described in the calibration procedure), each time-window corresponding to the descending part of a modulation period. These modulation parameters allow the scan of around one FSR of the optical fiber loop. Each step of the staircase has a duration of $130~\mu$s during which the laser frequency is kept constant. At the beginning of each laser step, we wait $5~\mu$s in order for the system to reach a steady-state before acquiring data. 

The coupling between the different frequency eigenmodes that underlies the synthetic dimension scheme is ensured by driving the EOMs with a signal formed from Fourier components with frequencies equal to the frequency splittings between each pairs of modes to be coupled and amplitudes proportional to the corresponding hopping amplitude. The general formula describing the signal sent to the EOMs is the following:
\begin{equation}
\label{eq:mod_signals}
\begin{aligned}
        V^{(cw/ccw)}(t) = &\pm \frac{1}{2} [V_{nn}^{(1)} e^{-i\delta t} + V_{nn}^{(2)} e^{-i((\Omega-\delta)t+\phi_{shift})}\\
     &+ V_{nn}^{(3)} e^{-i(10\Omega+\delta)t}]\\
     &+\frac{V_{nnn}}{2}e^{-i(\Omega t\pm \phi_{mod})}+\mathbf{c.c}
\end{aligned}
\end{equation}
with $+$ and $-$ corresponding to the modulation of the CW and CCW modes respectively. Specifically, $V^{(i)}_{nn}$ with $i=1,2,3$ represent the nearest-neighbor hoppings of the honeycomb lattice; in particular the coupling $V^{(3)}_{nn}$ to the $21^{st}$ neighbor is used to simulate the second dimension. Components $V_{nnn}$ are linked to the next-nearest-neighbor hoppings of the Haldane model; $V_{nnn}$ is taken to be zero when emulating graphene-like and hBN-like Hamiltonians. The hBN-like Hamiltonian is realized by introducing slight detunings of the NN couplings: 
\begin{align}
\label{eq:hbn_frequencies}
\begin{split}
    &\delta\rightarrow\delta+\Delta\\
    &\Omega-\delta\rightarrow\Omega-\delta-\Delta\\
    &M\Omega+\delta\rightarrow M\Omega+\delta+\Delta.
\end{split}
\end{align}
with $\Delta/2\pi=200$~kHz.

Throughout each frequency step of the laser scan, we measure the time-resolved transmitted signal with a high-bandwidth photodiode. This signal associated to the resonances in the lower ($+$) and upper ($-$) bands follows the relation (see theoretical derivation of the time-resolved transmission in the Supplementary Materials \cite{SM}):
\begin{equation}
\begin{aligned}
\label{eq:transmitted_intensity}
    \frac{I^{(+)}(t)}{|F|^{2}} & \sim\left.1-2\kappa\Re{\frac{f_{\kk}^{(+)}(t)}{\frac{\gamma}{2}-i\left(\Delta\omega_{L}-\epsilon_{\kk}^{(+)}\right)}} \right\rvert_{(k_x,k_y)}\\
    \frac{I^{(-)}(t)}{|F|^{2}} & \sim\left.1-2\kappa\Re{\frac{f_{\kk}^{(-)}(t)}{\frac{\gamma}{2}-i\left(\Delta\omega_{L}-\epsilon_{\kk}^{(-)}\right)}} \right\rvert_{(k_x,k_y)}   
\end{aligned}
\end{equation}
with numerators $f_{k}^{(\pm)}$ given by:
\begin{align}
\label{eq:trans_int_numerator}
\begin{split}
    f_{\kk}^{(+)} &= U^{2}\cos^{2}{\left(\frac{\theta_{\kk}}{2}\right)}+
    UV\sin{\left(\frac{\theta_{\kk}}{2}\right)}\,\cos{\left(\frac{\theta_{\kk}}{2}\right)}
    e^{-i\delta t}e^{i\phi_\kk}\\
    f_{k}^{(-)} &= U^{2}\sin^{2}{\left(\frac{\theta_{\kk}}{2}\right)}-%
    UV\cos{\left(\frac{\theta_{\kk}}{2}\right)}\,\sin{\left(\frac{\theta_{\kk}}{2}\right)}
    e^{-i\delta t}e^{i\phi_\kk}       
\end{split}
\end{align}
where we have used the shorthand $\theta_\kk=\theta^{(+)}_\kk$ and $\phi_\kk=\phi_\kk^{(+)}$ and we have taken into account that $\theta^{(-)}_\kk=\pi-\theta^{(+)}_\kk$ and $\phi^{(-)}_\kk=\pi+\phi^{(+)}_\kk$. These quantities carry information on the polar and azimuthal angles $\theta^{(\pm)}_{\mathbf{k}},\phi^{(\pm)}_{\mathbf{k}}$ of the Bloch eigenstates that is useful for extracting the Berry curvature (see Appendix C: Experimental Protocol, Section B. Berry Curvature Measurement).

In Eqs.~\ref{eq:transmitted_intensity} and \ref{eq:trans_int_numerator}, $F$ and $\Delta\omega_{L}$ are the driving field amplitude and detuning with respect to the loop's closest eigenmodes, $\gamma$ and $\kappa$ are the loop's total decay rate and input-output-coupling ratio, and each time $t$ is linked to a $\kk$-point in the effective BZ using the relations:
\begin{equation}
\begin{aligned}
\label{eq:param_2D}
    t &\mapsto (k_x,k_y)\\
    k_x &= t+\frac{\phi_{shift}}{\Omega}\mod{\frac{2\pi}{\Omega}}\\
    k_y &= t \mod{\frac{2\pi}{10\Omega}}.
\end{aligned}
\end{equation}

The denominators in Eq.~\ref{eq:transmitted_intensity} therefore leads to Lorentzian intensity dips whenever the laser is resonant with the lower/upper band dispersions $\epsilon_{k}^{(\pm)}$ at crystal momentum $k_{x},k_y$. Hence, the effective BZ in the time domain has periodicity $\frac{2\pi}{\Omega}$ along $k_{x}$ and $\frac{2\pi}{10\Omega}$ along $k_{y}$. The measurement of intensity dips at time $t$ indicates that the laser is resonant with an eigenmode with effective crystal momentum $\textbf{k}=(k_{x},k_{y})$.

\begin{figure}
    \includegraphics[width=0.45\textwidth]{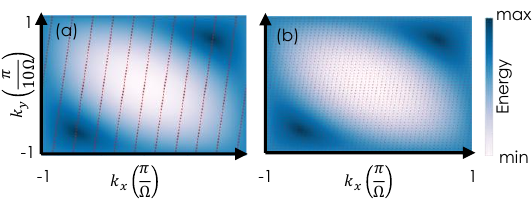}
    \caption{Points sampled on the complete 2D BZ. The colorplot in the background shows the theoretical dispersion of the lower band for graphene. The points sampled by using a single value of $\phi_{shift}$ value are depicted with red dots in (a) and the combined sampling using 5 different values of $\phi_{shift}$ is shown in (b).}
    \label{fig:sampling}
\end{figure}

\begin{figure}
    \includegraphics[width=0.45\textwidth]{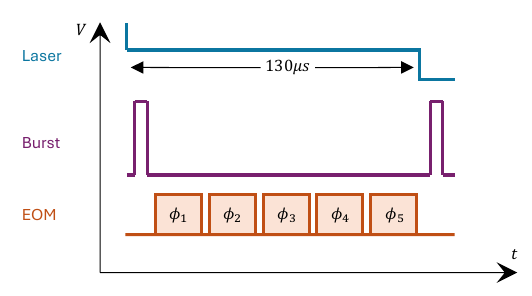}
    \caption{Electrical signals sent to modulate the frequency of the laser (blue), the EOMs (orange), and sent directly to the digitizer (purple) to realize band structure measurements. }
    \label{fig:functions_bandStructure}
\end{figure}

Moreover, the effect of adding a phase $\phi_{shift}$ to the Fourier component $V_{nn}^{(2)}$ results in a shift of the definition of $k_{x}$. This allows the time trace monitored on the photodiode to follow horizontally shifted trajectories in reciprocal space. This is best seen by considering the mapping of the time bins of each measurement on the photodiode onto the BZ: this is presented in Fig.~\ref{fig:sampling} (a) for $\phi_{shift}=0$ (each red dot represents a single time bin with a time separation dictated by the sampling rate of the digitizer), and in Fig.~\ref{fig:sampling} (b) for 5 distinct values $\phi_{shift}=1.25\frac{\pi}{10},\,0.625\frac{\pi}{10},\,0,\,-0.625\frac{\pi}{10},\,-1.25\frac{\pi}{10}$. In order to obtain such a denser covering of the Brillouin zone, the EOM signal through each laser frequency step is partitioned in 5 equally spaced time regions, separated by $3~\mu$s and lasting the equivalent of 150 BZs. In each region, the EOM signal has one specific value of the phase $\phi_{shift}$ among the 5 ones used in Fig.~\ref{fig:sampling} (b). A schematic of the pulse sequence used is depicted in Fig.~\ref{fig:functions_bandStructure}; a burst signal is also generated by the HDAWG at the beginning of each laser step and monitored on the digitizer to ensure that we can precisely identify the temporal position of each laser step in our data analysis.

For each $\phi_{shift}$ region of each laser step, we average the photodiode signal over the 150 trajectories through the BZ to obtain a single averaged trajectory across the BZ with a high signal-to-noise ratio. Then, by scanning the laser frequency and vertically stacking these averaged signals measured at each step, we can construct the full band structure. Figure~2 (a), (e), (g) of the main text show this construction when stacking only the signal associated to the region with $\phi_{shift}=0$. Figure~2 (c) of the main text presents the value of the lower band energy measured at each time bin in Fig.~\ref{fig:sampling} (b). Finally, Fig.~2 (d), (f), (h) of the main text present the data measured at specific time bins along the high-symmetry trajectory $M-K-\Gamma-M$.

\subsection{Berry Curvature Measurement}

\subsubsection{Extraction of $\phi_{\kk}$}

The extraction of $\phi_{\kk}$ is realized using the procedure introduced in Ref.~\cite{pellerin_wave-function_2024}. Let us assume we want to extract $\phi_{\kk}$ for the lower band. Equations~\ref{eq:transmitted_intensity} show that the transmitted intensity ($I^{(+)}_{pd}$) is time-modulated with frequency $\delta$ and phase $\phi_{\kk}$. Specifically, each of the 150 BZs on a given laser step will have transmission peaks happening at the same times with respect to the start of each BZ, but the intensity of these peaks will vary between different BZs. 

Hence, for each $\kk$-point, we want to measure how the transmitted intensity varies over the different BZs, and extract the phase of this modulation. To obtain a precise value of the transmitted intensity at each time bin corresponding to a given $\kk$-point, we sum the intensity measured at this specific time bin (with respect to the beginning of the step) over all the laser steps covering the lower band; this corresponds to integrating the measured signal for a given time bin over each laser detuning. This allows properly taking into account the spectral linewidth of each eigenmode. Then, for every $\kk$-point we evaluate how the transmitted intensity varies over the different BZs by taking the Fourier component at $\delta$ of the signal created from juxtaposing the 150 different intensity values measured at time bins corresponding to this $\kk$-point. Taking the  argument of this demodulated signal at every $\kk$ yields $\phi_{\kk}$.

\begin{figure}
    \includegraphics[width=0.45\textwidth]{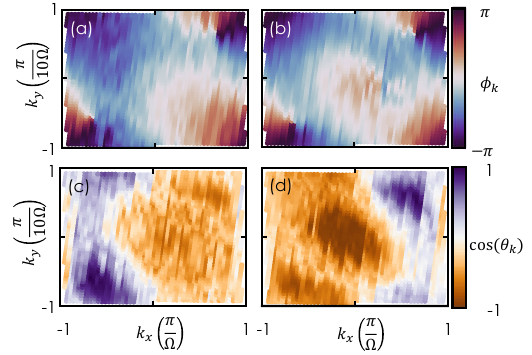}
    \caption{2D Mapping of extracted phases $\phi_{\kk}$ and $\theta_{\kk}$ that are used in figure 3 of main text for the Haldane-like model datasets with $\phi_{h}=\pm\frac{\pi}{2}$. (a) and (b) show the mapping of $\phi_{\kk}$ for $\phi_{h}=+\frac{\pi}{2}$ and $\phi_{h}=-\frac{\pi}{2}$ respectively, whereas (c) and (d) show the mapping of $\cos{\left(\theta_{\kk}\right)}$ with respect to the same data as (a) and (b) respectively. }
    \label{fig:phi_sigma_z}
\end{figure}

Examples of the extraction of $\phi_k$ for the Haldane-like model (presented in Fig.~3 of the main text) are presented in Fig.~\ref{fig:phi_sigma_z} (a) and (b).\\

\subsubsection{Extraction of $\theta_{k}$}

To extract $\theta_{k}$, we now consider the time-integrated transmitted intensity over the 150 BZs that is used in the band structure measurements (as in Fig.~2 of the main text). Doing so, the time-dependent oscillation at frequency $\delta$ averages out and the integrated dip in the transmitted intensity, when the frequency of the driven input field is resonant to the energy of a band, is given by:
\begin{equation}
\begin{aligned}
    T^{(+)}&\equiv 1-\frac{\bar{I}_{pd}^{(+)}}{|{F}|^{2}}\approx4\kappa\frac{U^{2}\cos^{2}{\left(\frac{\theta_{\kk}}{2}\right)}}{\gamma}\\
    T^{(-)}&\equiv1-\frac{\bar{I}_{pd}^{(-)}}{|{F}|^{2}}\approx4\kappa\frac{U^{2}\sin^{2}{\left(\frac{\theta_{\kk}}{2}\right)}}{\gamma}.
\end{aligned}
\end{equation}
Then, we consider the difference between the intensities of the two bands, normalized by their sum:
\begin{equation}
\begin{aligned}
    \frac{T^{(+)}-T^{(-)}}{T^{(+)}+T^{(-)}}&=\frac{\frac{4\kappa U^{2}}{\gamma}\left(\cos^{2}{\left(\frac{\theta_{\kk}}{2}\right)}-\sin^{2}{\left(\frac{\theta_{\kk}}{2}\right)}\right)}{\frac{4\kappa U^{2}}{\gamma}\left(\cos^{2}{\left(\frac{\theta_{\kk}}{2}\right)}+\sin^{2}{\left(\frac{\theta_{\kk}}{2}\right)}\right)}\\
    &=\cos^{2}{\left(\frac{\theta_{\kk}}{2}\right)}-\sin^{2}{\left(\frac{\theta_{\kk}}{2}\right)}\\
    &=\cos(\theta_{\kk}).
\end{aligned}
\end{equation}
Hence, we obtain that $\theta_{\kk} = \arccos{\left(\frac{T^{(+)}-T^{(-)}}{T^{(+)}+T^{(-)}}\right)}$. 

Examples of the extraction of $\cos(\theta_k)$ for the Haldane-like model (used in Fig.~3 of the main text) are presented in Fig.~\ref{fig:phi_sigma_z} (c) and (d).\\

\subsubsection{Mapping on the Bloch sphere}

Once we have extracted both polar and azimuthal angles using the methods described above, it is very straightforward to map the eigenstates on a Bloch sphere. The order with which the different points cover the Bloch sphere is chosen to be the ordering of all the points with respect to their value of  $k_{x}$.\\

\subsubsection{Extraction of the Berry curvature}

\begin{figure}
    \includegraphics[width=0.45\textwidth]{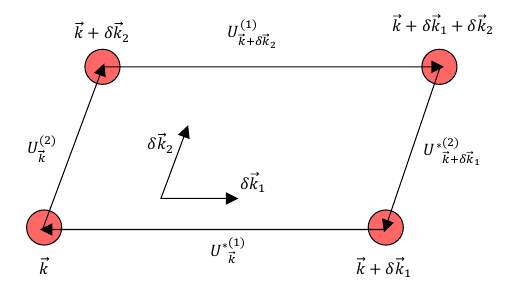}
    \caption{Schematic representation of the plaquette considered in the extraction of $\Omega_{k}$. The red circles represent sampled points.}
    \label{fig:plaquette}
\end{figure}
The expression for the Berry curvature is defined as:
\begin{equation}
    \Omega_{\kk} =  \nabla_{\kk} \times {\bra{\psi_{\kk}}i\nabla_{\kk}\ket{\psi_{\kk}}}
\end{equation}
with $\ket{\psi_{\kk}}=\cos{\left(\frac{\theta_{\kk}}{2}\right)}\ket{a} +\sin{\left(\frac{\theta_{\kk}}{2}\right)}e^{i\phi_{\kk}}\ket{b}$. This is difficult to evaluate from experimental data as it involves numerical derivatives over a discrete number of points presenting experimental fluctuations. Instead, we use a technique  developed by Fukui et al.\cite{fukui_chern_2005} for discretized BZs. We first consider the following quantity:
 \begin{equation}
     U_\textbf{k}^{(\mu)} = \frac{\braket{\psi_\textbf{k}|\psi_{\textbf{k}+\delta\textbf{k}_\mu}}}{|\braket{\psi_\textbf{k}|\psi_{\textbf{k}+\delta\textbf{k}_\mu}}|}      
 \end{equation}
with $\textbf{k}$ a discrete point sampled in the BZ (here a specific time bin) and $\delta\textbf{k}_{\mu}$ the vectors linking adjacent $\textbf{k}$-points.

This quantity is easy to calculate at each $\kk$ as we have extracted both polar and azimuthal angles at each point; thus, we know the eigenstate at each $\kk$. 

We can then refer to the 2D mapping described previously to determine the distribution of the $\kk$-points sampled across the BZ, partition this latter into plaquettes, and identify the pairs of $\kk$-points for which the scalar products have to be calculated.

The Berry curvature flux through a plaquette is given by:
\begin{equation}
    \Omega_{\kk} = \frac{1}{i}\ln{\left[U_\textbf{k}^{(2)}U_{\textbf{k}+\delta\textbf{k}_2}^{(1)}U_{\textbf{k}+\delta\textbf{k}_1}^{*(2)}U_\textbf{k}^{*(1)}\right]}
\end{equation}
where the plaquette described by this equation is schematically shown in Fig \ref{fig:plaquette}. Finally, the Chern number is then simply obtained by summing the extracted Berry curvature over all plaquettes forming the BZ:
\begin{equation}
    \mathcal{C} = \frac{1}{2\pi}\sum_{\kk\in BZ}\Omega_{\kk}.
\end{equation}

\subsection{Anomalous Transverse Displacement and quantized Hall drift}
\begin{figure}
    \includegraphics[width=0.45\textwidth]{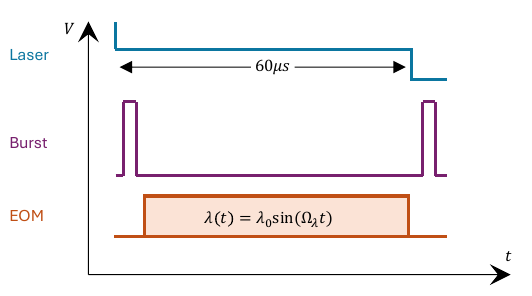}
    \caption{Electrical signals send to modulate the frequency of the laser(blue), to modulate the EOMs(orange), and sent directly to the digitizer(purple) to realize anomalous displacement measurements.}
    \label{fig:Anomalous_signals}
\end{figure}
To obtain anomalous transverse displacement measurements, we slightly change the signals sent to the laser and to the EOMs. The laser frequency is now modulated by a staircase triangular waveform of frequency 20 Hz and amplitude $0.75$~V, with steps of $60~\mu$s (see Fig.~\ref{fig:Anomalous_signals}). The general formula describing the signal sent to the EOMs is now the following:
\begin{equation}
\begin{aligned}
        V^{(cw/ccw)}(t) = &\pm \frac{1}{2} [V_{nn}^{(1)} e^{-i\delta t} + V_{nn}^{(2)} e^{-i((\Omega-\delta)t)}\\
     &+ V_{nn}^{(3)} e^{-i(10\Omega+\delta+\lambda_{t})t}] \\
    &+\frac{V_{nnn}}{2}e^{-i(\Omega t\pm \phi_{h})}+\mathbf{c.c}
\end{aligned}
\end{equation}
where $\lambda_{t} = \lambda_{0}\sin{\left(\Omega_{\lambda}t\right)}$
simulates an oscillating electric potential $V=\lambda_{t}y$ with amplitude $\lambda_{0}/2\pi=10$~kHz and frequency $\Omega_{\lambda}/2\pi=100$~kHz. This acts as a uniform electric field on the lattice along $\hat{y}$:
\begin{equation}
    \textbf{E}(t)=-\nabla V=-\lambda_{t}\hat{y}.
\end{equation}
The fact that this field is oscillating rather than constant is critically important for two main reasons. First, it enables a much better signal-to-noise ratio by mimicking the effect of a lock-in amplifier; by later demodulating our measured displacements with respect to the oscillation frequency, we are able to filter out most of the external noise that would pollute our signals. As the displacements observed are on the order of a tenth of a unit cell, this filtering is essential to our measurements. Second, as shown in numerical simulations (see Supplementary Material section \cite{SM} II.), very strong zero-field contributions to the displacement typically dominate the displacement when using a constant synthetic electric field. As such, by demodulating the extracted signal at the field's oscillating frequency, we can easily filter out such DC components that do not depend on the field. Finally, in order to mitigate inter-bands mixing, which would impact the quantization of the transverse drifts, the value of $\lambda_0$, the amplitude of the modulated field, is chosen to be an order of magnitude smaller than the band gaps ($\sim 2\pi \times \SI{100}{\kilo\hertz}$). 
\begin{figure}
    \includegraphics[width=0.45\textwidth]{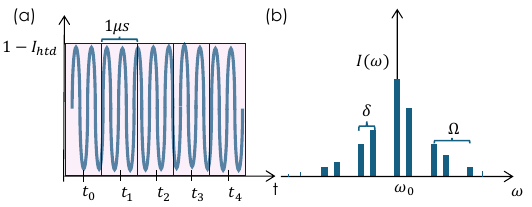}
    \caption{Schematic representation of the windowed Fourier transform process to measure the anomalous displacement. (a) depicts the partitioning of the heterodyne signal, $I_{htd}$, into different time windows, with the Fourier transform of a given window shown in (b). }
    \label{fig:window}
\end{figure}

In contrast with the band structure measurements, the EOM signal is now partitioned into a single time frame lasting the equivalent of 370 BZs per laser step ($\sim\SI{56}{\micro\second}$). Through each of these frequency steps scanned by the laser, we now consider the time-resolved heterodyne signal measured with two high-bandwidth photodiodes, each probing the signal radiating from both directions from the output coupler shown in Fig~\ref{fig:setup}. Each of these heterodyne signals is then split into different time windows of $1~\mu$s associated to a time value $t_{n}$ (see Fig.~\ref{fig:window}). For each time window, the Fourier transform gives a spectrum centered at $\omega_0/2\pi =200$~MHz -  the frequency of the LO - with peaks corresponding to the locations of the supermodes in the frequency space. We then define the displacement for a given $t_n$ as:
\begin{equation}
    \braket{\Delta x}(t_n) = \sum_{\omega}\frac{[\frac{(\omega - \omega_{0})}{\Omega}\,\textrm{mod}\,(M\Omega)] I_{\omega}}{\sum_{\omega}I_{\omega}},
\end{equation}
where $\omega$ indexes the unit cell such that $I_\omega$ represents the total intensity inside of a given unit cell. Following our encoding scheme sketched in Fig.~\ref{fig:transv_displacement} (a), taking the frequency modulo $M\Omega$ (with $M=10$ in our experiment) serves to discard the motion along $\hat{y}$ and isolate the displacement along the $\hat{x}$ direction. In the experiment, there are only 3 rows of the lattice that have enough light intensity to contribute non-negligibly to the displacement. 
Redoing this for each time window yields a time-dependent displacement $\braket{x}(t)$. We then demodulate this displacement with respect to the oscillating synthetic electric field's frequency:
\begin{equation}
   \braket{\Delta x} = \frac{1}{\xi}\int_{0}^{56\mu s} dt\braket{\Delta x}(t)\sin{\left(\Omega_{\lambda}t\right)}
\end{equation}
where $\xi = \int_{0}^{56\mu s}dt\sin^{2}{(\Omega_{\lambda}t)}$ is a normalization constant. We then repeat this process for every laser frequency step of a given band.

Finally, to obtain the displacement results presented in Fig.\ref{fig:transv_displacement}, we must remove spurious, non-Berry contributions to the transverse displacement which arise from the quantum metric and the asymmetric shape of the band dispersion. These terms are both significant and detrimental in the context of the quantization of transport (see Supplementary Materials \cite{SM}).

To eliminate these unwanted contributions, we first take a measurement with the desired value of $\phi_h$ then we subtract this measurement with a second one obtained with $-\phi_{h}$. Since the non-Berry terms are even under an inversion of the topological phase, whereas the Berry contributions are odd, taking the half-difference between the displacements from these two distinct measurements cancels out the non-Berry terms, isolating only the Berry curvature-dependent contributions.

\begin{figure}[h]
    \includegraphics[width=0.45\textwidth]{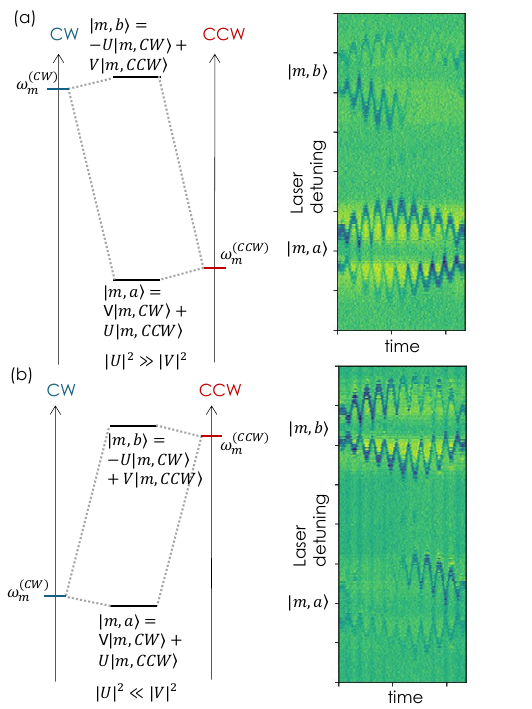}
    \caption{(a) Probing only the CCW direction, when thermal fluctuations make it so that $\abs{U}^2\gg\abs{V}^2$, we have that the band structure of the $\ket{a}$ site is more visible. (b) When $\abs{U}^2\ll\abs{V}^2$, the band structure of the $\ket{b}$ site is more visible instead. }
    \label{fig:Bands_more_visible}
\end{figure}

An example of this two-step procedure is presented in Fig.~\ref{fig:Bands_more_visible}. Panel (a) shows the conditions in which the first measurement is taken, in the regime $\abs{U}^2\gg\abs{V}^2$. As light is injected in the $CCW$ direction, in this regime the band structure emerging from the $\ket{a}$ sub-lattice (measured along the transmission line) is brighter. Therefore the first heterodyne measurement is performed when the drive is resonant with this sub-lattice. Panel (b) shows the conditions of the second measurement. We now work in a regime where $\abs{U}^2\ll\abs{V}^2$, leading to a bright band structure emerging from the $\ket{b}$ sub-lattice; the heterodyne spectrum is now measured when the drive is resonant with this sub-lattice. Both measurements are performed with the same $\phi_{mod}$, hence the reversal of the ratio $\abs{U}/\abs{V}$ leads to a change of sign of $\phi_h$ (see Eq.~(S.26) of the Supplementary Materials \cite{SM}) as required for canceling out non-Berry contributions.

As a result, if $\braket{\Delta x}_a $ and $\braket{\Delta x}_b $ denote the displacements obtained for both configurations (i.e with the laser aligned to a $\ket{a}$ or $\ket{b}$ mode respectively), then the contribution to the displacement coming from the Berry curvature is given by:
\begin{equation}
    \braket{\Delta x} =\frac{\braket{\Delta x}_a-\braket{\Delta x}_b}{2}.
\label{eq:average_displacement}
\end{equation}

\begin{figure}[h]
    \includegraphics[width=0.45\textwidth]{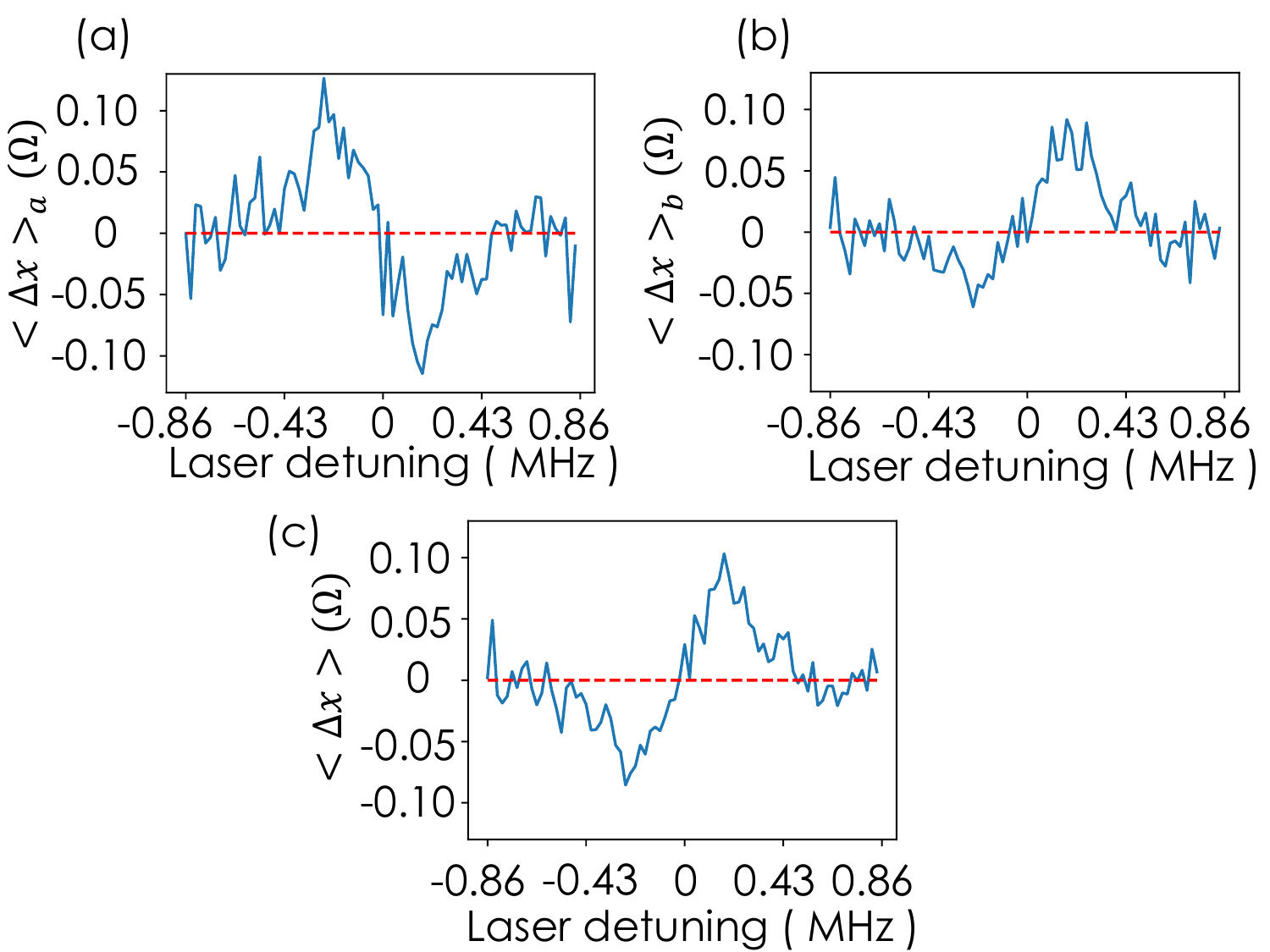}
    \caption{Example of measured displacement obtained when the pump laser is aligned to an $\ket{a}$ site (a), to a $\ket{b}$ site (b), and the resulting displacement when using equation \ref{eq:average_displacement} (c). These measurements were taken with $\phi_h = -\frac{\pi}{2}$. In (a), we see the non-Berry terms interfering positively with the Berry curvature-dependent contributions, leading to a greater displacement amplitude, whereas the non-Berry terms interfere destructively with the topology-dependent contributions in (b), leading to a smaller displacement amplitude. In (c), these non-Berry contributions are canceled and we observe only the anomalous displacement.}
    \label{fig:AandB_displacements}
\end{figure}

Figure~\ref{fig:AandB_displacements} (a) and (b) depicts the measurements obtained in each individual configuration, and Fig.~\ref{fig:AandB_displacements} (c) depicts the half-subtraction described above. These anomalous displacements $\braket{\Delta x}$ are the ones depicted in Fig.~\ref{fig:transv_displacement} and the ones used to extract Chern numbers. 

The Chern number for a given band is then extracted using the formula (see theoretical derivation of the anomalous displacement and its relation to the Chern number in the Supplementary Materials \cite{SM}):
\begin{equation}
\label{eq:Chern_extraction}
    \mathcal{C}=-\frac{\gamma^2}{4\pi^2\,\lambda_0}\,\int_{\rm band}\!d\omega_L\,\braket{\Delta x}_{\omega_{L}}\,\int \,\frac{\!d^2\kk}{(\epsilon_{\kk}-\omega_L)^2+(\gamma/2)^2}\, 
\end{equation}
where $\omega_{L}$ is the frequency corresponding to a step of the laser scan and the integral is performed across the spectral width of the band of interest; $\gamma$ is the average linewidth of the band; $\epsilon_{\kk}$ is the energy of the band; and $\braket{\Delta x}_{\omega_{L}}$ is the displacement measured above for a given laser step in units of the unit cell size. 

To experimentally extract $\gamma$ for a given measurement, we need to look at the corresponding band structure presented in Fig.~\ref{fig:Bands_more_visible}. Indeed, $\gamma$ is defined as the average linewidth of the band, which we can obtain by measuring the full width at half maximum (FWHM) at \textbf{k}-points where the band dispersion reaches an extremum and average these values. However, we cannot consider the transmission signal on the whole step of the laser scan  as the field modulation has the effect of mixing different eigenstates together which leads to an artificial broadening of the bands. Therefore, we instead consider the transmission peaks in time windows of $\SI{1}{\micro\second}$ where the synthetic electric field modulation is minimal. By adding the transmission peaks found in the Brillouin zones corresponding to these time windows for each laser step, we obtain band structures. From them, we can then straightforwardly extract $\gamma$ for a given measurement. This procedure is done independently on both experiments with the laser pump aligned to $\ket{a}$ and $\ket{b}$ sites, yielding two different $\gamma$ values for each band. The one used in Eq.(\ref{eq:Chern_extraction}) for a given band is then simply the average of those two values. 

In the measurements of the band structures in Fig 2. of main text and of the Berry curvatures in Fig 3. of main text, it was desirable to have the longest lifetime possible inside of the cavity. We therefore tuned the amplification level of the SOA to the maximum level compatible with the absence of lasing. In contrast, for the displacement measurements, we slightly reduced the amplification level of the SOA to a regime with greater loss in order to mitigate finite-lattice-size effects arising from the twisted boundary conditions along $x$ and reduce the impact of the modulation frequency $\Omega_\lambda$. As such, for the anomalous transverse displacements, we had $\gamma/2\pi \sim \SI{115}{\kilo\hertz}$; whereas we had $\gamma/2\pi \sim \SI{50}{\kilo\hertz}$ for the other kinds of measurements.  A similar procedure is also done to experimentally determine the corresponding $\epsilon_\textbf{k}$ of a displacement measurement. Finally, the integral over $d^2\kk$ is done experimentally by the change of variable $d^2\kk=2\pi\Omega d\tau$, where $d\tau=320$~ps is the digitizer's sampling rate (in this equation, $\kk$ is dimensionless as it is defined for a normalized unit cell).

To get an estimate of the uncertainty on these extracted Chern numbers, we first calculate the variance in the displacement for graphene-like lattice in Fig 4.(d) of main text. As we expect a theoretical transverse drift of 0 for this model, the variance in this measurement characterizes the uncertainty on the measured transverse displacement for each model. Then, we propagate this uncertainty by injecting the value found for the variance into Eq.(\ref{eq:Chern_extraction}). Finally, taking the square root of this yields the propagated standard deviation and an experimental uncertainty on the Chern number of $\pm0.10$. This estimate on the uncertainty seems reasonable when comparing to the standard deviation of $\sigma =0.14$ that is found when looking at all the experimental measurements taken in Fig 4.(f). 

It is important to note that our extraction procedure for the Chern number from the experimental data relies on a few approximations, namely a linear response in the amplitude of the applied force, and a slow modulation of this force. As such, it is likely that any discrepancies can be explained by these approximations adding systematic uncertainties to our measurements, making the true uncertainty on the measurements slightly greater.

\clearpage
\onecolumngrid

\section{Supplementary Material: Quantized Hall drift \\ in a frequency-encoded photonic Chern insulator}

\renewcommand{\topfraction}{0.85}
\renewcommand{\bottomfraction}{0.85}
\renewcommand{\textfraction}{0.15}
\renewcommand{\floatpagefraction}{0.7}
\renewcommand{\theequation}{S.\arabic{equation}}
\renewcommand{\thefigure}{S.\arabic{figure}}

\section{Theoretical framework}

\subsection{Derivation of the Haldane-like Hamiltonian}

The Hamiltonian describing the effect of the EOMs is given by:
\begin{equation}
\begin{aligned}
    H_{EOM} &= H_{CW} +H_{CCW}\\
            &= 2\eta \sum_{m,n}{\left[V_{CW}(t)d_{m,\downarrow}^{\dagger}d_{n,\downarrow}e^{2i\pi(n-m)\frac{l_{CW}}{L}}+V_{CCW}(t)d_{m,\uparrow}^{\dagger}d_{n,\uparrow}e^{2i\pi(n-m)\frac{l_{CCW}}{L}}\right]} 
\end{aligned}
\end{equation}
with $d_{m,\downarrow}$, $d_{m,\uparrow}$ the bosonic operators associated to the m-th CW and CCW modes, $l_{CW}$ and $l_{CCW}$ the position of the CW and CCW modulators with respect to the start of the loop, and $\eta$ the electro-optics coupling coefficient. We will pose $l_{CW} =l_{CCW} = \frac{L}{2}$ as the change of length due to thermal fluctuations ($\sim \SI{1}{\micro\meter}$) is negligible with respect to the total length of the cavity ($\sim \SI{40}{\meter}$). The expression for the Hamiltonian therefore becomes:
\begin{equation}
    H_{EOM} = 2\eta\sum_{m,n}{(-1)^{n-m}\left[V_{CW}(t)d_{m,\downarrow}^{\dagger}d_{n,\downarrow}+V_{CCW}(t)d_{m,\uparrow}^{\dagger}d_{n,\uparrow}\right]}.
\end{equation}
We then define the operators associated to the coupled basis:
\begin{equation}
\begin{aligned}
    a_{m} &= U d_{m,\uparrow} + V d_{m,\downarrow}\\
    b_{m} &= V d_{,m\uparrow} - U d_{m,\downarrow}
\end{aligned}
\end{equation}\
where we choose, without loss of generality, a gauge where $U,V \in \mathbb{R}$ .
In this basis, the Hamiltonian becomes:
\begin{equation}
\begin{aligned}
    H_{EOM} & = 2\eta\sum_{m,n}{(-1)^{n-m}\left[V_{CW}(t)(Va^{\dagger}_{m}-Ub^{\dagger}_{m})(Va_{n}-Ub_{n})+V_{CCW}(t)(Ua^{\dagger}_{m}+Vb^{\dagger}_{m})(Ua_{n}+Vb_{n})\right]}\\
    &=2\eta\sum_{m,n}{(-1)^{n-m}\left[V_{CW}(t)\left(V^{2}a^{\dagger}_{m}a_{n}+U^{2}b^{\dagger}_{m}b_{n}-UV(a^{\dagger}_{m}b_{n}+b^{\dagger}_{m}a_{n})\right)\right]}\\
    &+2\eta\sum_{m,n}(-1)^{n-m}\left[V_{CCW}(t)\left(U^{2}a^{\dagger}_{m}a_{n}+V^{2}b^{\dagger}_{m}b_{n}+UV(a^{\dagger}_{m}b_{n}+b^{\dagger}_{m}a_{n})\right)\right].
\end{aligned}
\end{equation}
Let us now consider the effect of the NN and NNN couplings seperately by posing
\begin{equation}
\begin{aligned}
    V_{CW}(t) &= V_{NN}(t) + V_{NNN}^{CW}(t)\\
    V_{CCW}(t)&=-V_{NN}(t) + V_{NNN}^{CCW}(t)
\end{aligned}
\end{equation}
such that
\begin{equation}
    H_{EOM} = H_{NN} + H_{NNN}.
\end{equation}
\subsubsection{Derivation of the NN Hamiltonian}
Let us first look at the NN couplings:
\begin{equation}
    H_{NN} = -2\eta V_{NN}(t)\sum_{m,n}{(-1)^{n-m}\left[(U^2-V^2)(a^{\dagger}_{m}a_{n}-b^{\dagger}_{m}b_{n})+2UV(a^{\dagger}_{m}b_{n}+b^{\dagger}_{m}a_{n})\right]}.
\end{equation}
We then move to the rotating frame using:
\begin{equation}
\begin{aligned}
    \Tilde{a}_{m} &= a_{m}e^{i\left(m\Omega-\frac{\delta}{2}\right)t}\\
    \Tilde{b}_{m} &= b_{m}e^{i\left(m\Omega+\frac{\delta}{2}\right)t}
\end{aligned}
\end{equation}
The Hamiltonian becomes:
\begin{equation}
    \Tilde{H}_{NN} = -2\eta V_{NN}(t)\sum_{m,n}{(-1)^{n-m}e^{-i(n-m)\Omega t}\left[\left(U^2-V^2\right)\left(\Tilde{a}^\dagger_{m}\Tilde{a}_{n}-\Tilde{b}^\dagger_{m}\Tilde{b}_{n}\right)+2UV\left(e^{-i\delta t}\Tilde{a}^\dagger_{m}\Tilde{b}_{n}+e^{i\delta t}\Tilde{b}^\dagger_{m}\Tilde{a}_{n}\right)\right]} .
\end{equation}
We then define 
\begin{equation}
    V_{NN}(t) = \left[\frac{V_{NN}^{(0)}}{2}e^{-i(\delta+\Delta)t}+\frac{V_{NN}^{(1)}}{2}e^{-i(\Omega-\delta-\Delta)t}+
    \frac{V_{NN}^{(2)}}{2}e^{-i(10\Omega+\delta+\Delta)t}\right] +\mathbf{c.c}.
\end{equation}
Injecting this expression in the Hamiltonian and using the Rotating Wave Approximation to neglect fast-oscillating terms, we obtain:
\begin{equation}
\begin{aligned}
    \Tilde{H}_{NN} &= -2\eta UV\sum_{m}\left[V_{NN}^{(0)}(e^{-i\Delta t}\Tilde{b}^\dagger_{m}\Tilde{a}_{m}+e^{i\Delta t}\Tilde{a}^\dagger_{m}\Tilde{b}_{m})-V_{NN}^{(1)}(e^{-i\Delta t}\Tilde{b}^{\dagger}_{m-1}\Tilde{a}_{m}+e^{i\Delta t}\Tilde{a}^\dagger_{m}\Tilde{b}_{m-1})\right.\\
    &+\left.V_{NN}^{(2)}(e^{-i\Delta t}\Tilde{b}^\dagger_{m}\Tilde{a}_{m-10}+e^{i\Delta t}\Tilde{a}^{\dagger}_{m-10}\Tilde{b}_{m})\right]
\end{aligned}
\end{equation}
Under a gauge transformation (see Ref. \cite{yuan_three-dimensional_2015}):  
\begin{equation}
\begin{aligned}
    \Tilde{a}_m &\rightarrow \Tilde{a}_m e^{-i\frac{\Delta}{2}t} \\
    \Tilde{b}_m &\rightarrow \Tilde{b}_m e^{i\frac{\Delta}{2}t} 
\end{aligned}
\end{equation}
we obtain the following equivalent Hamiltonian:
\begin{equation}
\begin{aligned}
    \Tilde{H}_{NN}&=\frac{\Delta}{2}\sum_{m}\left[\Tilde{a}^\dagger_{m}\Tilde{a}_{m}-\Tilde{b}^\dagger_{m}\Tilde{b}_{m}\right] \\
    &-2\eta UV\sum_{m}\left[V_{NN}^{(0)}(\Tilde{b}^\dagger_{m}\Tilde{a}_{m}+\Tilde{a}^\dagger_{m}\Tilde{b}_{m})-V_{NN}^{(1)}(\Tilde{b}^{\dagger}_{m-1}\Tilde{a}_{m}+\Tilde{a}^\dagger_{m}\Tilde{b}_{m-1})\right.\\
    &+\left.V_{NN}^{(2)}(\Tilde{b}^\dagger_{m}\Tilde{a}_{m-10}+\Tilde{a}^{\dagger}_{m-10}\Tilde{b}_{m})\right].
\end{aligned}
\label{eq:H-intersublatt}
\end{equation}
Using the periodicity of our system, we can move to reciprocal space using 
\begin{equation}
\begin{aligned}
    \Tilde{a}_{m} &= \int_{0}^{T} \frac{dk_x}{T}\int_{0}^{T'} \frac{dk_y}{T'} e^{im(k_x\Omega+k_y\Omega)}a_{\kk}\\
    \Tilde{b}_{m} &= \int_{0}^{T} \frac{dk_x}{T}\int_{0}^{T'} \frac{dk_y}{T'} e^{im(k_x\Omega+k_y\Omega)}b_{\kk}
\end{aligned}
\end{equation}
where $T=\frac{2\pi}{\Omega}$, $T'=\frac{2\pi}{10\Omega}$, and $k_{x,y}$ are effective crystal momenta with units of time and periodicity of $T$ and $T'$ respectively. This reflects the $\Omega$ and $10\Omega$ periodicity of the effective lattice along $x$ and $y$ respectively and yields the following Hamiltonian in momentum space:
\begin{equation}
    H_{NN}(\kk)=\int_{0}^{T} \frac{dk_x}{T}\int_{0}^{T'} \frac{dk_y}{T'}\left[\frac{\Delta}{4}\left(a^{\dagger}_{\kk}a_{\kk}-b^{\dagger}_{\kk}b_{\kk}\right)
    -2\eta UV\left(V_{NN}^{(0)}-V_{NN}^{(1)}e^{ik_x\Omega}+V_{NN}^{(2)}e^{-10ik_y\Omega}\right)b^{\dagger}_{\kk}a_{\kk}\right] +\mathbf{h.c}
\end{equation}
where we have used the definition of a Dirac comb
\begin{equation}
    D_{T} \left(t-a\right) = \frac{1}{T}\sum_{m}e^{i\left(\frac{2\pi}{T}\right)m\left(t-a\right)}
\end{equation}
to simplify the expression. This Hamiltonian can then be written in matrix form:
\begin{equation}
    H_{NN}(\kk) = \int_{0}^{T} \frac{dk_x}{T}\int_{0}^{T'} \frac{dk_y}{T'}\psi^{\dagger}_{\kk}\cdot h_{NN}(\kk)\cdot \psi_\kk
\end{equation}
\begin{equation}
    h_{NN}(\kk)= 
    \begin{bmatrix}
        -\frac{\Delta}{2} & g(\kk)\\
        g^{*}(\kk)&\frac{\Delta}{2}
    \end{bmatrix} ; \psi_{\kk} = 
    \begin{bmatrix}
        b_{\kk}\\
        a_{\kk}
    \end{bmatrix}
\end{equation}
with $g(\kk) = -2\eta UV\left(V_{NN}^{(0)}-V_{NN}^{(1)}e^{ik_x\Omega}+V_{NN}^{(2)}e^{-10ik_y\Omega}\right)$. 

This correctly describes the Hamiltonian of a particle hopping on a honeycomb lattice in the brickwall geometry with $\Delta$ providing a constant mass term.

\subsubsection{Derivation of the NNN Hamiltonian}
We will now consider the effect of the NNN couplings:
\begin{equation}
\begin{aligned}
    H_{NNN} = 2\eta\sum_{m,n}(-1)^{n-m}&\left[V_{NNN}^{CW}(t)\left(V^2a^{\dagger}_{m}a_{n}+U^2b^{\dagger}_{m}b_{n}-UV\left(a^{\dagger}_{m}b_{n}+b^{\dagger}_{m}a_{n}\right)\right)\right.\\
    &+\left.V_{NNN}^{CCW}(t)\left(U^2a^{\dagger}_{m}a_{n}+V^2b^{\dagger}_{m}b_{n}+UV\left(a^{\dagger}_{m}b_{n}+b^{\dagger}_{m}a_{n}\right)\right)\right].
\end{aligned}
\end{equation}
We move to the rotating frame:
\begin{equation}
\begin{aligned}
    \Tilde{H}_{NNN} = 2\eta\sum_{m,n}(-1)^{n-m}e^{-i(n-m)\Omega t}&\left[V_{NNN}^{CW}(t)\left(V^2\Tilde{a}^\dagger_{m}\Tilde{a}_{n}+U^2\Tilde{b}^\dagger_{m}\Tilde{b}_{n}-UV\left(e^{-i\delta t}\Tilde{a}^\dagger_{m}\Tilde{b}_{n}+e^{i\delta t}\Tilde{b}^\dagger_{m}\Tilde{a}_{n}\right)\right)\right.\\
    &\left.+V_{NNN}^{CCW}(t)\left(U^2\Tilde{a}^\dagger_{m}\Tilde{a}_{n}+V^2\Tilde{b}^\dagger_{m}\Tilde{b}_{n}+UV\left(e^{-i\delta t}\Tilde{a}^\dagger_{m}\Tilde{b}_{n}+e^{i\delta t}\Tilde{b}^\dagger_{m}\Tilde{a}_{n}\right)\right)\right].
\end{aligned}
\end{equation}
We then define:
\begin{equation}
\begin{aligned}
     V_{NNN}^{CW}(t) &= \frac{V_{NNN}}{2}e^{-i(\Omega t+\phi_{mod})} +\mathbf{c.c}\\
     V_{NNN}^{CCW}(t)&=\frac{V_{NNN}}{2}e^{-i(\Omega t-\phi_{mod})} +\mathbf{c.c}.
\end{aligned}
\end{equation}
Substituting these relations in the expression of the Hamiltonian and once again using the Rotating Wave Approximation to neglect fast oscillating terms, we get:
\begin{equation}
    \Tilde{H}_{NNN} = -\eta V_{NNN}\sum_{m}\left[\left(V^2e^{i\phi_{mod}}+U^2e^{-i\phi_{mod}}\right)\Tilde{a}^\dagger_{m}\Tilde{a}_{m-1}+\left(U^2e^{i\phi_{mod}}+V^2e^{-i\phi_{mod}}\right)\Tilde{b}^\dagger_{m}\Tilde{b}_{m-1}\right]+\mathbf{h.c}.
\end{equation}
This expression highlights a feature that is crucial to our implementation of the Haldane-like model: while the NN hoppings between different $a,b$ sublattices described in \eqref{eq:H-intersublatt} are all proportional to $UV$, the different NNN hoppings included in $\Tilde{H}_{NNN}$ involve different combinations of $U^2$ and $V^2$ and, thus, can have different NNN hopping phases for the $a$ and $b$ sublattices as required to realize the simplified Haldane-like model. In physical terms, the fact that, for example, the $a$ ($b$) sublattices involve a majority of CCW (CW)-propagating light allows to selectively address them in spite of the frequency difference being the same $\Omega$: their hopping is in fact mostly determined by the modulation applied to the EOM located in the corresponding CCW (CW) branch, which can be individually controlled. 

Going into reciprocal space and using the definition of a Dirac comb to simplify the expression, this becomes:
\begin{equation}
\begin{aligned}
    H_{NNN}(\kk)= -\eta V_{NNN} \int_{0}^{T} \frac{dk_x}{T}\int_{0}^{T'} \frac{dk_y}{T'}&\left[\left(V^2e^{i(k_x\Omega+\phi_{mod})}+U^2e^{i(k_x\Omega-\phi_{mod})}\right)a^\dagger_{\kk}a_{\kk}\right.\\
    &\left.+\left(U^2e^{i(k_x\Omega+\phi_{mod})}+V^2e^{i(k_x\Omega-\phi_{mod})}\right)b^\dagger_{\kk}b_{\kk}\right] +\mathbf{h.c}
\end{aligned}
\end{equation}
This can also be expressed in matrix form as:
\begin{equation}
H_{NNN}(\kk) = \int_{0}^{T} \frac{dk_x}{T}\int_{0}^{T'} \frac{dk_y}{T'}\psi^{\dagger}_{\kk}\cdot h_{NNN}(\kk)\cdot \psi_{\kk}
\end{equation}
\begin{equation}
\begin{aligned}
    h_{NNN}(\kk)&=-2\eta V_{NNN}
    \begin{bmatrix}
        U^2\cos{\left(k_x\Omega+\phi_{mod}\right)}+V^2\cos{\left(k_x\Omega-\phi_{mod}\right)} & 0\\
        0&V^2\cos{\left(k_x\Omega+\phi_{mod}\right)}+U^2\cos{\left(k_x\Omega-\phi_{mod}\right)}
    \end{bmatrix} \\ 
    \psi_{\kk} &= 
    \begin{bmatrix}
        b_{\kk}\\
        a_{\kk}
    \end{bmatrix}.
\end{aligned}
\end{equation}
Importantly, we can rewrite $h_{NNN}(\kk)$ as:
\begin{equation}
    h_{NNN}(\kk) = -2\eta V_{NNN}\left[\cos(k_x\Omega)\cos(\phi_{mod})\mathbb{I}+(U^2-V^2)\sin(k_x\Omega)\sin(\phi_{mod})\sigma_z\right]
\end{equation}
where we have used the normalization condition $U^2+V^2=1$. This Hamiltonian involves the linear combination of a component proportional to the identity matrix and a $\kk$-dependent $\sigma_z$ component, as expected in the conventional Haldane model. We further see that any phase $\phi_h$ in the topological phase diagram presented in Fig.~1 (b) of the main text can be implemented experimentally, provided that $U\neq V$. Namely, we have that
\begin{equation}
    \phi_h = \pm\phi_{mod},
    \label{eq:sign_phase}
\end{equation}
where the topological phase that we probe with respect to the modulation phase on the NNN coupling terms is determined by the ratio U/V. By linearity, adding the contributions of $h_{NN}(\kk)$ and $h_{NNN}(\kk)$ yields the desired, complete Haldane-like Hamiltonian.

Experimentally, we therefore need to work in a regime where $U^2 \gg V^2$ or $V^2\gg U^2$ in order to have non-negligible NNN couplings, but where the ratio $U/V$ is not too extreme so that NN couplings remain appreciable.

\subsection{Derivation of the time-resolved transmission measured by the photodiode}

The following derivation follows closely what is done in the supplementals of \cite{pellerin_wave-function_2024} without assuming perfect hybridization between the modes.

Assuming a CCW input field, the evolution of the electromagnetic field confined in the optical fibre loop cavity is given by the following coupled-mode equations:
\begin{equation}
\begin{aligned}
    \dot{d}_{m,\downarrow} &= i\left[H(t),d_{m,\downarrow}\right]-\frac{\gamma}{2}d_{m,\downarrow} \\
    \dot{d}_{m,\uparrow} &= i\left[H(t),d_{m,\uparrow}\right]-\frac{\gamma}{2}d_{m,\uparrow} - i\sqrt{\kappa}s_{in}(t) 
    \label{eq:Langevin}
\end{aligned}
\end{equation}
where $d_{m,\downarrow}$ and $d_{m,\uparrow}$ are annihilation operators describing the CW and CCW modes respectively, $H(t)$ is the sum of the EOM modulation and hybridization fibre coupler Hamiltonians, $\kappa$ is the input-output coupling strength that quantifies the radiative loss rate, $s_{in}$ is the input field given by $s_{in} = Fe^{-i\omega_{L}t}$, and $\gamma$ is the total decay rate of the field inside the cavity. 

We then move to the coupled basis : 
\begin{equation}
\begin{aligned}
    a_{m} &= U d_{m,\uparrow} + V d_{m,\downarrow}\\
    b_{m} &= V d_{m,\uparrow} - U d_{m,\downarrow}
\end{aligned}
\end{equation}
where we pose $U,V \in \mathbb{R}$ without loss of generality, the differential equations can be written in the rotating frame as:
\begin{equation}
\begin{aligned}
    \dot{\Tilde{a}}_{m} &= i\left[\Tilde{H}(t),\Tilde{a}_{m} \right] - \frac{\gamma}{2} \Tilde{a}_{m} -iU \sqrt{\kappa} e^{i\left(m\Omega-\frac{\delta}{2}\right)t}s_{in}\\
    \dot{\Tilde{b}}_{m} &= i\left[\Tilde{H}(t),\Tilde{b}_{m} \right] - \frac{\gamma}{2} \Tilde{b}_{m} -iV \sqrt{\kappa} e^{i\left(m\Omega+\frac{\delta}{2}\right)t}s_{in}
\end{aligned}
\end{equation}
In Fourier space, this transforms into:
\begin{equation}
\begin{aligned}
 \dot{a}_{\kk} &= i\left[ H(\kk),a_{\kk} \right]- \frac{\gamma}{2} a_{\kk}-i U\sqrt{\kappa} \sum_{m} e^{-i(k_x+k_y)m\Omega}e^{i\left(m\Omega-\frac{\delta}{2}\right)t}s_{in}\\
 \dot{b}_{\kk} &= i\left[ H(\kk),b_{\kk} \right]- \frac{\gamma}{2} b_{\kk}-i V\sqrt{\kappa} \sum_{m} e^{-i(k_x+k_y)m\Omega}e^{i\left(m\Omega+\frac{\delta}{2}\right)t}s_{in}
\end{aligned}
\end{equation}
We can then use the general basis that diagonalizes $H(\kk)$:
\begin{equation}
\begin{aligned}
    u_{\kk,+} &= \cos{\left(\frac{\theta_{\kk}}{2}\right)} a_{\kk} + \sin{\left(\frac{\theta_{\kk}}{2}\right)} e^{i\phi_{\kk}} b_{\kk}\\
    u_{\kk,-} &= \sin{\left(\frac{\theta_{\kk}}{2}\right)} a_{\kk} - \cos{\left(\frac{\theta_{\kk}}{2}\right)}e^{i\phi_{\kk}} b_{\kk}
\end{aligned}
\end{equation}
with associated eigenenergies $\omega_{\kk,\pm} = \epsilon_{\kk}^{\pm}$ in which the equations of motion become:
\begin{equation}
\begin{aligned}
      \dot{u}_{\kk,+} &= - i\epsilon_{\kk}^{+}u_{\kk,+} - \frac{\gamma}{2}u_{\kk,+} - i\sqrt{\kappa} s_{in} \left[ U\cos{\left(\frac{\theta_{\kk}}{2}\right)}e^{-i\frac{\delta}{2}t} +V\sin{\left(\frac{\theta_{\kk}}{2}\right)} e^{i\frac{\delta}{2}t + i\phi(\kk)} \right] \sum_{m} e^{im\Omega\left(t-k_x-k_y\right)} \\
      \dot{u}_{\kk,-} &= - i\epsilon_{\kk}^{-}u_{\kk,-} - \frac{\gamma}{2}u_{\kk,-} - i\sqrt{\kappa} s_{in} \left[ U\sin{\left(\frac{\theta_{\kk}}{2}\right)}e^{-i\frac{\delta}{2}t} -V\cos{\left(\frac{\theta_{\kk}}{2}\right)} e^{i\frac{\delta}{2}t + i\phi(\kk)} \right] \sum_{m} e^{im\Omega\left(t-k_x-k_y\right)}
\end{aligned}
\end{equation}
Integrating over time, we obtain (assuming initial conditions $u_{\kk,\pm}(0)=0$):
\begin{equation}
\begin{aligned}
    u_{\kk,+}(t) &= -i\sqrt{\kappa} \int_{0}^{t}dt' e^{( i\epsilon_{\kk}^{+}+ \gamma/2)(t'-t)} s_{in}(t') \left[ U\cos{\left(\frac{\theta_{\kk}}{2}\right)}e^{-i\frac{\delta}{2}t'} +V\sin{\left(\frac{\theta_{\kk}}{2}\right)} e^{i\frac{\delta}{2}t'+i\phi(\kk)}  \right]\sum_{m} e^{im\Omega\left(t'-k_x-k_y\right)}\\
    u_{\kk,-}(t) &= -i\sqrt{\kappa} \int_{0}^{t}dt' e^{( i\epsilon_{\kk}^{-}+ \gamma/2)(t'-t)} s_{in}(t') \left[ U\sin{\left(\frac{\theta_{\kk}}{2}\right)}e^{-i\frac{\delta}{2}t'} -V\cos{\left(\frac{\theta_{\kk}}{2}\right)} e^{i\frac{\delta}{2}t'+i\phi(\kk)} \right]\sum_{m} e^{im\Omega\left(t'-k_x-k_y\right)}
\end{aligned}
\end{equation}
Injecting the definition of $s_{in}$ in these equations and doing the change of variable $t''=t-t'$, they become:
\begin{equation}
\begin{aligned}
    u_{\kk,+} &=-i\sqrt{\kappa}\left[U\cos{\left(\frac{\theta_{\kk}}{2}\right)}C^{+}_{1}+V\sin{\left(\frac{\theta_{\kk}}{2}\right)}C^{+}_{2}\right]\\
    u_{\kk,-} &=-i\sqrt{\kappa}\left[U\sin{\left(\frac{\theta_{\kk}}{2}\right)}C^{-}_{1}-V\cos{\left(\frac{\theta_{\kk}}{2}\right)}C^{-}_{2}\right]
\end{aligned}
\end{equation}
where we define
\begin{equation}
\begin{aligned}
    C_{1}^{(\pm)}(t) &= Fe^{-i\omega_{L}t}e^{-i\frac{\delta}{2}t} \sum_{m}\left[e^{im\Omega(t-k_x-k_y)} \int_{0}^{t} dt''e^{-\frac{\gamma}{2}t''} e^{i(\omega_{L} - (m\Omega -\frac{\delta}{2}+\epsilon_{\kk}^{\pm}))t''}
    \right]\\
    C_{2}^{(\pm)}(t) &= Fe^{-i\omega_{L}t} e^{+i\frac{\delta}{2}t} e^{+i\phi(\kk)} \sum_{m}\left[e^{im\Omega(t-k_x-k_y)} \int_{0}^{t} dt''e^{-\frac{\gamma}{2}t''} e^{i(\omega_{L} - (m\Omega +\frac{\delta}{2}+\epsilon_{\kk}^{\pm}))t''}
    \right].
\end{aligned}
\end{equation}
Under the assumption that the cavity's lifetime is much smaller than the time over which we integrate ($\gamma^{-1} \ll t$) - which is reasonable as we wait $\SI{3}{\micro\second}$ before each experiment for the field to reach steady-state - these integrals are solved to : 
\begin{equation}
\begin{aligned}
    C_{1}^{(\pm)}(t) &= -Fe^{-i\omega_{L}t}e^{-i\frac{\delta}{2}t} \sum_{m}\left[e^{im\Omega(t-k_x-k_y)} \frac{1}{i(\omega_{L} - (m\Omega -\frac{\delta}{2}+\epsilon_{\kk}^{\pm})) - \frac{\gamma}{2}}
    \right] \\
    C_{2}^{(\pm)}(t) &= -Fe^{-i\omega_{L}t}e^{+i\frac{\delta}{2}t} e^{i\phi(\kk)} \sum_{m}\left[e^{im\Omega(t-k_x-k_y)} \frac{1}{i(\omega_{L} - (m\Omega +\frac{\delta}{2}+\epsilon_{\kk}^{\pm})) - \frac{\gamma}{2}}
    \right].
\end{aligned}
\end{equation}
These functions therefore describe Lorentz distributions with evident resonance peaks happening at frequencies $\omega_{L}=m\Omega-\frac{\delta}{2}+\epsilon_{\kk}^{\pm}$ and $\omega_{L}=m\Omega+\frac{\delta}{2}+\epsilon_{\kk}^{\pm}$ with linewidth $\frac{\gamma}{2}$.

Let us now consider the transmitted field amplitude at the photodiode assuming detection of the CCW modes only:
\begin{equation}
\begin{aligned}
    s_{pd} &= s_{in} - i\sqrt{\kappa}\sum_{m}{d_{m,\uparrow}}\\
    &= s_{in} -i\sqrt{\kappa}\sum_m \left[U\Tilde{a}_{m}e^{-i(m\Omega -\frac{\delta}{2})t} +V \Tilde{b}_{m} e^{-i(m\Omega +\frac{\delta}{2})t} \right]\\
    &= s_{in} -i\sqrt{\kappa}
           \sum_m \int_0^T \frac{dk_x}{T}\int_0^{T'} \frac{dk_y}{T'}
           \left[Ua_{\kk} e^{-i(m\Omega -\frac{\delta}{2})t} e^{i(k_x+k_y)m\Omega} + Vb_{\kk} e^{-i(m\Omega +\frac{\delta}{2})t} e^{i(k_x+k_y)m\Omega} \right]\\
    &= s_{in}-i\sqrt{\kappa}
           \sum_m \int_0^T \frac{dk_x}{T}\int_0^{T'} \frac{dk_y}{T'}\left[U\left(\cos{\left(\frac{\theta_{\kk}}{2}\right)}u_{\kk,+}+\sin{\left(\frac{\theta_{\kk}}{2}\right)}u_{\kk,-}\right)e^{-i(m\Omega -\frac{\delta}{2})t} e^{i(k_x+k_y)m\Omega}\right]\\
           &-i\sqrt{\kappa}\sum_m \int_0^T \frac{dk_x}{T}\int_0^{T'} \frac{dk_y}{T'}\left[V\left(\sin{\left(\frac{\theta_{\kk}}{2}\right)}u_{\kk,+}-\cos{\left(\frac{\theta_{\kk}}{2}\right)}u_{\kk,-}\right) e^{-i(m\Omega +\frac{\delta}{2})t} e^{i(k_x+k_y)m\Omega} \right].
\end{aligned}
\end{equation}
This expression can be considerably simplified under the following approximations: first, we pose that the drive input field is resonant with the $\bar{m}^{th}$ symmetric mode, i.e $\omega_{L}\approx \bar{m}\Omega-\frac{\delta}{2}$. Because we are in a regime where $\gamma\ll\delta$, this leads to only the $C_{1}^{\pm}$ terms contributing significantly in $u_{\kk,\pm}$, letting us neglect $C_{2}^{\pm}$(which would correspond to a drive resonant with the $\bar{m}^{th}$ anti-symmetric mode). Furthermore, every $m\neq\bar{m}$ in the relations of the field amplitudes can be neglected. This gives rise to the simplified expressions :
\begin{equation}
\begin{aligned}
    u_{\kk,+}(t) &= - i\frac{\sqrt{\kappa}U\cos{\left(\frac{\theta_{\kk}}{2}\right)}F e^{-i\omega_{L}t}e^{-i(k_x+k_y)\bar{m}\Omega}e^{i(\Bar{m}\Omega -\frac{\delta}{2}) t}e^{i\phi(\kk)}}{\gamma/2 -i(\Delta\omega_{L}-\epsilon_{\kk}^{+}))}\\
     u_{\kk,-}(t) &= - i\frac{\sqrt{\kappa}U\sin{\left(\frac{\theta_{\kk}}{2}\right)}F e^{-i\omega_{L}t}e^{-i(k_x+k_y)\bar{m}\Omega}e^{i(\Bar{m}\Omega -\frac{\delta}{2}) t}e^{i\phi(\kk)}}{\gamma/2 -i(\Delta\omega_{L}-\epsilon_{\kk}^{-}))}.
\end{aligned}
\end{equation}
where $\Delta\omega_{L}=\omega_{L}-(\bar{m}\Omega-\frac{\delta}{2})$. Second, we consider that the driving field is resonant with either the upper band or the lower band at a given time, but not with both. This is reasonable for most cases as we are usually in a regime where $\gamma\ll\abs{\epsilon_{\kk}^{\pm}}$; thus, the condition $\abs{\Delta\omega_{L}-\epsilon_{\kk}^{\pm}}\ll\abs{\epsilon_{\kk}^{\pm}}$ is only true for one band at a time. This enables us to neglect the contribution of the lower/upper band when considering the amplitude on the photodiodes of the upper/lower band independently. This approximation is however less accurate for $\bold{k}$-points where the linewidth is comparable to the size of the gap, and even less so at the positions of the Dirac points where $\epsilon_{\kk}^{+}=\epsilon_{\kk}^{-}$. This causes small deviations in the value of the phases extracted at these points.

Under these two approximations, the expressions for the transmitted amplitude becomes:
\begin{equation}
\begin{aligned}
    s^{+}_{pd} &= S_{in{}} - i\sqrt{\kappa}\sum_{m}\int_0^T \frac{dk_x}{T}\int_0^{T'} \frac{dk_y}{T'}e^{i(k_x+k_y)m\Omega}\left[U\cos{\left(\frac{\theta_{\kk}}{2}\right)}e^{-i\left(m\Omega-\frac{\delta}{2}\right)t}e^{-i\phi_{\kk}}+V\sin{\left(\frac{\theta_{\kk}}{2}\right)}e^{-i\left(m\Omega+\frac{\delta}{2}\right)t}\right]u_{\kk,+}\\
    s^{-}_{pd} &= S_{in{}} - i\sqrt{\kappa}\sum_{m}\int_0^T \frac{dk_x}{T}\int_0^{T'} \frac{dk_y}{T'}e^{i(k_x+k_y)m\Omega}\left[U\sin{\left(\frac{\theta_{\kk}}{2}\right)}e^{-i\left(m\Omega-\frac{\delta}{2}\right)t}e^{-i\phi_{\kk}}-V\cos{\left(\frac{\theta_{\kk}}{2}\right)}e^{-i\left(m\Omega+\frac{\delta}{2}\right)t}\right]u_{\kk,-}.
\end{aligned}
\end{equation}
We now inject the simplified expressions for $u_{\kk,\pm}$ in the above relations so that:
\begin{equation}
\begin{aligned}
    s^{+}_{pd} = Fe^{-i\omega_{L}t}\left[1-\kappa\sum_{m'}\int_0^T \frac{dk_x}{T}\int_0^{T'} \frac{dk_y}{T'}\frac{e^{im'\Omega\left(t-k_x-k_y\right)}}{\frac{\gamma}{2}-i\left(\Delta\omega_{L}-\epsilon_{\kk}^{+}\right)}\left(U^{2}\cos^{2}{\left(\frac{\theta_{\kk}}{2}\right)}+\frac{UV}{2}\sin{\left(\theta_{\kk}\right)}e^{-i\delta t}e^{i\phi_\kk}\right)\right]\\
    s^{-}_{pd} = Fe^{-i\omega_{L}t}\left[1-\kappa\sum_{m'}\int_0^T \frac{dk_x}{T}\int_0^{T'} \frac{dk_y}{T'}\frac{e^{im'\Omega\left(t-k_x-k_y\right)}}{\frac{\gamma}{2}-i\left(\Delta\omega_{L}-\epsilon_{\kk}^{-}\right)}\left(U^{2}\sin^{2}{\left(\frac{\theta_{\kk}}{2}\right)}-\frac{UV}{2}\sin{\left(\theta_{\kk}\right)}e^{-i\delta t}e^{i\phi_\kk}\right)\right]
\end{aligned}
\end{equation}
where we have defined $m'=\bar{m}-m$. Now, we note that $\Omega\equiv\frac{2\pi}{T}$ by definition. Hence, we can use the identity 
\begin{equation}
    D_{T} \left(t-a\right) = \frac{1}{T}\sum_{m}e^{i\left(\frac{2\pi}{T}\right)m\left(t-a\right)}
\end{equation}
with $D_{T}$ being a Dirac comb of period $T$ that is defined as
\begin{equation}
    D_{T} \left(t\right) = \sum_{n=-\infty}^{\infty}\delta\left(t-nT\right)
\end{equation}
to further simplify the expressions for the transmitted amplitudes.

Integrating over $\kk$ is then very straightforward and yields:
\begin{equation}
\begin{aligned}
    s^{+}_{pd} &= Fe^{-i\omega_{L}t}\left.\left[1-\kappa\frac{U^{2}\cos^{2}{\left(\frac{\theta_{\kk}}{2}\right)}+\frac{UV}{2}\sin{\left(\theta_{\kk}\right)}e^{-i\delta t}e^{i\phi_\kk}}{\frac{\gamma}{2}-i\left(\Delta\omega_{L}-\epsilon_{\kk}^{+}\right)} \right]\right\rvert_{(k_x,k_y)= (t\,\mathrm{mod}\,T,t\,\mathrm{mod}\,T')}\\
    s^{-}_{pd} &= Fe^{-i\omega_{L}t}\left.\left[1-\kappa\frac{U^{2}\sin^{2}{\left(\frac{\theta_{\kk}}{2}\right)}-\frac{UV}{2}\sin{\left(\theta_{\kk}\right)}e^{-i\delta t}e^{i\phi_\kk}}{\frac{\gamma}{2}-i\left(\Delta\omega_{L}-\epsilon_{\kk}^{-}\right)}\right]\right\rvert_{(k_x,k_y)= (t\,\mathrm{mod}\,T,t\,\mathrm{mod}\,T')}.
\end{aligned}
\end{equation}
The transmitted intensity measured by the photodiode is therefore given by:
\begin{equation}
\begin{aligned}
    I^{+}_{pd} &= \abs{s^{+}_{pd}}^{2}\approx\abs{F}^{2}\left.\left[1-2\kappa\Re{\frac{U^{2}\cos^{2}{\left(\frac{\theta_{\kk}}{2}\right)}+\frac{UV}{2}\sin{\left(\theta_{\kk}\right)}e^{-i\delta t}e^{i\phi_\kk}}{\frac{\gamma}{2}-i\left(\Delta\omega_{L}-\epsilon_{\kk}^{+}\right)}} \right\rvert_{(k_x,k_y)= (t\,\mathrm{mod}\,T,t\,\mathrm{mod}\,T')} + \mathcal{O}\left(\left(\frac{\kappa}{\gamma}\right)^{2}\right)\right]\\
    I^{-}_{pd} &= \abs{s^{-}_{pd}}^{2}\approx\abs{F}^{2}\left.\left[1-2\kappa\Re{\frac{U^{2}\sin^{2}{\left(\frac{\theta_{\kk}}{2}\right)}-\frac{UV}{2}\sin{\left(\theta_{\kk}\right)}e^{-i\delta t}e^{i\phi_\kk}}{\frac{\gamma}{2}-i\left(\Delta\omega_{L}-\epsilon_{\kk}^{-}\right)}} \right\rvert_{(k_x,k_y)= (t\,\mathrm{mod}\,T,t\,\mathrm{mod}\,T')} + \mathcal{O}\left(\left(\frac{\kappa}{\gamma}\right)^{2}\right)\right]
\end{aligned}
\end{equation}
where we neglect the quadratic term as we work in an under-coupling regime with $\gamma>\kappa$. 
As the wavevector is related to the time variable via the relations indicated in the subscript, the transmitted intensity exhibits dips at those times at which the laser is resonant with the eigenmode at the associated effective momentum.


\subsection{Anomalous transverse displacement and Chern number extraction}
Our strategy to measure the photonic analogue of the transverse anomalous Hall current and then extract the Chern number of the topological bands is based on recent theoretical works~\cite{ozawa_anomalous_2014,ozawa_steady-state_2018}. In these works, a link between the Berry curvature and the displacement of the light intensity profile under the effect of a synthetic electric field, i.e. a potential gradient in the photonic lattice, was put forward for driven-dissipative systems, generalizing to a non-equilibrium context well-known results of semi-classical electronic transport~\cite{xiao_berry_2010}. In particular, our experiment corresponds to the anomalous Hall regime of~\cite{ozawa_anomalous_2014} where the linewidth of the photonic modes is much smaller than the overall width of the energy dispersion $\epsilon_{\kk}$ of the photonic band and smaller than the band gaps.

For lattices with equivalent sublattices, one has the relation 
\begin{equation}
\delta r_x \simeq \frac{\int d^2\mathbf{k}\, \Omega(\kk)\,n(\mathbf{k})^2}{\int d^2\mathbf{k}\, n(\mathbf{k})} \frac{\gamma\,\lambda}{2}\, \simeq \frac{\bar{\Omega}(\omega_L)\,\lambda}{\gamma}
\end{equation}
for a small force $\lambda$ along the $y$ direction. Here, $\delta \mathbf{r}$ is the spatial shift of the center-of-mass of the intensity distribution
\begin{equation}
\delta \mathbf{r} = \frac{\sum_{\mathbf{r}} {\mathbf{r}} \, I^{(\lambda)}_{\mathbf{r}}}{\sum_{\mathbf{r}} I^{(\lambda)}_{\mathbf{r}}}-\frac{\sum_{\mathbf{r}} {\mathbf{r}}\, I^{(\lambda=0)}_{\mathbf{r}} }{\sum_{\mathbf{r}} I^{(\lambda=0)}_{\mathbf{r}}}\,.
\end{equation}
Note that in this formula, the position $\mathbf{r}$ runs over the unit cells and, for each value of the applied force $\lambda$, $I^\lambda_{\mathbf{r}}=\sum_l I^\lambda_{\mathbf{r},l}$ is the total intensity in the unit cell, obtained by summing over the sub-sites $l$ as if these were located at the same position $\mathbf{r}$. 
Then, $n(\mathbf{k})=[(\epsilon_{\kk}-\omega_L)^2+(\gamma/2)^2]^{-1}$ is the population of the different $\mathbf{k}$-space modes and $\bar{\Omega}(\omega_L)$ is the average value of the Berry curvature over all resonant modes $\epsilon_{\kk}=\omega_L$. The coherent field at frequency $\omega_L$ is assumed to drive a single site of a single unit cell, so to drive in an effectively uniform way to states all $\kk$ points within the Brillouin zone. For the assumed small value of the linewidth $\gamma$ of the photonic modes, only a single band is effectively excited by the drive and the other bands are far enough in energy to be neglected.

If the photonic lattice has two non-equivalent sites per unit cell, a slightly more sophisticated approach must be used, first derived in~\cite{ozawa_steady-state_2018}.
An averaged displacement is defined here as
\begin{equation}
\delta \rr = \frac{\sum_m \sum_\mathbf{r} \mathbf{r}\, I_{\mathbf{r}}^{(m,\lambda)}}{\sum_m \sum_r I_{\mathbf{r}}^{(m,\lambda)}} - \frac{\sum_m \sum_r\mathbf{r}\, I_{\mathbf{r}}^{(m,\lambda=0)}}{\sum_m \sum_r I_{\mathbf{r}}^{(m,\lambda=0)}}
\label{eq:delta_r_noneq}
\end{equation}
where $I_{\mathbf{r}}^{(m,\lambda)}$ refers again to the total intensity within a unit cell and the index $m$ corresponds to the specific (single) sub-site on which the drive is located.
This quantity is then related to the geometry of the Bloch bands by the generalized formula
\begin{equation}
\delta r_i=\frac{1}{ \int d^2\mathbf{k}\,n(\mathbf{k})}\,\sum_j \left\{ -\frac{\gamma}{2}\,\int d^2\mathbf{k}\,n(\mathbf{k})^2 \Omega_{ij}(\mathbf{k})
+
\int d^2\mathbf{k}\,(\omega_L-\epsilon_{\mathbf{k}})\,[2 n(\kk)^2\,g_{ij}(\mathbf{k})+2n(\mathbf{k})^3\, \partial_i \epsilon_{\mathbf{k}}\,\partial_j \epsilon_{\mathbf{k}}]  \right\} \, E_j
\label{eq:deltax1}
\end{equation}
where $i,j=\{x,y\}$ run over the Cartesian directions, $\Omega_{ij}=\varepsilon_{ij}\,\Omega(\kk)$ is the (antisymmetric) Berry curvature tensor, $g_{ij}$ is the (symmetric) quantum metric tensor. The other terms present in Eq.~(23) of the SM of~\cite{ozawa_steady-state_2018} vanish in our case of a wide gap between the bands compared to their linewidth and of a pump localized on a single site, which drives in an effectively uniform way all $\kk$ states within the Brillouin zone.
Remarkably, in the absence of any staggered on-site potential $\Delta=0$, the contributions of the different sub-lattices $m$ to \eqref{eq:delta_r_noneq} are equal within linear response theory for a given $\phi_h$, which allows to only evaluate the quantities for a single $m$ (this can be seen by acting with a $\pi$-rotation of the whole system).




The main goal of our work is the experimental measurement of the Chern number. To extract this quantity, one focuses on the transverse response $\delta r_x$ under a force $E_y$ and then integrate the Berry curvature across the Brillouin zone. This can be performed by repeating the measurement of the transverse displacement for different values of the driving laser frequency $\omega_L$ and summing up the results. One key hurdle in doing so is that Eq.~(\ref{eq:deltax1}) includes non-Berry contributions to the anomalous transport, notably those associated to the quantum metric $g_{ij}(\textbf{k})$ and the bands' dispersion ($\partial_{i}\epsilon_{\textbf{k}}\partial_{j}\epsilon_{\textbf{k}}$). These spurious contributions may disturb our extraction of the Chern number that is solely dependent on the bands' Berry curvature. In theory, the integral over $\omega_L$ is sufficient to remove the non-Berry terms that average out to zero over the integral. Since these latter may be quantitatively large, this procedure is prone to experimental error: in the next section, we show how in the experiment we managed to suppress all non-Berry contributions, notably through subtracting the system's response for topological phases with opposite Chern number.   



Using these techniques, we are only left with the first integral in Eq.~(\ref{eq:deltax1}). In practice, we call $\delta x_{\omega_L}$ the displacement experimentally measured for a coherent drive at $\omega_L$ and a force along $y$ of strength $\lambda$, and correspondingly $n(\mathbf{k},\omega_L)$ the resonant population of the mode at $\mathbf{k}$. From \eqref{eq:deltax1}, we then have
\begin{equation}
\delta x_{\omega_L}\, \int d^2\mathbf{k}\,n(\mathbf{k},\omega_L)= -\frac{\lambda \gamma}{2}\int d^2\mathbf{k}\,\Omega(\mathbf{k})\,n(\mathbf{k},\omega_0)^2\,.
\end{equation}
Integrating both sides over $\omega_L$ within the band of interest, we obtain
\begin{multline}
\int_{\rm band}\! d\omega_L\, \delta x_{\omega_L}\,\int d^2\mathbf{k}\,n(\mathbf{k},\omega_L)=-\frac{\lambda \gamma}{2}\int_{\rm band}\!  d\omega_L \int d^2\mathbf{k}\,\Omega(\mathbf{k})\,n(\mathbf{k},\omega_L)^2=\\ =-\frac{\lambda \gamma}{2} \int d^2\mathbf{k}\,\Omega(\mathbf{k})\,\int_{\rm band}\!  d\omega_L\,n(\mathbf{k},\omega_L)^2=-\frac{\lambda \gamma}{2} \int d^2\mathbf{k}\,\Omega(\mathbf{k})\, \frac{4\pi}{\gamma^3} = -\frac{4\pi^2\lambda}{\gamma^2}\,\mathcal{C}
\end{multline}
where on the right-hand side we have exchanged the order of the integrals and, then, analytically performed the one of the squared resonant function over $\omega_L$: the remaining integral over $\mathbf{k}$ is the definition of Chern number $\mathcal{C}$.

This formula provides a straightforward way to extract the Chern number from a suitable average of the experimentally observed displacement $\delta x_{\omega_0}$ weighted by a factor $n(\mathbf{k},\omega_L)=[(\epsilon_\kk-\omega_L)^2+(\gamma/2)^2]^{-1}$ which is also experimentally extracted from the measured band dispersion,
\begin{equation}
\mathcal{C}=-\frac{\gamma^2}{4\pi^2\,\lambda}\,\int_{\rm band}\!d\omega_L\,\delta x_{\omega_L}\,\int \!d^2\kk\,n(\kk,\omega_L)\,.
\end{equation}
The values of the Chern number indicated in Fig.~4 of the main text are obtained using this procedure.

\section{Numerical simulation of the anomalous displacement}
In the context of photonic systems, probing an anomalous transport requires coherently driving the lattice with a localized pump and probing how the spatial distribution of the field intensity in the lattice varies in response to the synthetic electric field. Upon integrating the response for various detunings of the pump (as described in the previous section), we can extract the value of the Chern number. As discussed in the main text and in the previous Section, the main challenge in doing so, for driven-dissipative systems, is that there are many non-topological contributions to the anomalous transport that may disturb an accurate extraction of the Chern number. In this Section, we present several numerical simulations of our systems that illustrate the effect of such non-topological contributions and show how the technique we implemented in the experiment indeed allows canceling them out. More specifically, we will show how implementing an oscillating force and subtracting the spatial profiles obtained for opposite topological phases allow canceling spurious contributions to the transverse displacement.

We performed various simulations comparing the effects of having no field, a temporally constant field, and a temporally modulated one. In all cases, these simulations were done by numerically solving the driven-dissipative coupled-mode equations in Eq.~(\ref{eq:Langevin}), where we added a term $\lambda_{row}$ to the main diagonal of the Haldane-like Hamiltonian to mimic the effect of an electric potential; this term is constant with respect to all the frequency sites corresponding to the same lattice row, and grows linearly between different rows (see Fig.~4(a) of main text). For the modulated field simulations, this term was replaced by a time-varying term $\lambda_{row}\sin(\Omega_{\lambda}t)$. 
For the no-field and constant-field cases, we calculated the time-independent steady state of the system, i.e. solving Eqs.~(\ref{eq:Langevin}) for $\dot{d}_i = 0$. For each value of the laser detuning, we then extract $\langle\Delta x\rangle$ by using Eq.~(20) of Appendix C. The simulations for the modulated potential required a bit more care as the Hamiltonian now has a time-dependent contribution coming from the modulation of the field. They were done by first calculating the steady state of the system with $\lambda_{row}=0$ to define a zero-field initial state. Then, we introduced $\lambda_{row}\neq0$ in the Hamiltonian and iteratively propagated in time the zero-field initial state using the differential equations in Eq.~(S.28). This allowed extracting the time evolution of the 
field amplitudes in the different modes $d_{m,\uparrow/\downarrow}(t)$ and, from these, the time-dependent anomalous displacement. 
\begin{figure}
    \centering
    \includegraphics[width=\linewidth]{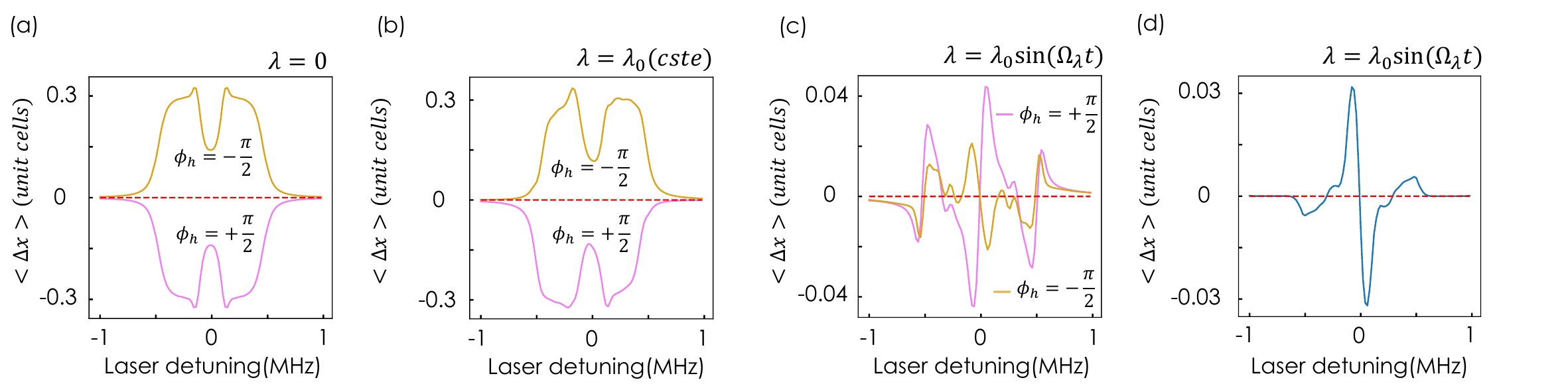}
    \caption{Simulated displacements obtained when using no synthetic electric field (a), a constant field (b), and a modulated field (c) in the simulation for both topological phases. Driving is on the $\ket{a}$ mode. The value for the amplitude of the field was chosen to be $\lambda_0/2\pi=\SI{10}{\kilo\hertz}$ to match the experimental value. We see that using a modulated electric field completely washes out the zero-field contribution that dominates the transverse displacement in (a) and (b) and enables us to retrieve a displacement which is opposite between the two bands. We however see from the non-equivalent drifts in (c) that non-Berry terms are still present. (d) Taking half the difference between the displacement curves of (c) allows us to recover the topologically-induced anomalous transverse displacement.}
    \label{fig:modulated_field}
\end{figure}

First, when doing simulations with a constant synthetic electric field, while driving a single $\ket{a}$ mode (see Fig. \ref{fig:modulated_field} (b) for simulations with $\phi_h = \pm\pi/2$), we found that the dominating contribution to the transverse drift is a large displacement going in the same direction for both bands but in opposite directions when changing the topological phase. This contribution dominates the quantized anomalous transverse displacements that we are interested in probing. Furthermore, this contribution does not depend on the field's orientation or amplitude. Indeed, when doing simulations with no external field (Fig \ref{fig:modulated_field} (a)), it remains unchanged. This large zero-field contribution can be suppressed by modulating the field at frequency $\Omega_{\lambda}$ and then demodulating the time-dependent displacements at this specific frequency. By doing so, all DC contributions that are not related to the field cancel out and this can be clearly seen in Fig. \ref{fig:modulated_field} (c) that depicts such a demodulation procedure. This shows the importance of using a modulated field to remove such DC contributions.
However, Fig.~\ref{fig:modulated_field} (c) includes all contributions to the anomalous transport in Eq.~\ref{eq:deltax1}: both the Berry and non-Berry contributions. To further cancel out these latter contributions, we must subtract the system's response upon changing the topological phase from $\phi_h = +\pi/2$ to $\phi_h = -\pi/2$: when subtracting these last two curves, as only the Berry contributions change sign when reversing the sign of $\phi_h$, these are the only one left.

\begin{figure}[h]
    \centering
    \includegraphics[width=\linewidth]{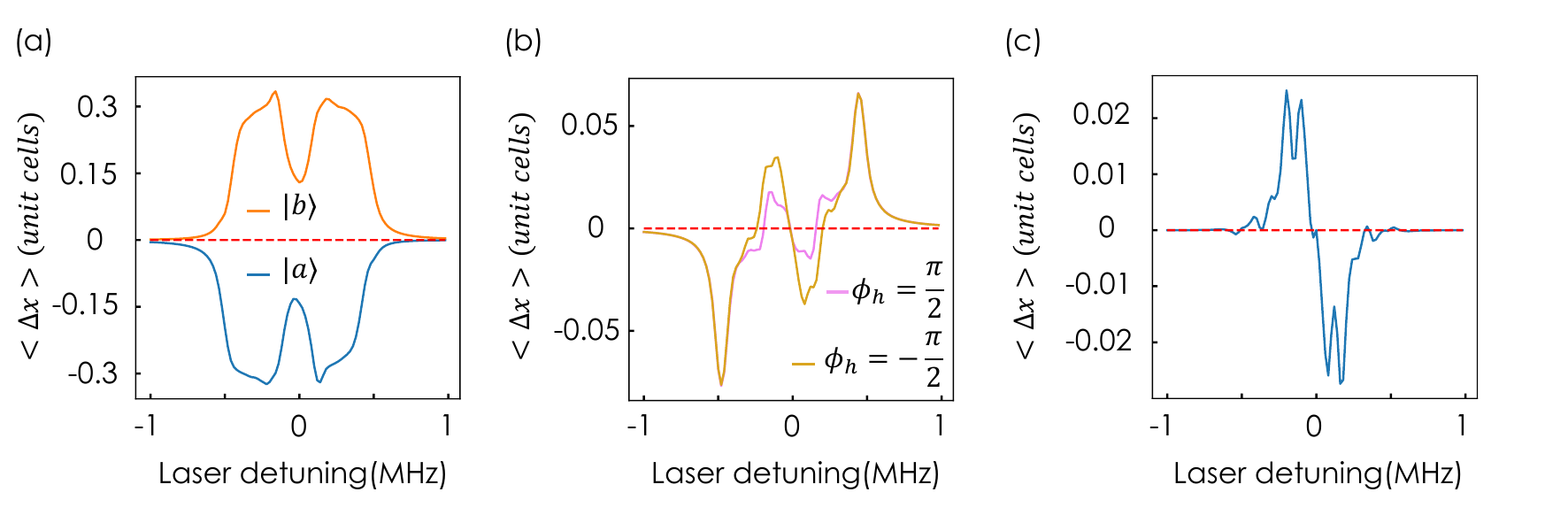}
    \caption{(a) Simulated displacement under a constant field: when the pump goes from being aligned to a $\ket{a}$ mode to a $\ket{b}$ mode, the direction of the zero-field contribution to the displacement flips sign. (b) When we average these two contributions together  for each of the two phases of $\phi_h = \frac{\pi}{2}$ and $\phi_h=-\frac{\pi}{2}$, we observe the displacements go in the same direction in both cases. This indicates that non-Berry terms dominate. This is further indicated by the small difference in drift amplitude between the two topological phases. (c) Taking half the difference between the displacement curves of (b) allows us to recover the topologically-induced anomalous transverse displacement.}
    \label{fig:nonBerry}
\end{figure}

As a final confirmation that the extraction of the Chern number is not affected by the temporal modulation of the field, a similar subtraction procedure can also be done on simulations with a constant field. However, such procedure would be impractical in a real experiment. Indeed, these very strong zero-field contributions are never exactly the same in consecutive experiments and don't necessarily completely cancel out. 
This is the reason why a modulated field was used in the experiment. On the other hand, this is not a difficulty for numerical simulations and such procedure is summarized in Fig.~\ref{fig:nonBerry}. In Panel (a), we see the displacement extracted for a drive localized on either the $\ket{a}$ or the $\ket{b}$ sub-lattice; both simulations are realized with a non-modulated field and $\phi_h = +\pi/2$. We observe a strong zero-field contribution in opposite directions as expected. Panel (b) shows the results when summing up these two curves, for $\phi_h=+\pi/2$ and $\phi_h=-\pi/2$. The sum effectively removes the zero-field term, so we are left, again, with only the Berry and non-Berry contributions to the displacement. Finally, Panel (c) exhibit the result when taking the half-difference of these last two curves. The remaining contribution arising from the Berry curvature is comparable to that obtained with a modulated field (Fig.~\ref{fig:modulated_field} (d)). This suggests that the temporal modulation of the field has a negligible effect on the quantization of the displacement.


\clearpage


\begin{thebibliography}{80}%
\makeatletter
\providecommand \@ifxundefined [1]{%
 \@ifx{#1\undefined}
}%
\providecommand \@ifnum [1]{%
 \ifnum #1\expandafter \@firstoftwo
 \else \expandafter \@secondoftwo
 \fi
}%
\providecommand \@ifx [1]{%
 \ifx #1\expandafter \@firstoftwo
 \else \expandafter \@secondoftwo
 \fi
}%
\providecommand \natexlab [1]{#1}%
\providecommand \enquote  [1]{``#1''}%
\providecommand \bibnamefont  [1]{#1}%
\providecommand \bibfnamefont [1]{#1}%
\providecommand \citenamefont [1]{#1}%
\providecommand \href@noop [0]{\@secondoftwo}%
\providecommand \href [0]{\begingroup \@sanitize@url \@href}%
\providecommand \@href[1]{\@@startlink{#1}\@@href}%
\providecommand \@@href[1]{\endgroup#1\@@endlink}%
\providecommand \@sanitize@url [0]{\catcode `\\12\catcode `\$12\catcode `\&12\catcode `\#12\catcode `\^12\catcode `\_12\catcode `\%12\relax}%
\providecommand \@@startlink[1]{}%
\providecommand \@@endlink[0]{}%
\providecommand \url  [0]{\begingroup\@sanitize@url \@url }%
\providecommand \@url [1]{\endgroup\@href {#1}{\urlprefix }}%
\providecommand \urlprefix  [0]{URL }%
\providecommand \Eprint [0]{\href }%
\providecommand \doibase [0]{https://doi.org/}%
\providecommand \selectlanguage [0]{\@gobble}%
\providecommand \bibinfo  [0]{\@secondoftwo}%
\providecommand \bibfield  [0]{\@secondoftwo}%
\providecommand \translation [1]{[#1]}%
\providecommand \BibitemOpen [0]{}%
\providecommand \bibitemStop [0]{}%
\providecommand \bibitemNoStop [0]{.\EOS\space}%
\providecommand \EOS [0]{\spacefactor3000\relax}%
\providecommand \BibitemShut  [1]{\csname bibitem#1\endcsname}%
\let\auto@bib@innerbib\@empty
\bibitem [{\citenamefont {Hasan}\ and\ \citenamefont {Kane}(2010)}]{hasan_colloquium_2010}%
  \BibitemOpen
  \bibfield  {author} {\bibinfo {author} {\bibfnamefont {M.~Z.}\ \bibnamefont {Hasan}}\ and\ \bibinfo {author} {\bibfnamefont {C.~L.}\ \bibnamefont {Kane}},\ }\bibfield  {title} {\bibinfo {title} {Colloquium: {Topological} insulators},\ }\href {https://doi.org/10.1103/RevModPhys.82.3045} {\bibfield  {journal} {\bibinfo  {journal} {Reviews of Modern Physics}\ }\textbf {\bibinfo {volume} {82}},\ \bibinfo {pages} {3045} (\bibinfo {year} {2010})},\ \bibinfo {note} {publisher: American Physical Society}\BibitemShut {NoStop}%
\bibitem [{\citenamefont {Qi}\ and\ \citenamefont {Zhang}(2011)}]{qi_topological_2011}%
  \BibitemOpen
  \bibfield  {author} {\bibinfo {author} {\bibfnamefont {X.-L.}\ \bibnamefont {Qi}}\ and\ \bibinfo {author} {\bibfnamefont {S.-C.}\ \bibnamefont {Zhang}},\ }\bibfield  {title} {\bibinfo {title} {Topological insulators and superconductors},\ }\href {https://doi.org/10.1103/RevModPhys.83.1057} {\bibfield  {journal} {\bibinfo  {journal} {Reviews of Modern Physics}\ }\textbf {\bibinfo {volume} {83}},\ \bibinfo {pages} {1057} (\bibinfo {year} {2011})},\ \bibinfo {note} {publisher: American Physical Society}\BibitemShut {NoStop}%
\bibitem [{\citenamefont {Thouless}\ \emph {et~al.}(1982)\citenamefont {Thouless}, \citenamefont {Kohmoto}, \citenamefont {Nightingale},\ and\ \citenamefont {den Nijs}}]{thouless_quantized_1982}%
  \BibitemOpen
  \bibfield  {author} {\bibinfo {author} {\bibfnamefont {D.~J.}\ \bibnamefont {Thouless}}, \bibinfo {author} {\bibfnamefont {M.}~\bibnamefont {Kohmoto}}, \bibinfo {author} {\bibfnamefont {M.~P.}\ \bibnamefont {Nightingale}},\ and\ \bibinfo {author} {\bibfnamefont {M.}~\bibnamefont {den Nijs}},\ }\bibfield  {title} {\bibinfo {title} {Quantized {Hall} {Conductance} in a {Two}-{Dimensional} {Periodic} {Potential}},\ }\href {https://doi.org/10.1103/PhysRevLett.49.405} {\bibfield  {journal} {\bibinfo  {journal} {Physical Review Letters}\ }\textbf {\bibinfo {volume} {49}},\ \bibinfo {pages} {405} (\bibinfo {year} {1982})},\ \bibinfo {note} {publisher: American Physical Society}\BibitemShut {NoStop}%
\bibitem [{\citenamefont {Prange}\ and\ \citenamefont {Girvin}(1990)}]{prange_quantum_1990}%
  \BibitemOpen
  \bibinfo {editor} {\bibfnamefont {R.~E.}\ \bibnamefont {Prange}}\ and\ \bibinfo {editor} {\bibfnamefont {S.~M.}\ \bibnamefont {Girvin}},\ eds.,\ \href@noop {} {\emph {\bibinfo {title} {The {Quantum} {Hall} effect}}},\ \bibinfo {edition} {2nd}\ ed.,\ Graduate texts in contemporary physics\ (\bibinfo  {publisher} {Springer-Verlag},\ \bibinfo {address} {New York},\ \bibinfo {year} {1990})\BibitemShut {NoStop}%
\bibitem [{\citenamefont {von Klitzing}(2019)}]{von_klitzing_essay_2019}%
  \BibitemOpen
  \bibfield  {author} {\bibinfo {author} {\bibfnamefont {K.}~\bibnamefont {von Klitzing}},\ }\bibfield  {title} {\bibinfo {title} {Essay: {Quantum} {Hall} {Effect} and the {New} {International} {System} of {Units}},\ }\href {https://doi.org/10.1103/PhysRevLett.122.200001} {\bibfield  {journal} {\bibinfo  {journal} {Physical Review Letters}\ }\textbf {\bibinfo {volume} {122}},\ \bibinfo {pages} {200001} (\bibinfo {year} {2019})},\ \bibinfo {note} {publisher: American Physical Society}\BibitemShut {NoStop}%
\bibitem [{\citenamefont {Nayak}\ \emph {et~al.}(2008)\citenamefont {Nayak}, \citenamefont {Simon}, \citenamefont {Stern}, \citenamefont {Freedman},\ and\ \citenamefont {Das~Sarma}}]{nayak_non-abelian_2008}%
  \BibitemOpen
  \bibfield  {author} {\bibinfo {author} {\bibfnamefont {C.}~\bibnamefont {Nayak}}, \bibinfo {author} {\bibfnamefont {S.~H.}\ \bibnamefont {Simon}}, \bibinfo {author} {\bibfnamefont {A.}~\bibnamefont {Stern}}, \bibinfo {author} {\bibfnamefont {M.}~\bibnamefont {Freedman}},\ and\ \bibinfo {author} {\bibfnamefont {S.}~\bibnamefont {Das~Sarma}},\ }\bibfield  {title} {\bibinfo {title} {Non-{Abelian} anyons and topological quantum computation},\ }\href {https://doi.org/10.1103/RevModPhys.80.1083} {\bibfield  {journal} {\bibinfo  {journal} {Reviews of Modern Physics}\ }\textbf {\bibinfo {volume} {80}},\ \bibinfo {pages} {1083} (\bibinfo {year} {2008})},\ \bibinfo {note} {publisher: American Physical Society}\BibitemShut {NoStop}%
\bibitem [{\citenamefont {Zangeneh-Nejad}\ \emph {et~al.}(2021)\citenamefont {Zangeneh-Nejad}, \citenamefont {Sounas}, \citenamefont {Alù},\ and\ \citenamefont {Fleury}}]{zangeneh-nejad_analogue_2021}%
  \BibitemOpen
  \bibfield  {author} {\bibinfo {author} {\bibfnamefont {F.}~\bibnamefont {Zangeneh-Nejad}}, \bibinfo {author} {\bibfnamefont {D.~L.}\ \bibnamefont {Sounas}}, \bibinfo {author} {\bibfnamefont {A.}~\bibnamefont {Alù}},\ and\ \bibinfo {author} {\bibfnamefont {R.}~\bibnamefont {Fleury}},\ }\bibfield  {title} {\bibinfo {title} {Analogue computing with metamaterials},\ }\href {https://doi.org/10.1038/s41578-020-00243-2} {\bibfield  {journal} {\bibinfo  {journal} {Nature Reviews Materials}\ }\textbf {\bibinfo {volume} {6}},\ \bibinfo {pages} {207} (\bibinfo {year} {2021})},\ \bibinfo {note} {publisher: Nature Publishing Group}\BibitemShut {NoStop}%
\bibitem [{\citenamefont {Raghu}\ and\ \citenamefont {Haldane}(2008)}]{raghu_analogs_2008}%
  \BibitemOpen
  \bibfield  {author} {\bibinfo {author} {\bibfnamefont {S.}~\bibnamefont {Raghu}}\ and\ \bibinfo {author} {\bibfnamefont {F.~D.~M.}\ \bibnamefont {Haldane}},\ }\bibfield  {title} {\bibinfo {title} {Analogs of quantum-{Hall}-effect edge states in photonic crystals},\ }\href {https://doi.org/10.1103/PhysRevA.78.033834} {\bibfield  {journal} {\bibinfo  {journal} {Physical Review A}\ }\textbf {\bibinfo {volume} {78}},\ \bibinfo {pages} {033834} (\bibinfo {year} {2008})},\ \bibinfo {note} {publisher: American Physical Society}\BibitemShut {NoStop}%
\bibitem [{\citenamefont {Ozawa}\ and\ \citenamefont {Price}(2019)}]{ozawa_topological_2019-1}%
  \BibitemOpen
  \bibfield  {author} {\bibinfo {author} {\bibfnamefont {T.}~\bibnamefont {Ozawa}}\ and\ \bibinfo {author} {\bibfnamefont {H.~M.}\ \bibnamefont {Price}},\ }\bibfield  {title} {\bibinfo {title} {Topological quantum matter in synthetic dimensions},\ }\href {https://doi.org/10.1038/s42254-019-0045-3} {\bibfield  {journal} {\bibinfo  {journal} {Nature Reviews Physics}\ }\textbf {\bibinfo {volume} {1}},\ \bibinfo {pages} {349} (\bibinfo {year} {2019})},\ \bibinfo {note} {publisher: Nature Publishing Group}\BibitemShut {NoStop}%
\bibitem [{\citenamefont {Price}\ \emph {et~al.}(2022)\citenamefont {Price}, \citenamefont {Chong}, \citenamefont {Khanikaev}, \citenamefont {Schomerus}, \citenamefont {Maczewsky}, \citenamefont {Kremer}, \citenamefont {Heinrich}, \citenamefont {Szameit}, \citenamefont {Zilberberg}, \citenamefont {Yang}, \citenamefont {Zhang}, \citenamefont {Alù}, \citenamefont {Thomale}, \citenamefont {Carusotto}, \citenamefont {St-Jean}, \citenamefont {Amo}, \citenamefont {Dutt}, \citenamefont {Yuan}, \citenamefont {Fan}, \citenamefont {Yin}, \citenamefont {Peng}, \citenamefont {Ozawa},\ and\ \citenamefont {Blanco-Redondo}}]{price_roadmap_2022}%
  \BibitemOpen
  \bibfield  {author} {\bibinfo {author} {\bibfnamefont {H.}~\bibnamefont {Price}}, \bibinfo {author} {\bibfnamefont {Y.}~\bibnamefont {Chong}}, \bibinfo {author} {\bibfnamefont {A.}~\bibnamefont {Khanikaev}}, \bibinfo {author} {\bibfnamefont {H.}~\bibnamefont {Schomerus}}, \bibinfo {author} {\bibfnamefont {L.~J.}\ \bibnamefont {Maczewsky}}, \bibinfo {author} {\bibfnamefont {M.}~\bibnamefont {Kremer}}, \bibinfo {author} {\bibfnamefont {M.}~\bibnamefont {Heinrich}}, \bibinfo {author} {\bibfnamefont {A.}~\bibnamefont {Szameit}}, \bibinfo {author} {\bibfnamefont {O.}~\bibnamefont {Zilberberg}}, \bibinfo {author} {\bibfnamefont {Y.}~\bibnamefont {Yang}}, \bibinfo {author} {\bibfnamefont {B.}~\bibnamefont {Zhang}}, \bibinfo {author} {\bibfnamefont {A.}~\bibnamefont {Alù}}, \bibinfo {author} {\bibfnamefont {R.}~\bibnamefont {Thomale}}, \bibinfo {author} {\bibfnamefont {I.}~\bibnamefont {Carusotto}}, \bibinfo {author} {\bibfnamefont {P.}~\bibnamefont {St-Jean}}, \bibinfo {author} {\bibfnamefont {A.}~\bibnamefont
  {Amo}}, \bibinfo {author} {\bibfnamefont {A.}~\bibnamefont {Dutt}}, \bibinfo {author} {\bibfnamefont {L.}~\bibnamefont {Yuan}}, \bibinfo {author} {\bibfnamefont {S.}~\bibnamefont {Fan}}, \bibinfo {author} {\bibfnamefont {X.}~\bibnamefont {Yin}}, \bibinfo {author} {\bibfnamefont {C.}~\bibnamefont {Peng}}, \bibinfo {author} {\bibfnamefont {T.}~\bibnamefont {Ozawa}},\ and\ \bibinfo {author} {\bibfnamefont {A.}~\bibnamefont {Blanco-Redondo}},\ }\bibfield  {title} {\bibinfo {title} {Roadmap on topological photonics},\ }\href {https://doi.org/10.1088/2515-7647/ac4ee4} {\bibfield  {journal} {\bibinfo  {journal} {Journal of Physics: Photonics}\ }\textbf {\bibinfo {volume} {4}},\ \bibinfo {pages} {032501} (\bibinfo {year} {2022})},\ \bibinfo {note} {publisher: IOP Publishing}\BibitemShut {NoStop}%
\bibitem [{\citenamefont {Bahari}\ \emph {et~al.}(2017)\citenamefont {Bahari}, \citenamefont {Ndao}, \citenamefont {Vallini}, \citenamefont {El~Amili}, \citenamefont {Fainman},\ and\ \citenamefont {Kanté}}]{bahari_nonreciprocal_2017}%
  \BibitemOpen
  \bibfield  {author} {\bibinfo {author} {\bibfnamefont {B.}~\bibnamefont {Bahari}}, \bibinfo {author} {\bibfnamefont {A.}~\bibnamefont {Ndao}}, \bibinfo {author} {\bibfnamefont {F.}~\bibnamefont {Vallini}}, \bibinfo {author} {\bibfnamefont {A.}~\bibnamefont {El~Amili}}, \bibinfo {author} {\bibfnamefont {Y.}~\bibnamefont {Fainman}},\ and\ \bibinfo {author} {\bibfnamefont {B.}~\bibnamefont {Kanté}},\ }\bibfield  {title} {\bibinfo {title} {Nonreciprocal lasing in topological cavities of arbitrary geometries},\ }\href {https://doi.org/10.1126/science.aao4551} {\bibfield  {journal} {\bibinfo  {journal} {Science}\ }\textbf {\bibinfo {volume} {358}},\ \bibinfo {pages} {636} (\bibinfo {year} {2017})},\ \bibinfo {note} {publisher: American Association for the Advancement of Science}\BibitemShut {NoStop}%
\bibitem [{\citenamefont {Klembt}\ \emph {et~al.}(2018)\citenamefont {Klembt}, \citenamefont {Harder}, \citenamefont {Egorov}, \citenamefont {Winkler}, \citenamefont {Ge}, \citenamefont {Bandres}, \citenamefont {Emmerling}, \citenamefont {Worschech}, \citenamefont {Liew}, \citenamefont {Segev}, \citenamefont {Schneider},\ and\ \citenamefont {Höfling}}]{klembt_exciton-polariton_2018}%
  \BibitemOpen
  \bibfield  {author} {\bibinfo {author} {\bibfnamefont {S.}~\bibnamefont {Klembt}}, \bibinfo {author} {\bibfnamefont {T.~H.}\ \bibnamefont {Harder}}, \bibinfo {author} {\bibfnamefont {O.~A.}\ \bibnamefont {Egorov}}, \bibinfo {author} {\bibfnamefont {K.}~\bibnamefont {Winkler}}, \bibinfo {author} {\bibfnamefont {R.}~\bibnamefont {Ge}}, \bibinfo {author} {\bibfnamefont {M.~A.}\ \bibnamefont {Bandres}}, \bibinfo {author} {\bibfnamefont {M.}~\bibnamefont {Emmerling}}, \bibinfo {author} {\bibfnamefont {L.}~\bibnamefont {Worschech}}, \bibinfo {author} {\bibfnamefont {T.~C.~H.}\ \bibnamefont {Liew}}, \bibinfo {author} {\bibfnamefont {M.}~\bibnamefont {Segev}}, \bibinfo {author} {\bibfnamefont {C.}~\bibnamefont {Schneider}},\ and\ \bibinfo {author} {\bibfnamefont {S.}~\bibnamefont {Höfling}},\ }\bibfield  {title} {\bibinfo {title} {Exciton-polariton topological insulator},\ }\href {https://doi.org/10.1038/s41586-018-0601-5} {\bibfield  {journal} {\bibinfo  {journal} {Nature}\ }\textbf {\bibinfo {volume} {562}},\
  \bibinfo {pages} {552} (\bibinfo {year} {2018})},\ \bibinfo {note} {publisher: Nature Publishing Group}\BibitemShut {NoStop}%
\bibitem [{\citenamefont {Kraus}\ \emph {et~al.}(2012)\citenamefont {Kraus}, \citenamefont {Lahini}, \citenamefont {Ringel}, \citenamefont {Verbin},\ and\ \citenamefont {Zilberberg}}]{kraus_topological_2012}%
  \BibitemOpen
  \bibfield  {author} {\bibinfo {author} {\bibfnamefont {Y.~E.}\ \bibnamefont {Kraus}}, \bibinfo {author} {\bibfnamefont {Y.}~\bibnamefont {Lahini}}, \bibinfo {author} {\bibfnamefont {Z.}~\bibnamefont {Ringel}}, \bibinfo {author} {\bibfnamefont {M.}~\bibnamefont {Verbin}},\ and\ \bibinfo {author} {\bibfnamefont {O.}~\bibnamefont {Zilberberg}},\ }\bibfield  {title} {\bibinfo {title} {Topological {States} and {Adiabatic} {Pumping} in {Quasicrystals}},\ }\href {https://doi.org/10.1103/PhysRevLett.109.106402} {\bibfield  {journal} {\bibinfo  {journal} {Physical Review Letters}\ }\textbf {\bibinfo {volume} {109}},\ \bibinfo {pages} {106402} (\bibinfo {year} {2012})},\ \bibinfo {note} {publisher: American Physical Society}\BibitemShut {NoStop}%
\bibitem [{\citenamefont {Rechtsman}\ \emph {et~al.}(2013)\citenamefont {Rechtsman}, \citenamefont {Zeuner}, \citenamefont {Plotnik}, \citenamefont {Lumer}, \citenamefont {Podolsky}, \citenamefont {Dreisow}, \citenamefont {Nolte}, \citenamefont {Segev},\ and\ \citenamefont {Szameit}}]{rechtsman_photonic_2013}%
  \BibitemOpen
  \bibfield  {author} {\bibinfo {author} {\bibfnamefont {M.~C.}\ \bibnamefont {Rechtsman}}, \bibinfo {author} {\bibfnamefont {J.~M.}\ \bibnamefont {Zeuner}}, \bibinfo {author} {\bibfnamefont {Y.}~\bibnamefont {Plotnik}}, \bibinfo {author} {\bibfnamefont {Y.}~\bibnamefont {Lumer}}, \bibinfo {author} {\bibfnamefont {D.}~\bibnamefont {Podolsky}}, \bibinfo {author} {\bibfnamefont {F.}~\bibnamefont {Dreisow}}, \bibinfo {author} {\bibfnamefont {S.}~\bibnamefont {Nolte}}, \bibinfo {author} {\bibfnamefont {M.}~\bibnamefont {Segev}},\ and\ \bibinfo {author} {\bibfnamefont {A.}~\bibnamefont {Szameit}},\ }\bibfield  {title} {\bibinfo {title} {Photonic {Floquet} topological insulators},\ }\href {https://doi.org/10.1038/nature12066} {\bibfield  {journal} {\bibinfo  {journal} {Nature}\ }\textbf {\bibinfo {volume} {496}},\ \bibinfo {pages} {196} (\bibinfo {year} {2013})},\ \bibinfo {note} {publisher: Nature Publishing Group}\BibitemShut {NoStop}%
\bibitem [{\citenamefont {Hafezi}\ \emph {et~al.}(2013)\citenamefont {Hafezi}, \citenamefont {Mittal}, \citenamefont {Fan}, \citenamefont {Migdall},\ and\ \citenamefont {Taylor}}]{hafezi_imaging_2013}%
  \BibitemOpen
  \bibfield  {author} {\bibinfo {author} {\bibfnamefont {M.}~\bibnamefont {Hafezi}}, \bibinfo {author} {\bibfnamefont {S.}~\bibnamefont {Mittal}}, \bibinfo {author} {\bibfnamefont {J.}~\bibnamefont {Fan}}, \bibinfo {author} {\bibfnamefont {A.}~\bibnamefont {Migdall}},\ and\ \bibinfo {author} {\bibfnamefont {J.~M.}\ \bibnamefont {Taylor}},\ }\bibfield  {title} {\bibinfo {title} {Imaging topological edge states in silicon photonics},\ }\href {https://doi.org/10.1038/nphoton.2013.274} {\bibfield  {journal} {\bibinfo  {journal} {Nature Photonics}\ }\textbf {\bibinfo {volume} {7}},\ \bibinfo {pages} {1001} (\bibinfo {year} {2013})},\ \bibinfo {note} {publisher: Nature Publishing Group}\BibitemShut {NoStop}%
\bibitem [{\citenamefont {Wu}\ and\ \citenamefont {Hu}(2015)}]{wu_scheme_2015}%
  \BibitemOpen
  \bibfield  {author} {\bibinfo {author} {\bibfnamefont {L.-H.}\ \bibnamefont {Wu}}\ and\ \bibinfo {author} {\bibfnamefont {X.}~\bibnamefont {Hu}},\ }\bibfield  {title} {\bibinfo {title} {Scheme for {Achieving} a {Topological} {Photonic} {Crystal} by {Using} {Dielectric} {Material}},\ }\href {https://doi.org/10.1103/PhysRevLett.114.223901} {\bibfield  {journal} {\bibinfo  {journal} {Physical Review Letters}\ }\textbf {\bibinfo {volume} {114}},\ \bibinfo {pages} {223901} (\bibinfo {year} {2015})},\ \bibinfo {note} {publisher: American Physical Society}\BibitemShut {NoStop}%
\bibitem [{\citenamefont {Ma}\ and\ \citenamefont {Shvets}(2016)}]{ma_all-si_2016}%
  \BibitemOpen
  \bibfield  {author} {\bibinfo {author} {\bibfnamefont {T.}~\bibnamefont {Ma}}\ and\ \bibinfo {author} {\bibfnamefont {G.}~\bibnamefont {Shvets}},\ }\bibfield  {title} {\bibinfo {title} {All-{Si} valley-{Hall} photonic topological insulator},\ }\href {https://doi.org/10.1088/1367-2630/18/2/025012} {\bibfield  {journal} {\bibinfo  {journal} {New Journal of Physics}\ }\textbf {\bibinfo {volume} {18}},\ \bibinfo {pages} {025012} (\bibinfo {year} {2016})},\ \bibinfo {note} {publisher: IOP Publishing}\BibitemShut {NoStop}%
\bibitem [{\citenamefont {Noh}\ \emph {et~al.}(2018)\citenamefont {Noh}, \citenamefont {Huang}, \citenamefont {Chen},\ and\ \citenamefont {Rechtsman}}]{noh_observation_2018}%
  \BibitemOpen
  \bibfield  {author} {\bibinfo {author} {\bibfnamefont {J.}~\bibnamefont {Noh}}, \bibinfo {author} {\bibfnamefont {S.}~\bibnamefont {Huang}}, \bibinfo {author} {\bibfnamefont {K.~P.}\ \bibnamefont {Chen}},\ and\ \bibinfo {author} {\bibfnamefont {M.~C.}\ \bibnamefont {Rechtsman}},\ }\bibfield  {title} {\bibinfo {title} {Observation of {Photonic} {Topological} {Valley} {Hall} {Edge} {States}},\ }\href {https://doi.org/10.1103/PhysRevLett.120.063902} {\bibfield  {journal} {\bibinfo  {journal} {Physical Review Letters}\ }\textbf {\bibinfo {volume} {120}},\ \bibinfo {pages} {063902} (\bibinfo {year} {2018})},\ \bibinfo {note} {publisher: American Physical Society}\BibitemShut {NoStop}%
\bibitem [{\citenamefont {Rosiek}\ \emph {et~al.}(2023)\citenamefont {Rosiek}, \citenamefont {Arregui}, \citenamefont {Vladimirova}, \citenamefont {Albrechtsen}, \citenamefont {Vosoughi~Lahijani}, \citenamefont {Christiansen},\ and\ \citenamefont {Stobbe}}]{rosiek_observation_2023}%
  \BibitemOpen
  \bibfield  {author} {\bibinfo {author} {\bibfnamefont {C.~A.}\ \bibnamefont {Rosiek}}, \bibinfo {author} {\bibfnamefont {G.}~\bibnamefont {Arregui}}, \bibinfo {author} {\bibfnamefont {A.}~\bibnamefont {Vladimirova}}, \bibinfo {author} {\bibfnamefont {M.}~\bibnamefont {Albrechtsen}}, \bibinfo {author} {\bibfnamefont {B.}~\bibnamefont {Vosoughi~Lahijani}}, \bibinfo {author} {\bibfnamefont {R.~E.}\ \bibnamefont {Christiansen}},\ and\ \bibinfo {author} {\bibfnamefont {S.}~\bibnamefont {Stobbe}},\ }\bibfield  {title} {\bibinfo {title} {Observation of strong backscattering in valley-{Hall} photonic topological interface modes},\ }\href {https://doi.org/10.1038/s41566-023-01189-x} {\bibfield  {journal} {\bibinfo  {journal} {Nature Photonics}\ }\textbf {\bibinfo {volume} {17}},\ \bibinfo {pages} {386} (\bibinfo {year} {2023})},\ \bibinfo {note} {publisher: Nature Publishing Group}\BibitemShut {NoStop}%
\bibitem [{\citenamefont {Lu}\ \emph {et~al.}(2016)\citenamefont {Lu}, \citenamefont {Joannopoulos},\ and\ \citenamefont {Soljačić}}]{lu_topological_2016}%
  \BibitemOpen
  \bibfield  {author} {\bibinfo {author} {\bibfnamefont {L.}~\bibnamefont {Lu}}, \bibinfo {author} {\bibfnamefont {J.~D.}\ \bibnamefont {Joannopoulos}},\ and\ \bibinfo {author} {\bibfnamefont {M.}~\bibnamefont {Soljačić}},\ }\bibfield  {title} {\bibinfo {title} {Topological states in photonic systems},\ }\href {https://doi.org/10.1038/nphys3796} {\bibfield  {journal} {\bibinfo  {journal} {Nature Physics}\ }\textbf {\bibinfo {volume} {12}},\ \bibinfo {pages} {626} (\bibinfo {year} {2016})},\ \bibinfo {note} {publisher: Nature Publishing Group}\BibitemShut {NoStop}%
\bibitem [{\citenamefont {D’Errico}\ \emph {et~al.}(2020)\citenamefont {D’Errico}, \citenamefont {Cardano}, \citenamefont {Maffei}, \citenamefont {Dauphin}, \citenamefont {Barboza}, \citenamefont {Esposito}, \citenamefont {Piccirillo}, \citenamefont {Lewenstein}, \citenamefont {Massignan},\ and\ \citenamefont {Marrucci}}]{derrico_two-dimensional_2020}%
  \BibitemOpen
  \bibfield  {author} {\bibinfo {author} {\bibfnamefont {A.}~\bibnamefont {D’Errico}}, \bibinfo {author} {\bibfnamefont {F.}~\bibnamefont {Cardano}}, \bibinfo {author} {\bibfnamefont {M.}~\bibnamefont {Maffei}}, \bibinfo {author} {\bibfnamefont {A.}~\bibnamefont {Dauphin}}, \bibinfo {author} {\bibfnamefont {R.}~\bibnamefont {Barboza}}, \bibinfo {author} {\bibfnamefont {C.}~\bibnamefont {Esposito}}, \bibinfo {author} {\bibfnamefont {B.}~\bibnamefont {Piccirillo}}, \bibinfo {author} {\bibfnamefont {M.}~\bibnamefont {Lewenstein}}, \bibinfo {author} {\bibfnamefont {P.}~\bibnamefont {Massignan}},\ and\ \bibinfo {author} {\bibfnamefont {L.}~\bibnamefont {Marrucci}},\ }\bibfield  {title} {\bibinfo {title} {Two-dimensional topological quantum walks in the momentum space of structured light},\ }\href {https://doi.org/10.1364/OPTICA.365028} {\bibfield  {journal} {\bibinfo  {journal} {Optica}\ }\textbf {\bibinfo {volume} {7}},\ \bibinfo {pages} {108} (\bibinfo {year} {2020})},\ \bibinfo {note} {publisher: Optica
  Publishing Group}\BibitemShut {NoStop}%
\bibitem [{\citenamefont {Ozawa}\ and\ \citenamefont {Carusotto}(2014)}]{ozawa_anomalous_2014}%
  \BibitemOpen
  \bibfield  {author} {\bibinfo {author} {\bibfnamefont {T.}~\bibnamefont {Ozawa}}\ and\ \bibinfo {author} {\bibfnamefont {I.}~\bibnamefont {Carusotto}},\ }\bibfield  {title} {\bibinfo {title} {Anomalous and {Quantum} {Hall} {Effects} in {Lossy} {Photonic} {Lattices}},\ }\href {https://doi.org/10.1103/PhysRevLett.112.133902} {\bibfield  {journal} {\bibinfo  {journal} {Physical Review Letters}\ }\textbf {\bibinfo {volume} {112}},\ \bibinfo {pages} {133902} (\bibinfo {year} {2014})},\ \bibinfo {note} {publisher: American Physical Society}\BibitemShut {NoStop}%
\bibitem [{\citenamefont {Haldane}(1988)}]{haldane_model_1988}%
  \BibitemOpen
  \bibfield  {author} {\bibinfo {author} {\bibfnamefont {F.~D.~M.}\ \bibnamefont {Haldane}},\ }\bibfield  {title} {\bibinfo {title} {Model for a {Quantum} {Hall} {Effect} without {Landau} {Levels}: {Condensed}-{Matter} {Realization} of the "{Parity} {Anomaly}"},\ }\href {https://doi.org/10.1103/PhysRevLett.61.2015} {\bibfield  {journal} {\bibinfo  {journal} {Physical Review Letters}\ }\textbf {\bibinfo {volume} {61}},\ \bibinfo {pages} {2015} (\bibinfo {year} {1988})},\ \bibinfo {note} {publisher: American Physical Society}\BibitemShut {NoStop}%
\bibitem [{\citenamefont {Jotzu}\ \emph {et~al.}(2014)\citenamefont {Jotzu}, \citenamefont {Messer}, \citenamefont {Desbuquois}, \citenamefont {Lebrat}, \citenamefont {Uehlinger}, \citenamefont {Greif},\ and\ \citenamefont {Esslinger}}]{jotzu_experimental_2014}%
  \BibitemOpen
  \bibfield  {author} {\bibinfo {author} {\bibfnamefont {G.}~\bibnamefont {Jotzu}}, \bibinfo {author} {\bibfnamefont {M.}~\bibnamefont {Messer}}, \bibinfo {author} {\bibfnamefont {R.}~\bibnamefont {Desbuquois}}, \bibinfo {author} {\bibfnamefont {M.}~\bibnamefont {Lebrat}}, \bibinfo {author} {\bibfnamefont {T.}~\bibnamefont {Uehlinger}}, \bibinfo {author} {\bibfnamefont {D.}~\bibnamefont {Greif}},\ and\ \bibinfo {author} {\bibfnamefont {T.}~\bibnamefont {Esslinger}},\ }\bibfield  {title} {\bibinfo {title} {Experimental realization of the topological {Haldane} model with ultracold fermions},\ }\href {https://doi.org/10.1038/nature13915} {\bibfield  {journal} {\bibinfo  {journal} {Nature}\ }\textbf {\bibinfo {volume} {515}},\ \bibinfo {pages} {237} (\bibinfo {year} {2014})},\ \bibinfo {note} {publisher: Nature Publishing Group}\BibitemShut {NoStop}%
\bibitem [{\citenamefont {Fläschner}\ \emph {et~al.}(2018)\citenamefont {Fläschner}, \citenamefont {Vogel}, \citenamefont {Tarnowski}, \citenamefont {Rem}, \citenamefont {Lühmann}, \citenamefont {Heyl}, \citenamefont {Budich}, \citenamefont {Mathey}, \citenamefont {Sengstock},\ and\ \citenamefont {Weitenberg}}]{flaschner_observation_2018}%
  \BibitemOpen
  \bibfield  {author} {\bibinfo {author} {\bibfnamefont {N.}~\bibnamefont {Fläschner}}, \bibinfo {author} {\bibfnamefont {D.}~\bibnamefont {Vogel}}, \bibinfo {author} {\bibfnamefont {M.}~\bibnamefont {Tarnowski}}, \bibinfo {author} {\bibfnamefont {B.~S.}\ \bibnamefont {Rem}}, \bibinfo {author} {\bibfnamefont {D.-S.}\ \bibnamefont {Lühmann}}, \bibinfo {author} {\bibfnamefont {M.}~\bibnamefont {Heyl}}, \bibinfo {author} {\bibfnamefont {J.~C.}\ \bibnamefont {Budich}}, \bibinfo {author} {\bibfnamefont {L.}~\bibnamefont {Mathey}}, \bibinfo {author} {\bibfnamefont {K.}~\bibnamefont {Sengstock}},\ and\ \bibinfo {author} {\bibfnamefont {C.}~\bibnamefont {Weitenberg}},\ }\bibfield  {title} {\bibinfo {title} {Observation of dynamical vortices after quenches in a system with topology},\ }\href {https://doi.org/10.1038/s41567-017-0013-8} {\bibfield  {journal} {\bibinfo  {journal} {Nature Physics}\ }\textbf {\bibinfo {volume} {14}},\ \bibinfo {pages} {265} (\bibinfo {year} {2018})},\ \bibinfo {note} {publisher: Nature
  Publishing Group}\BibitemShut {NoStop}%
\bibitem [{\citenamefont {Mitra}\ \emph {et~al.}(2024)\citenamefont {Mitra}, \citenamefont {Jiménez-Galán}, \citenamefont {Aulich}, \citenamefont {Neuhaus}, \citenamefont {Silva}, \citenamefont {Pervak}, \citenamefont {Kling},\ and\ \citenamefont {Biswas}}]{mitra_light-wave-controlled_2024}%
  \BibitemOpen
  \bibfield  {author} {\bibinfo {author} {\bibfnamefont {S.}~\bibnamefont {Mitra}}, \bibinfo {author} {\bibfnamefont {A.}~\bibnamefont {Jiménez-Galán}}, \bibinfo {author} {\bibfnamefont {M.}~\bibnamefont {Aulich}}, \bibinfo {author} {\bibfnamefont {M.}~\bibnamefont {Neuhaus}}, \bibinfo {author} {\bibfnamefont {R.~E.~F.}\ \bibnamefont {Silva}}, \bibinfo {author} {\bibfnamefont {V.}~\bibnamefont {Pervak}}, \bibinfo {author} {\bibfnamefont {M.~F.}\ \bibnamefont {Kling}},\ and\ \bibinfo {author} {\bibfnamefont {S.}~\bibnamefont {Biswas}},\ }\bibfield  {title} {\bibinfo {title} {Light-wave-controlled {Haldane} model in monolayer hexagonal boron nitride},\ }\href {https://doi.org/10.1038/s41586-024-07244-z} {\bibfield  {journal} {\bibinfo  {journal} {Nature}\ }\textbf {\bibinfo {volume} {628}},\ \bibinfo {pages} {752} (\bibinfo {year} {2024})},\ \bibinfo {note} {publisher: Nature Publishing Group}\BibitemShut {NoStop}%
\bibitem [{\citenamefont {Zhao}\ \emph {et~al.}(2024)\citenamefont {Zhao}, \citenamefont {Kang}, \citenamefont {Zhang}, \citenamefont {Knüppel}, \citenamefont {Tao}, \citenamefont {Li}, \citenamefont {Tschirhart}, \citenamefont {Redekop}, \citenamefont {Watanabe}, \citenamefont {Taniguchi}, \citenamefont {Young}, \citenamefont {Shan},\ and\ \citenamefont {Mak}}]{zhao_realization_2024}%
  \BibitemOpen
  \bibfield  {author} {\bibinfo {author} {\bibfnamefont {W.}~\bibnamefont {Zhao}}, \bibinfo {author} {\bibfnamefont {K.}~\bibnamefont {Kang}}, \bibinfo {author} {\bibfnamefont {Y.}~\bibnamefont {Zhang}}, \bibinfo {author} {\bibfnamefont {P.}~\bibnamefont {Knüppel}}, \bibinfo {author} {\bibfnamefont {Z.}~\bibnamefont {Tao}}, \bibinfo {author} {\bibfnamefont {L.}~\bibnamefont {Li}}, \bibinfo {author} {\bibfnamefont {C.~L.}\ \bibnamefont {Tschirhart}}, \bibinfo {author} {\bibfnamefont {E.}~\bibnamefont {Redekop}}, \bibinfo {author} {\bibfnamefont {K.}~\bibnamefont {Watanabe}}, \bibinfo {author} {\bibfnamefont {T.}~\bibnamefont {Taniguchi}}, \bibinfo {author} {\bibfnamefont {A.~F.}\ \bibnamefont {Young}}, \bibinfo {author} {\bibfnamefont {J.}~\bibnamefont {Shan}},\ and\ \bibinfo {author} {\bibfnamefont {K.~F.}\ \bibnamefont {Mak}},\ }\bibfield  {title} {\bibinfo {title} {Realization of the {Haldane} {Chern} insulator in a moiré lattice},\ }\href {https://doi.org/10.1038/s41567-023-02284-0} {\bibfield  {journal}
  {\bibinfo  {journal} {Nature Physics}\ }\textbf {\bibinfo {volume} {20}},\ \bibinfo {pages} {275} (\bibinfo {year} {2024})},\ \bibinfo {note} {publisher: Nature Publishing Group}\BibitemShut {NoStop}%
\bibitem [{\citenamefont {Cai}\ \emph {et~al.}(2023)\citenamefont {Cai}, \citenamefont {Anderson}, \citenamefont {Wang}, \citenamefont {Zhang}, \citenamefont {Liu}, \citenamefont {Holtzmann}, \citenamefont {Zhang}, \citenamefont {Fan}, \citenamefont {Taniguchi}, \citenamefont {Watanabe}, \citenamefont {Ran}, \citenamefont {Cao}, \citenamefont {Fu}, \citenamefont {Xiao}, \citenamefont {Yao},\ and\ \citenamefont {Xu}}]{cai_signatures_2023}%
  \BibitemOpen
  \bibfield  {author} {\bibinfo {author} {\bibfnamefont {J.}~\bibnamefont {Cai}}, \bibinfo {author} {\bibfnamefont {E.}~\bibnamefont {Anderson}}, \bibinfo {author} {\bibfnamefont {C.}~\bibnamefont {Wang}}, \bibinfo {author} {\bibfnamefont {X.}~\bibnamefont {Zhang}}, \bibinfo {author} {\bibfnamefont {X.}~\bibnamefont {Liu}}, \bibinfo {author} {\bibfnamefont {W.}~\bibnamefont {Holtzmann}}, \bibinfo {author} {\bibfnamefont {Y.}~\bibnamefont {Zhang}}, \bibinfo {author} {\bibfnamefont {F.}~\bibnamefont {Fan}}, \bibinfo {author} {\bibfnamefont {T.}~\bibnamefont {Taniguchi}}, \bibinfo {author} {\bibfnamefont {K.}~\bibnamefont {Watanabe}}, \bibinfo {author} {\bibfnamefont {Y.}~\bibnamefont {Ran}}, \bibinfo {author} {\bibfnamefont {T.}~\bibnamefont {Cao}}, \bibinfo {author} {\bibfnamefont {L.}~\bibnamefont {Fu}}, \bibinfo {author} {\bibfnamefont {D.}~\bibnamefont {Xiao}}, \bibinfo {author} {\bibfnamefont {W.}~\bibnamefont {Yao}},\ and\ \bibinfo {author} {\bibfnamefont {X.}~\bibnamefont {Xu}},\ }\bibfield  {title}
  {\bibinfo {title} {Signatures of fractional quantum anomalous {Hall} states in twisted {MoTe2}},\ }\href {https://doi.org/10.1038/s41586-023-06289-w} {\bibfield  {journal} {\bibinfo  {journal} {Nature}\ }\textbf {\bibinfo {volume} {622}},\ \bibinfo {pages} {63} (\bibinfo {year} {2023})},\ \bibinfo {note} {publisher: Nature Publishing Group}\BibitemShut {NoStop}%
\bibitem [{\citenamefont {He}\ \emph {et~al.}(2019)\citenamefont {He}, \citenamefont {Addison}, \citenamefont {Jin}, \citenamefont {Mele}, \citenamefont {Johnson},\ and\ \citenamefont {Zhen}}]{he_floquet_2019}%
  \BibitemOpen
  \bibfield  {author} {\bibinfo {author} {\bibfnamefont {L.}~\bibnamefont {He}}, \bibinfo {author} {\bibfnamefont {Z.}~\bibnamefont {Addison}}, \bibinfo {author} {\bibfnamefont {J.}~\bibnamefont {Jin}}, \bibinfo {author} {\bibfnamefont {E.~J.}\ \bibnamefont {Mele}}, \bibinfo {author} {\bibfnamefont {S.~G.}\ \bibnamefont {Johnson}},\ and\ \bibinfo {author} {\bibfnamefont {B.}~\bibnamefont {Zhen}},\ }\bibfield  {title} {\bibinfo {title} {Floquet {Chern} insulators of light},\ }\href {https://doi.org/10.1038/s41467-019-12231-4} {\bibfield  {journal} {\bibinfo  {journal} {Nature Communications}\ }\textbf {\bibinfo {volume} {10}},\ \bibinfo {pages} {4194} (\bibinfo {year} {2019})},\ \bibinfo {note} {publisher: Nature Publishing Group}\BibitemShut {NoStop}%
\bibitem [{\citenamefont {Liu}\ \emph {et~al.}(2021)\citenamefont {Liu}, \citenamefont {Jung}, \citenamefont {Parto}, \citenamefont {Christodoulides},\ and\ \citenamefont {Khajavikhan}}]{liu_gain-induced_2021}%
  \BibitemOpen
  \bibfield  {author} {\bibinfo {author} {\bibfnamefont {Y.~G.~N.}\ \bibnamefont {Liu}}, \bibinfo {author} {\bibfnamefont {P.~S.}\ \bibnamefont {Jung}}, \bibinfo {author} {\bibfnamefont {M.}~\bibnamefont {Parto}}, \bibinfo {author} {\bibfnamefont {D.~N.}\ \bibnamefont {Christodoulides}},\ and\ \bibinfo {author} {\bibfnamefont {M.}~\bibnamefont {Khajavikhan}},\ }\bibfield  {title} {\bibinfo {title} {Gain-induced topological response via tailored long-range interactions},\ }\href {https://doi.org/10.1038/s41567-021-01185-4} {\bibfield  {journal} {\bibinfo  {journal} {Nature Physics}\ }\textbf {\bibinfo {volume} {17}},\ \bibinfo {pages} {704} (\bibinfo {year} {2021})},\ \bibinfo {note} {publisher: Nature Publishing Group}\BibitemShut {NoStop}%
\bibitem [{\citenamefont {Sridhar}\ \emph {et~al.}(2024)\citenamefont {Sridhar}, \citenamefont {Ghosh}, \citenamefont {Srinivasan}, \citenamefont {Miller},\ and\ \citenamefont {Dutt}}]{sridhar_quantized_2024}%
  \BibitemOpen
  \bibfield  {author} {\bibinfo {author} {\bibfnamefont {S.~K.}\ \bibnamefont {Sridhar}}, \bibinfo {author} {\bibfnamefont {S.}~\bibnamefont {Ghosh}}, \bibinfo {author} {\bibfnamefont {D.}~\bibnamefont {Srinivasan}}, \bibinfo {author} {\bibfnamefont {A.~R.}\ \bibnamefont {Miller}},\ and\ \bibinfo {author} {\bibfnamefont {A.}~\bibnamefont {Dutt}},\ }\bibfield  {title} {\bibinfo {title} {Quantized topological pumping in {Floquet} synthetic dimensions with a driven dissipative photonic molecule},\ }\href {https://doi.org/10.1038/s41567-024-02413-3} {\bibfield  {journal} {\bibinfo  {journal} {Nature Physics}\ }\textbf {\bibinfo {volume} {20}},\ \bibinfo {pages} {843} (\bibinfo {year} {2024})},\ \bibinfo {note} {publisher: Nature Publishing Group}\BibitemShut {NoStop}%
\bibitem [{\citenamefont {Ozawa}\ \emph {et~al.}(2019)\citenamefont {Ozawa}, \citenamefont {Price}, \citenamefont {Amo}, \citenamefont {Goldman}, \citenamefont {Hafezi}, \citenamefont {Lu}, \citenamefont {Rechtsman}, \citenamefont {Schuster}, \citenamefont {Simon}, \citenamefont {Zilberberg},\ and\ \citenamefont {Carusotto}}]{ozawa_topological_2019}%
  \BibitemOpen
  \bibfield  {author} {\bibinfo {author} {\bibfnamefont {T.}~\bibnamefont {Ozawa}}, \bibinfo {author} {\bibfnamefont {H.~M.}\ \bibnamefont {Price}}, \bibinfo {author} {\bibfnamefont {A.}~\bibnamefont {Amo}}, \bibinfo {author} {\bibfnamefont {N.}~\bibnamefont {Goldman}}, \bibinfo {author} {\bibfnamefont {M.}~\bibnamefont {Hafezi}}, \bibinfo {author} {\bibfnamefont {L.}~\bibnamefont {Lu}}, \bibinfo {author} {\bibfnamefont {M.~C.}\ \bibnamefont {Rechtsman}}, \bibinfo {author} {\bibfnamefont {D.}~\bibnamefont {Schuster}}, \bibinfo {author} {\bibfnamefont {J.}~\bibnamefont {Simon}}, \bibinfo {author} {\bibfnamefont {O.}~\bibnamefont {Zilberberg}},\ and\ \bibinfo {author} {\bibfnamefont {I.}~\bibnamefont {Carusotto}},\ }\bibfield  {title} {\bibinfo {title} {Topological photonics},\ }\href {https://doi.org/10.1103/RevModPhys.91.015006} {\bibfield  {journal} {\bibinfo  {journal} {Reviews of Modern Physics}\ }\textbf {\bibinfo {volume} {91}},\ \bibinfo {pages} {015006} (\bibinfo {year} {2019})},\ \bibinfo {note}
  {publisher: American Physical Society}\BibitemShut {NoStop}%
\bibitem [{\citenamefont {Yuan}\ \emph {et~al.}(2018{\natexlab{a}})\citenamefont {Yuan}, \citenamefont {Lin}, \citenamefont {Xiao},\ and\ \citenamefont {Fan}}]{yuan_synthetic_2018-1}%
  \BibitemOpen
  \bibfield  {author} {\bibinfo {author} {\bibfnamefont {L.}~\bibnamefont {Yuan}}, \bibinfo {author} {\bibfnamefont {Q.}~\bibnamefont {Lin}}, \bibinfo {author} {\bibfnamefont {M.}~\bibnamefont {Xiao}},\ and\ \bibinfo {author} {\bibfnamefont {S.}~\bibnamefont {Fan}},\ }\bibfield  {title} {\bibinfo {title} {Synthetic dimension in photonics},\ }\href {https://doi.org/10.1364/OPTICA.5.001396} {\bibfield  {journal} {\bibinfo  {journal} {Optica}\ }\textbf {\bibinfo {volume} {5}},\ \bibinfo {pages} {1396} (\bibinfo {year} {2018}{\natexlab{a}})},\ \bibinfo {note} {publisher: Optica Publishing Group}\BibitemShut {NoStop}%
\bibitem [{\citenamefont {Parriaux}\ \emph {et~al.}(2020)\citenamefont {Parriaux}, \citenamefont {Hammani},\ and\ \citenamefont {Millot}}]{parriaux_electro-optic_2020}%
  \BibitemOpen
  \bibfield  {author} {\bibinfo {author} {\bibfnamefont {A.}~\bibnamefont {Parriaux}}, \bibinfo {author} {\bibfnamefont {K.}~\bibnamefont {Hammani}},\ and\ \bibinfo {author} {\bibfnamefont {G.}~\bibnamefont {Millot}},\ }\bibfield  {title} {\bibinfo {title} {Electro-optic frequency combs},\ }\href {https://doi.org/10.1364/AOP.382052} {\bibfield  {journal} {\bibinfo  {journal} {Advances in Optics and Photonics}\ }\textbf {\bibinfo {volume} {12}},\ \bibinfo {pages} {223} (\bibinfo {year} {2020})},\ \bibinfo {note} {publisher: Optica Publishing Group}\BibitemShut {NoStop}%
\bibitem [{\citenamefont {Ozawa}\ \emph {et~al.}(2016)\citenamefont {Ozawa}, \citenamefont {Price}, \citenamefont {Goldman}, \citenamefont {Zilberberg},\ and\ \citenamefont {Carusotto}}]{ozawa_synthetic_2016}%
  \BibitemOpen
  \bibfield  {author} {\bibinfo {author} {\bibfnamefont {T.}~\bibnamefont {Ozawa}}, \bibinfo {author} {\bibfnamefont {H.~M.}\ \bibnamefont {Price}}, \bibinfo {author} {\bibfnamefont {N.}~\bibnamefont {Goldman}}, \bibinfo {author} {\bibfnamefont {O.}~\bibnamefont {Zilberberg}},\ and\ \bibinfo {author} {\bibfnamefont {I.}~\bibnamefont {Carusotto}},\ }\bibfield  {title} {\bibinfo {title} {Synthetic dimensions in integrated photonics: {From} optical isolation to four-dimensional quantum {Hall} physics},\ }\href {https://doi.org/10.1103/PhysRevA.93.043827} {\bibfield  {journal} {\bibinfo  {journal} {Physical Review A}\ }\textbf {\bibinfo {volume} {93}},\ \bibinfo {pages} {043827} (\bibinfo {year} {2016})},\ \bibinfo {note} {publisher: American Physical Society}\BibitemShut {NoStop}%
\bibitem [{\citenamefont {Dutt}\ \emph {et~al.}(2019)\citenamefont {Dutt}, \citenamefont {Minkov}, \citenamefont {Lin}, \citenamefont {Yuan}, \citenamefont {Miller},\ and\ \citenamefont {Fan}}]{dutt_experimental_2019}%
  \BibitemOpen
  \bibfield  {author} {\bibinfo {author} {\bibfnamefont {A.}~\bibnamefont {Dutt}}, \bibinfo {author} {\bibfnamefont {M.}~\bibnamefont {Minkov}}, \bibinfo {author} {\bibfnamefont {Q.}~\bibnamefont {Lin}}, \bibinfo {author} {\bibfnamefont {L.}~\bibnamefont {Yuan}}, \bibinfo {author} {\bibfnamefont {D.~A.~B.}\ \bibnamefont {Miller}},\ and\ \bibinfo {author} {\bibfnamefont {S.}~\bibnamefont {Fan}},\ }\bibfield  {title} {\bibinfo {title} {Experimental band structure spectroscopy along a synthetic dimension},\ }\href {https://doi.org/10.1038/s41467-019-11117-9} {\bibfield  {journal} {\bibinfo  {journal} {Nature Communications}\ }\textbf {\bibinfo {volume} {10}},\ \bibinfo {pages} {3122} (\bibinfo {year} {2019})},\ \bibinfo {note} {publisher: Nature Publishing Group}\BibitemShut {NoStop}%
\bibitem [{\citenamefont {Sriram}\ \emph {et~al.}(2025)\citenamefont {Sriram}, \citenamefont {Sridhar},\ and\ \citenamefont {Dutt}}]{sriram_quantized_2025}%
  \BibitemOpen
  \bibfield  {author} {\bibinfo {author} {\bibfnamefont {S.}~\bibnamefont {Sriram}}, \bibinfo {author} {\bibfnamefont {S.~K.}\ \bibnamefont {Sridhar}},\ and\ \bibinfo {author} {\bibfnamefont {A.}~\bibnamefont {Dutt}},\ }\bibfield  {title} {\bibinfo {title} {Quantized topological phases beyond square lattices in {Floquet} synthetic dimensions [{Invited}]},\ }\href {https://doi.org/10.1364/OME.546801} {\bibfield  {journal} {\bibinfo  {journal} {Optical Materials Express}\ }\textbf {\bibinfo {volume} {15}},\ \bibinfo {pages} {272} (\bibinfo {year} {2025})},\ \bibinfo {note} {publisher: Optica Publishing Group}\BibitemShut {NoStop}%
\bibitem [{\citenamefont {Xiao}\ \emph {et~al.}(2010)\citenamefont {Xiao}, \citenamefont {Chang},\ and\ \citenamefont {Niu}}]{xiao_berry_2010}%
  \BibitemOpen
  \bibfield  {author} {\bibinfo {author} {\bibfnamefont {D.}~\bibnamefont {Xiao}}, \bibinfo {author} {\bibfnamefont {M.-C.}\ \bibnamefont {Chang}},\ and\ \bibinfo {author} {\bibfnamefont {Q.}~\bibnamefont {Niu}},\ }\bibfield  {title} {\bibinfo {title} {Berry phase effects on electronic properties},\ }\href {https://doi.org/10.1103/RevModPhys.82.1959} {\bibfield  {journal} {\bibinfo  {journal} {Reviews of Modern Physics}\ }\textbf {\bibinfo {volume} {82}},\ \bibinfo {pages} {1959} (\bibinfo {year} {2010})},\ \bibinfo {note} {publisher: American Physical Society}\BibitemShut {NoStop}%
\bibitem [{\citenamefont {Dutt}\ \emph {et~al.}(2020)\citenamefont {Dutt}, \citenamefont {Lin}, \citenamefont {Yuan}, \citenamefont {Minkov}, \citenamefont {Xiao},\ and\ \citenamefont {Fan}}]{dutt_single_2020}%
  \BibitemOpen
  \bibfield  {author} {\bibinfo {author} {\bibfnamefont {A.}~\bibnamefont {Dutt}}, \bibinfo {author} {\bibfnamefont {Q.}~\bibnamefont {Lin}}, \bibinfo {author} {\bibfnamefont {L.}~\bibnamefont {Yuan}}, \bibinfo {author} {\bibfnamefont {M.}~\bibnamefont {Minkov}}, \bibinfo {author} {\bibfnamefont {M.}~\bibnamefont {Xiao}},\ and\ \bibinfo {author} {\bibfnamefont {S.}~\bibnamefont {Fan}},\ }\bibfield  {title} {\bibinfo {title} {A single photonic cavity with two independent physical synthetic dimensions},\ }\href {https://doi.org/10.1126/science.aaz3071} {\bibfield  {journal} {\bibinfo  {journal} {Science}\ }\textbf {\bibinfo {volume} {367}},\ \bibinfo {pages} {59} (\bibinfo {year} {2020})},\ \bibinfo {note} {publisher: American Association for the Advancement of Science}\BibitemShut {NoStop}%
\bibitem [{\citenamefont {Yuan}\ \emph {et~al.}(2018{\natexlab{b}})\citenamefont {Yuan}, \citenamefont {Xiao}, \citenamefont {Lin},\ and\ \citenamefont {Fan}}]{yuan_synthetic_2018}%
  \BibitemOpen
  \bibfield  {author} {\bibinfo {author} {\bibfnamefont {L.}~\bibnamefont {Yuan}}, \bibinfo {author} {\bibfnamefont {M.}~\bibnamefont {Xiao}}, \bibinfo {author} {\bibfnamefont {Q.}~\bibnamefont {Lin}},\ and\ \bibinfo {author} {\bibfnamefont {S.}~\bibnamefont {Fan}},\ }\bibfield  {title} {\bibinfo {title} {Synthetic space with arbitrary dimensions in a few rings undergoing dynamic modulation},\ }\href {https://doi.org/10.1103/PhysRevB.97.104105} {\bibfield  {journal} {\bibinfo  {journal} {Physical Review B}\ }\textbf {\bibinfo {volume} {97}},\ \bibinfo {pages} {104105} (\bibinfo {year} {2018}{\natexlab{b}})},\ \bibinfo {note} {publisher: American Physical Society}\BibitemShut {NoStop}%
\bibitem [{\citenamefont {Senanian}\ \emph {et~al.}(2023)\citenamefont {Senanian}, \citenamefont {Wright}, \citenamefont {Wade}, \citenamefont {Doyle},\ and\ \citenamefont {McMahon}}]{senanian_programmable_2023}%
  \BibitemOpen
  \bibfield  {author} {\bibinfo {author} {\bibfnamefont {A.}~\bibnamefont {Senanian}}, \bibinfo {author} {\bibfnamefont {L.~G.}\ \bibnamefont {Wright}}, \bibinfo {author} {\bibfnamefont {P.~F.}\ \bibnamefont {Wade}}, \bibinfo {author} {\bibfnamefont {H.~K.}\ \bibnamefont {Doyle}},\ and\ \bibinfo {author} {\bibfnamefont {P.~L.}\ \bibnamefont {McMahon}},\ }\bibfield  {title} {\bibinfo {title} {Programmable large-scale simulation of bosonic transport in optical synthetic frequency lattices},\ }\href {https://doi.org/10.1038/s41567-023-02075-7} {\bibfield  {journal} {\bibinfo  {journal} {Nature Physics}\ }\textbf {\bibinfo {volume} {19}},\ \bibinfo {pages} {1333} (\bibinfo {year} {2023})},\ \bibinfo {note} {publisher: Nature Publishing Group}\BibitemShut {NoStop}%
\bibitem [{\citenamefont {Cheng}\ \emph {et~al.}(2023)\citenamefont {Cheng}, \citenamefont {Lustig}, \citenamefont {Wang},\ and\ \citenamefont {Fan}}]{cheng_multi-dimensional_2023}%
  \BibitemOpen
  \bibfield  {author} {\bibinfo {author} {\bibfnamefont {D.}~\bibnamefont {Cheng}}, \bibinfo {author} {\bibfnamefont {E.}~\bibnamefont {Lustig}}, \bibinfo {author} {\bibfnamefont {K.}~\bibnamefont {Wang}},\ and\ \bibinfo {author} {\bibfnamefont {S.}~\bibnamefont {Fan}},\ }\bibfield  {title} {\bibinfo {title} {Multi-dimensional band structure spectroscopy in the synthetic frequency dimension},\ }\href {https://doi.org/10.1038/s41377-023-01196-1} {\bibfield  {journal} {\bibinfo  {journal} {Light: Science \& Applications}\ }\textbf {\bibinfo {volume} {12}},\ \bibinfo {pages} {158} (\bibinfo {year} {2023})},\ \bibinfo {note} {publisher: Nature Publishing Group}\BibitemShut {NoStop}%
\bibitem{SM}
See Supplemental Material for details on the derivations of the effective Hamiltonian, time-resolved signals, Chern number extraction, and numerical simulations of the anomalous transverse displacement, including Ref.~[44]. 
\bibitem [{\citenamefont {Yuan}\ and\ \citenamefont {Fan}(2015)}]{yuan_three-dimensional_2015}%
  \BibitemOpen
  \bibfield  {author} {\bibinfo {author} {\bibfnamefont {L.}~\bibnamefont {Yuan}}\ and\ \bibinfo {author} {\bibfnamefont {S.}~\bibnamefont {Fan}},\ }\bibfield  {title} {\bibinfo {title} {Three-{Dimensional} {Dynamic} {Localization} of {Light} from a {Time}-{Dependent} {Effective} {Gauge} {Field} for {Photons}},\ }\href {https://doi.org/10.1103/PhysRevLett.114.243901} {\bibfield  {journal} {\bibinfo  {journal} {Physical Review Letters}\ }\textbf {\bibinfo {volume} {114}},\ \bibinfo {pages} {243901} (\bibinfo {year} {2015})},\ \bibinfo {note} {publisher: American Physical Society}\BibitemShut {NoStop}%
\bibitem [{\citenamefont {Mittal}\ \emph {et~al.}(2019)\citenamefont {Mittal}, \citenamefont {Orre}, \citenamefont {Leykam}, \citenamefont {Chong},\ and\ \citenamefont {Hafezi}}]{mittal_photonic_2019}%
  \BibitemOpen
  \bibfield  {author} {\bibinfo {author} {\bibfnamefont {S.}~\bibnamefont {Mittal}}, \bibinfo {author} {\bibfnamefont {V.~V.}\ \bibnamefont {Orre}}, \bibinfo {author} {\bibfnamefont {D.}~\bibnamefont {Leykam}}, \bibinfo {author} {\bibfnamefont {Y.}~\bibnamefont {Chong}},\ and\ \bibinfo {author} {\bibfnamefont {M.}~\bibnamefont {Hafezi}},\ }\bibfield  {title} {\bibinfo {title} {Photonic {Anomalous} {Quantum} {Hall} {Effect}},\ }\href {https://doi.org/10.1103/PhysRevLett.123.043201} {\bibfield  {journal} {\bibinfo  {journal} {Physical Review Letters}\ }\textbf {\bibinfo {volume} {123}},\ \bibinfo {pages} {043201} (\bibinfo {year} {2019})},\ \bibinfo {note} {publisher: American Physical Society}\BibitemShut {NoStop}%
\bibitem [{\citenamefont {Pellerin}\ \emph {et~al.}(2024)\citenamefont {Pellerin}, \citenamefont {Houvenaghel}, \citenamefont {Coish}, \citenamefont {Carusotto},\ and\ \citenamefont {St-Jean}}]{pellerin_wave-function_2024}%
  \BibitemOpen
  \bibfield  {author} {\bibinfo {author} {\bibfnamefont {F.}~\bibnamefont {Pellerin}}, \bibinfo {author} {\bibfnamefont {R.}~\bibnamefont {Houvenaghel}}, \bibinfo {author} {\bibfnamefont {W.}~\bibnamefont {Coish}}, \bibinfo {author} {\bibfnamefont {I.}~\bibnamefont {Carusotto}},\ and\ \bibinfo {author} {\bibfnamefont {P.}~\bibnamefont {St-Jean}},\ }\bibfield  {title} {\bibinfo {title} {Wave-{Function} {Tomography} of {Topological} {Dimer} {Chains} with {Long}-{Range} {Couplings}},\ }\href {https://doi.org/10.1103/PhysRevLett.132.183802} {\bibfield  {journal} {\bibinfo  {journal} {Physical Review Letters}\ }\textbf {\bibinfo {volume} {132}},\ \bibinfo {pages} {183802} (\bibinfo {year} {2024})},\ \bibinfo {note} {publisher: American Physical Society}\BibitemShut {NoStop}%
\bibitem [{\citenamefont {Gianfrate}\ \emph {et~al.}(2020)\citenamefont {Gianfrate}, \citenamefont {Bleu}, \citenamefont {Dominici}, \citenamefont {Ardizzone}, \citenamefont {De~Giorgi}, \citenamefont {Ballarini}, \citenamefont {Lerario}, \citenamefont {West}, \citenamefont {Pfeiffer}, \citenamefont {Solnyshkov}, \citenamefont {Sanvitto},\ and\ \citenamefont {Malpuech}}]{gianfrate_measurement_2020}%
  \BibitemOpen
  \bibfield  {author} {\bibinfo {author} {\bibfnamefont {A.}~\bibnamefont {Gianfrate}}, \bibinfo {author} {\bibfnamefont {O.}~\bibnamefont {Bleu}}, \bibinfo {author} {\bibfnamefont {L.}~\bibnamefont {Dominici}}, \bibinfo {author} {\bibfnamefont {V.}~\bibnamefont {Ardizzone}}, \bibinfo {author} {\bibfnamefont {M.}~\bibnamefont {De~Giorgi}}, \bibinfo {author} {\bibfnamefont {D.}~\bibnamefont {Ballarini}}, \bibinfo {author} {\bibfnamefont {G.}~\bibnamefont {Lerario}}, \bibinfo {author} {\bibfnamefont {K.~W.}\ \bibnamefont {West}}, \bibinfo {author} {\bibfnamefont {L.~N.}\ \bibnamefont {Pfeiffer}}, \bibinfo {author} {\bibfnamefont {D.~D.}\ \bibnamefont {Solnyshkov}}, \bibinfo {author} {\bibfnamefont {D.}~\bibnamefont {Sanvitto}},\ and\ \bibinfo {author} {\bibfnamefont {G.}~\bibnamefont {Malpuech}},\ }\bibfield  {title} {\bibinfo {title} {Measurement of the quantum geometric tensor and of the anomalous {Hall} drift},\ }\href {https://doi.org/10.1038/s41586-020-1989-2} {\bibfield  {journal} {\bibinfo  {journal}
  {Nature}\ }\textbf {\bibinfo {volume} {578}},\ \bibinfo {pages} {381} (\bibinfo {year} {2020})},\ \bibinfo {note} {publisher: Nature Publishing Group}\BibitemShut {NoStop}%
\bibitem [{\citenamefont {Fukui}\ \emph {et~al.}(2005)\citenamefont {Fukui}, \citenamefont {Hatsugai},\ and\ \citenamefont {Suzuki}}]{fukui_chern_2005}%
  \BibitemOpen
  \bibfield  {author} {\bibinfo {author} {\bibfnamefont {T.}~\bibnamefont {Fukui}}, \bibinfo {author} {\bibfnamefont {Y.}~\bibnamefont {Hatsugai}},\ and\ \bibinfo {author} {\bibfnamefont {H.}~\bibnamefont {Suzuki}},\ }\bibfield  {title} {\bibinfo {title} {Chern {Numbers} in {Discretized} {Brillouin} {Zone}: {Efficient} {Method} of {Computing} ({Spin}) {Hall} {Conductances}},\ }\href {https://doi.org/10.1143/JPSJ.74.1674} {\bibfield  {journal} {\bibinfo  {journal} {Journal of the Physical Society of Japan}\ }\textbf {\bibinfo {volume} {74}},\ \bibinfo {pages} {1674} (\bibinfo {year} {2005})},\ \bibinfo {note} {publisher: The Physical Society of Japan}\BibitemShut {NoStop}%
\bibitem [{\citenamefont {Chang}\ \emph {et~al.}(2023)\citenamefont {Chang}, \citenamefont {Liu},\ and\ \citenamefont {MacDonald}}]{chang_colloquium_2023}%
  \BibitemOpen
  \bibfield  {author} {\bibinfo {author} {\bibfnamefont {C.-Z.}\ \bibnamefont {Chang}}, \bibinfo {author} {\bibfnamefont {C.-X.}\ \bibnamefont {Liu}},\ and\ \bibinfo {author} {\bibfnamefont {A.~H.}\ \bibnamefont {MacDonald}},\ }\bibfield  {title} {\bibinfo {title} {Colloquium: {Quantum} anomalous {Hall} effect},\ }\href {https://doi.org/10.1103/RevModPhys.95.011002} {\bibfield  {journal} {\bibinfo  {journal} {Reviews of Modern Physics}\ }\textbf {\bibinfo {volume} {95}},\ \bibinfo {pages} {011002} (\bibinfo {year} {2023})},\ \bibinfo {note} {publisher: American Physical Society}\BibitemShut {NoStop}%
\bibitem [{\citenamefont {Li}\ \emph {et~al.}(2021)\citenamefont {Li}, \citenamefont {Zheng}, \citenamefont {Dutt}, \citenamefont {Yu}, \citenamefont {Shan}, \citenamefont {Liu}, \citenamefont {Yuan}, \citenamefont {Fan},\ and\ \citenamefont {Chen}}]{li_dynamic_2021}%
  \BibitemOpen
  \bibfield  {author} {\bibinfo {author} {\bibfnamefont {G.}~\bibnamefont {Li}}, \bibinfo {author} {\bibfnamefont {Y.}~\bibnamefont {Zheng}}, \bibinfo {author} {\bibfnamefont {A.}~\bibnamefont {Dutt}}, \bibinfo {author} {\bibfnamefont {D.}~\bibnamefont {Yu}}, \bibinfo {author} {\bibfnamefont {Q.}~\bibnamefont {Shan}}, \bibinfo {author} {\bibfnamefont {S.}~\bibnamefont {Liu}}, \bibinfo {author} {\bibfnamefont {L.}~\bibnamefont {Yuan}}, \bibinfo {author} {\bibfnamefont {S.}~\bibnamefont {Fan}},\ and\ \bibinfo {author} {\bibfnamefont {X.}~\bibnamefont {Chen}},\ }\bibfield  {title} {\bibinfo {title} {Dynamic band structure measurement in the synthetic space},\ }\href {https://doi.org/10.1126/sciadv.abe4335} {\bibfield  {journal} {\bibinfo  {journal} {Science Advances}\ }\textbf {\bibinfo {volume} {7}},\ \bibinfo {pages} {eabe4335} (\bibinfo {year} {2021})},\ \bibinfo {note} {publisher: American Association for the Advancement of Science}\BibitemShut {NoStop}%
\bibitem [{\citenamefont {Oliver}\ \emph {et~al.}(2023)\citenamefont {Oliver}, \citenamefont {Smith}, \citenamefont {Easton}, \citenamefont {Salerno}, \citenamefont {Guarrera}, \citenamefont {Goldman}, \citenamefont {Barontini},\ and\ \citenamefont {Price}}]{oliver_bloch_2023}%
  \BibitemOpen
  \bibfield  {author} {\bibinfo {author} {\bibfnamefont {C.}~\bibnamefont {Oliver}}, \bibinfo {author} {\bibfnamefont {A.}~\bibnamefont {Smith}}, \bibinfo {author} {\bibfnamefont {T.}~\bibnamefont {Easton}}, \bibinfo {author} {\bibfnamefont {G.}~\bibnamefont {Salerno}}, \bibinfo {author} {\bibfnamefont {V.}~\bibnamefont {Guarrera}}, \bibinfo {author} {\bibfnamefont {N.}~\bibnamefont {Goldman}}, \bibinfo {author} {\bibfnamefont {G.}~\bibnamefont {Barontini}},\ and\ \bibinfo {author} {\bibfnamefont {H.~M.}\ \bibnamefont {Price}},\ }\bibfield  {title} {\bibinfo {title} {Bloch oscillations along a synthetic dimension of atomic trap states},\ }\href {https://doi.org/10.1103/PhysRevResearch.5.033001} {\bibfield  {journal} {\bibinfo  {journal} {Physical Review Research}\ }\textbf {\bibinfo {volume} {5}},\ \bibinfo {pages} {033001} (\bibinfo {year} {2023})},\ \bibinfo {note} {publisher: American Physical Society}\BibitemShut {NoStop}%
\bibitem [{\citenamefont {Ozawa}(2018)}]{ozawa_steady-state_2018}%
  \BibitemOpen
  \bibfield  {author} {\bibinfo {author} {\bibfnamefont {T.}~\bibnamefont {Ozawa}},\ }\bibfield  {title} {\bibinfo {title} {Steady-state {Hall} response and quantum geometry of driven-dissipative lattices},\ }\href {https://doi.org/10.1103/PhysRevB.97.041108} {\bibfield  {journal} {\bibinfo  {journal} {Physical Review B}\ }\textbf {\bibinfo {volume} {97}},\ \bibinfo {pages} {041108} (\bibinfo {year} {2018})},\ \bibinfo {note} {publisher: American Physical Society}\BibitemShut {NoStop}%
\bibitem [{\citenamefont {Wimmer}\ \emph {et~al.}(2017)\citenamefont {Wimmer}, \citenamefont {Price}, \citenamefont {Carusotto},\ and\ \citenamefont {Peschel}}]{wimmer_experimental_2017}%
  \BibitemOpen
  \bibfield  {author} {\bibinfo {author} {\bibfnamefont {M.}~\bibnamefont {Wimmer}}, \bibinfo {author} {\bibfnamefont {H.~M.}\ \bibnamefont {Price}}, \bibinfo {author} {\bibfnamefont {I.}~\bibnamefont {Carusotto}},\ and\ \bibinfo {author} {\bibfnamefont {U.}~\bibnamefont {Peschel}},\ }\bibfield  {title} {\bibinfo {title} {Experimental measurement of the {Berry} curvature from anomalous transport},\ }\href {https://doi.org/10.1038/nphys4050} {\bibfield  {journal} {\bibinfo  {journal} {Nature Physics}\ }\textbf {\bibinfo {volume} {13}},\ \bibinfo {pages} {545} (\bibinfo {year} {2017})},\ \bibinfo {note} {publisher: Nature Publishing Group}\BibitemShut {NoStop}%
\bibitem [{\citenamefont {Rosen}\ \emph {et~al.}(2024)\citenamefont {Rosen}, \citenamefont {Muschinske}, \citenamefont {Barrett}, \citenamefont {Chatterjee}, \citenamefont {Hays}, \citenamefont {DeMarco}, \citenamefont {Karamlou}, \citenamefont {Rower}, \citenamefont {Das}, \citenamefont {Kim}, \citenamefont {Niedzielski}, \citenamefont {Schuldt}, \citenamefont {Serniak}, \citenamefont {Schwartz}, \citenamefont {Yoder}, \citenamefont {Grover},\ and\ \citenamefont {Oliver}}]{rosen_synthetic_2024}%
  \BibitemOpen
  \bibfield  {author} {\bibinfo {author} {\bibfnamefont {I.~T.}\ \bibnamefont {Rosen}}, \bibinfo {author} {\bibfnamefont {S.}~\bibnamefont {Muschinske}}, \bibinfo {author} {\bibfnamefont {C.~N.}\ \bibnamefont {Barrett}}, \bibinfo {author} {\bibfnamefont {A.}~\bibnamefont {Chatterjee}}, \bibinfo {author} {\bibfnamefont {M.}~\bibnamefont {Hays}}, \bibinfo {author} {\bibfnamefont {M.~A.}\ \bibnamefont {DeMarco}}, \bibinfo {author} {\bibfnamefont {A.~H.}\ \bibnamefont {Karamlou}}, \bibinfo {author} {\bibfnamefont {D.~A.}\ \bibnamefont {Rower}}, \bibinfo {author} {\bibfnamefont {R.}~\bibnamefont {Das}}, \bibinfo {author} {\bibfnamefont {D.~K.}\ \bibnamefont {Kim}}, \bibinfo {author} {\bibfnamefont {B.~M.}\ \bibnamefont {Niedzielski}}, \bibinfo {author} {\bibfnamefont {M.}~\bibnamefont {Schuldt}}, \bibinfo {author} {\bibfnamefont {K.}~\bibnamefont {Serniak}}, \bibinfo {author} {\bibfnamefont {M.~E.}\ \bibnamefont {Schwartz}}, \bibinfo {author} {\bibfnamefont {J.~L.}\ \bibnamefont {Yoder}}, \bibinfo {author}
  {\bibfnamefont {J.~A.}\ \bibnamefont {Grover}},\ and\ \bibinfo {author} {\bibfnamefont {W.~D.}\ \bibnamefont {Oliver}},\ }\bibfield  {title} {\bibinfo {title} {A synthetic magnetic vector potential in a {2D} superconducting qubit array},\ }\href {https://doi.org/10.1038/s41567-024-02661-3} {\bibfield  {journal} {\bibinfo  {journal} {Nature Physics}\ ,\ \bibinfo {pages} {1}} (\bibinfo {year} {2024})},\ \bibinfo {note} {publisher: Nature Publishing Group}\BibitemShut {NoStop}%
\bibitem [{\citenamefont {Cheng}\ \emph {et~al.}(2025)\citenamefont {Cheng}, \citenamefont {Wang}, \citenamefont {Roques-Carmes}, \citenamefont {Lustig}, \citenamefont {Long}, \citenamefont {Wang},\ and\ \citenamefont {Fan}}]{cheng_non-abelian_2025}%
  \BibitemOpen
  \bibfield  {author} {\bibinfo {author} {\bibfnamefont {D.}~\bibnamefont {Cheng}}, \bibinfo {author} {\bibfnamefont {K.}~\bibnamefont {Wang}}, \bibinfo {author} {\bibfnamefont {C.}~\bibnamefont {Roques-Carmes}}, \bibinfo {author} {\bibfnamefont {E.}~\bibnamefont {Lustig}}, \bibinfo {author} {\bibfnamefont {O.~Y.}\ \bibnamefont {Long}}, \bibinfo {author} {\bibfnamefont {H.}~\bibnamefont {Wang}},\ and\ \bibinfo {author} {\bibfnamefont {S.}~\bibnamefont {Fan}},\ }\bibfield  {title} {\bibinfo {title} {Non-{Abelian} lattice gauge fields in photonic synthetic frequency dimensions},\ }\href {https://doi.org/10.1038/s41586-024-08259-2} {\bibfield  {journal} {\bibinfo  {journal} {Nature}\ }\textbf {\bibinfo {volume} {637}},\ \bibinfo {pages} {52} (\bibinfo {year} {2025})},\ \bibinfo {note} {publisher: Nature Publishing Group}\BibitemShut {NoStop}%
\bibitem [{\citenamefont {Bello}\ \emph {et~al.}(2019)\citenamefont {Bello}, \citenamefont {Platero}, \citenamefont {Cirac},\ and\ \citenamefont {González-Tudela}}]{bello_unconventional_2019}%
  \BibitemOpen
  \bibfield  {author} {\bibinfo {author} {\bibfnamefont {M.}~\bibnamefont {Bello}}, \bibinfo {author} {\bibfnamefont {G.}~\bibnamefont {Platero}}, \bibinfo {author} {\bibfnamefont {J.~I.}\ \bibnamefont {Cirac}},\ and\ \bibinfo {author} {\bibfnamefont {A.}~\bibnamefont {González-Tudela}},\ }\bibfield  {title} {\bibinfo {title} {Unconventional quantum optics in topological waveguide {QED}},\ }\href {https://doi.org/10.1126/sciadv.aaw0297} {\bibfield  {journal} {\bibinfo  {journal} {Science Advances}\ }\textbf {\bibinfo {volume} {5}},\ \bibinfo {pages} {eaaw0297} (\bibinfo {year} {2019})},\ \bibinfo {note} {publisher: American Association for the Advancement of Science}\BibitemShut {NoStop}%
\bibitem [{\citenamefont {Owens}\ \emph {et~al.}(2022)\citenamefont {Owens}, \citenamefont {Panetta}, \citenamefont {Saxberg}, \citenamefont {Roberts}, \citenamefont {Chakram}, \citenamefont {Ma}, \citenamefont {Vrajitoarea}, \citenamefont {Simon},\ and\ \citenamefont {Schuster}}]{owens_chiral_2022}%
  \BibitemOpen
  \bibfield  {author} {\bibinfo {author} {\bibfnamefont {J.~C.}\ \bibnamefont {Owens}}, \bibinfo {author} {\bibfnamefont {M.~G.}\ \bibnamefont {Panetta}}, \bibinfo {author} {\bibfnamefont {B.}~\bibnamefont {Saxberg}}, \bibinfo {author} {\bibfnamefont {G.}~\bibnamefont {Roberts}}, \bibinfo {author} {\bibfnamefont {S.}~\bibnamefont {Chakram}}, \bibinfo {author} {\bibfnamefont {R.}~\bibnamefont {Ma}}, \bibinfo {author} {\bibfnamefont {A.}~\bibnamefont {Vrajitoarea}}, \bibinfo {author} {\bibfnamefont {J.}~\bibnamefont {Simon}},\ and\ \bibinfo {author} {\bibfnamefont {D.~I.}\ \bibnamefont {Schuster}},\ }\bibfield  {title} {\bibinfo {title} {Chiral cavity quantum electrodynamics},\ }\href {https://doi.org/10.1038/s41567-022-01671-3} {\bibfield  {journal} {\bibinfo  {journal} {Nature Physics}\ }\textbf {\bibinfo {volume} {18}},\ \bibinfo {pages} {1048} (\bibinfo {year} {2022})},\ \bibinfo {note} {publisher: Nature Publishing Group}\BibitemShut {NoStop}%
\bibitem [{\citenamefont {De~Bernardis}\ \emph {et~al.}(2023)\citenamefont {De~Bernardis}, \citenamefont {Piccioli}, \citenamefont {Rabl},\ and\ \citenamefont {Carusotto}}]{de_bernardis_chiral_2023}%
  \BibitemOpen
  \bibfield  {author} {\bibinfo {author} {\bibfnamefont {D.}~\bibnamefont {De~Bernardis}}, \bibinfo {author} {\bibfnamefont {F.~S.}\ \bibnamefont {Piccioli}}, \bibinfo {author} {\bibfnamefont {P.}~\bibnamefont {Rabl}},\ and\ \bibinfo {author} {\bibfnamefont {I.}~\bibnamefont {Carusotto}},\ }\bibfield  {title} {\bibinfo {title} {Chiral {Quantum} {Optics} in the {Bulk} of {Photonic} {Quantum} {Hall} {Systems}},\ }\href {https://doi.org/10.1103/PRXQuantum.4.030306} {\bibfield  {journal} {\bibinfo  {journal} {PRX Quantum}\ }\textbf {\bibinfo {volume} {4}},\ \bibinfo {pages} {030306} (\bibinfo {year} {2023})},\ \bibinfo {note} {publisher: American Physical Society}\BibitemShut {NoStop}%
\bibitem [{\citenamefont {Dutt}\ \emph {et~al.}(2022)\citenamefont {Dutt}, \citenamefont {Yuan}, \citenamefont {Yang}, \citenamefont {Wang}, \citenamefont {Buddhiraju}, \citenamefont {Vučković},\ and\ \citenamefont {Fan}}]{dutt_creating_2022}%
  \BibitemOpen
  \bibfield  {author} {\bibinfo {author} {\bibfnamefont {A.}~\bibnamefont {Dutt}}, \bibinfo {author} {\bibfnamefont {L.}~\bibnamefont {Yuan}}, \bibinfo {author} {\bibfnamefont {K.~Y.}\ \bibnamefont {Yang}}, \bibinfo {author} {\bibfnamefont {K.}~\bibnamefont {Wang}}, \bibinfo {author} {\bibfnamefont {S.}~\bibnamefont {Buddhiraju}}, \bibinfo {author} {\bibfnamefont {J.}~\bibnamefont {Vučković}},\ and\ \bibinfo {author} {\bibfnamefont {S.}~\bibnamefont {Fan}},\ }\bibfield  {title} {\bibinfo {title} {Creating boundaries along a synthetic frequency dimension},\ }\href {https://doi.org/10.1038/s41467-022-31140-7} {\bibfield  {journal} {\bibinfo  {journal} {Nature Communications}\ }\textbf {\bibinfo {volume} {13}},\ \bibinfo {pages} {3377} (\bibinfo {year} {2022})},\ \bibinfo {note} {publisher: Nature Publishing Group}\BibitemShut {NoStop}%
\bibitem [{\citenamefont {Reid}\ \emph {et~al.}(2025)\citenamefont {Reid}, \citenamefont {Oliver}, \citenamefont {Regan}, \citenamefont {Smith}, \citenamefont {Easton}, \citenamefont {Salerno}, \citenamefont {Barontini}, \citenamefont {Goldman},\ and\ \citenamefont {Price}}]{reid_topological_2025}%
  \BibitemOpen
  \bibfield  {author} {\bibinfo {author} {\bibfnamefont {D.~G.}\ \bibnamefont {Reid}}, \bibinfo {author} {\bibfnamefont {C.}~\bibnamefont {Oliver}}, \bibinfo {author} {\bibfnamefont {P.}~\bibnamefont {Regan}}, \bibinfo {author} {\bibfnamefont {A.}~\bibnamefont {Smith}}, \bibinfo {author} {\bibfnamefont {T.}~\bibnamefont {Easton}}, \bibinfo {author} {\bibfnamefont {G.}~\bibnamefont {Salerno}}, \bibinfo {author} {\bibfnamefont {G.}~\bibnamefont {Barontini}}, \bibinfo {author} {\bibfnamefont {N.}~\bibnamefont {Goldman}},\ and\ \bibinfo {author} {\bibfnamefont {H.~M.}\ \bibnamefont {Price}},\ }\bibfield  {title} {\bibinfo {title} {Topological chiral edge states in a synthetic dimension of atomic trap states},\ }\href {https://doi.org/10.1103/PhysRevA.111.033301} {\bibfield  {journal} {\bibinfo  {journal} {Physical Review A}\ }\textbf {\bibinfo {volume} {111}},\ \bibinfo {pages} {033301} (\bibinfo {year} {2025})},\ \bibinfo {note} {publisher: American Physical Society}\BibitemShut {NoStop}%
\bibitem [{\citenamefont {Zhu}\ \emph {et~al.}(2024)\citenamefont {Zhu}, \citenamefont {Gächter}, \citenamefont {Walter}, \citenamefont {Viebahn},\ and\ \citenamefont {Esslinger}}]{zhu_reversal_2024}%
  \BibitemOpen
  \bibfield  {author} {\bibinfo {author} {\bibfnamefont {Z.}~\bibnamefont {Zhu}}, \bibinfo {author} {\bibfnamefont {M.}~\bibnamefont {Gächter}}, \bibinfo {author} {\bibfnamefont {A.-S.}\ \bibnamefont {Walter}}, \bibinfo {author} {\bibfnamefont {K.}~\bibnamefont {Viebahn}},\ and\ \bibinfo {author} {\bibfnamefont {T.}~\bibnamefont {Esslinger}},\ }\bibfield  {title} {\bibinfo {title} {Reversal of quantized {Hall} drifts at noninteracting and interacting topological boundaries},\ }\href {https://doi.org/10.1126/science.adg3848} {\bibfield  {journal} {\bibinfo  {journal} {Science}\ }\textbf {\bibinfo {volume} {384}},\ \bibinfo {pages} {317} (\bibinfo {year} {2024})},\ \bibinfo {note} {publisher: American Association for the Advancement of Science}\BibitemShut {NoStop}%
\bibitem [{\citenamefont {Chiu}\ \emph {et~al.}(2016)\citenamefont {Chiu}, \citenamefont {Teo}, \citenamefont {Schnyder},\ and\ \citenamefont {Ryu}}]{chiu_classification_2016}%
  \BibitemOpen
  \bibfield  {author} {\bibinfo {author} {\bibfnamefont {C.-K.}\ \bibnamefont {Chiu}}, \bibinfo {author} {\bibfnamefont {J.~C.}\ \bibnamefont {Teo}}, \bibinfo {author} {\bibfnamefont {A.~P.}\ \bibnamefont {Schnyder}},\ and\ \bibinfo {author} {\bibfnamefont {S.}~\bibnamefont {Ryu}},\ }\bibfield  {title} {\bibinfo {title} {Classification of topological quantum matter with symmetries},\ }\href {https://doi.org/10.1103/RevModPhys.88.035005} {\bibfield  {journal} {\bibinfo  {journal} {Reviews of Modern Physics}\ }\textbf {\bibinfo {volume} {88}},\ \bibinfo {pages} {035005} (\bibinfo {year} {2016})},\ \bibinfo {note} {publisher: American Physical Society}\BibitemShut {NoStop}%
\bibitem [{\citenamefont {Carusotto}\ \emph {et~al.}(2020)\citenamefont {Carusotto}, \citenamefont {Houck}, \citenamefont {Kollár}, \citenamefont {Roushan}, \citenamefont {Schuster},\ and\ \citenamefont {Simon}}]{carusotto_photonic_2020}%
  \BibitemOpen
  \bibfield  {author} {\bibinfo {author} {\bibfnamefont {I.}~\bibnamefont {Carusotto}}, \bibinfo {author} {\bibfnamefont {A.~A.}\ \bibnamefont {Houck}}, \bibinfo {author} {\bibfnamefont {A.~J.}\ \bibnamefont {Kollár}}, \bibinfo {author} {\bibfnamefont {P.}~\bibnamefont {Roushan}}, \bibinfo {author} {\bibfnamefont {D.~I.}\ \bibnamefont {Schuster}},\ and\ \bibinfo {author} {\bibfnamefont {J.}~\bibnamefont {Simon}},\ }\bibfield  {title} {\bibinfo {title} {Photonic materials in circuit quantum electrodynamics},\ }\href {https://doi.org/10.1038/s41567-020-0815-y} {\bibfield  {journal} {\bibinfo  {journal} {Nature Physics}\ }\textbf {\bibinfo {volume} {16}},\ \bibinfo {pages} {268} (\bibinfo {year} {2020})},\ \bibinfo {note} {publisher: Nature Publishing Group}\BibitemShut {NoStop}%
\bibitem [{\citenamefont {Wang}\ \emph {et~al.}(2009)\citenamefont {Wang}, \citenamefont {Chong}, \citenamefont {Joannopoulos},\ and\ \citenamefont {Soljačić}}]{wang_observation_2009}%
  \BibitemOpen
  \bibfield  {author} {\bibinfo {author} {\bibfnamefont {Z.}~\bibnamefont {Wang}}, \bibinfo {author} {\bibfnamefont {Y.}~\bibnamefont {Chong}}, \bibinfo {author} {\bibfnamefont {J.~D.}\ \bibnamefont {Joannopoulos}},\ and\ \bibinfo {author} {\bibfnamefont {M.}~\bibnamefont {Soljačić}},\ }\bibfield  {title} {\bibinfo {title} {Observation of unidirectional backscattering-immune topological electromagnetic states},\ }\href {https://doi.org/10.1038/nature08293} {\bibfield  {journal} {\bibinfo  {journal} {Nature}\ }\textbf {\bibinfo {volume} {461}},\ \bibinfo {pages} {772} (\bibinfo {year} {2009})},\ \bibinfo {note} {publisher: Nature Publishing Group}\BibitemShut {NoStop}%
\bibitem [{\citenamefont {Villa}\ \emph {et~al.}(2024)\citenamefont {Villa}, \citenamefont {Carusotto},\ and\ \citenamefont {Ozawa}}]{villa_mean-chiral_2024}%
  \BibitemOpen
  \bibfield  {author} {\bibinfo {author} {\bibfnamefont {G.}~\bibnamefont {Villa}}, \bibinfo {author} {\bibfnamefont {I.}~\bibnamefont {Carusotto}},\ and\ \bibinfo {author} {\bibfnamefont {T.}~\bibnamefont {Ozawa}},\ }\bibfield  {title} {\bibinfo {title} {Mean-chiral displacement in coherently driven photonic lattices and its application to synthetic frequency dimensions},\ }\href {https://doi.org/10.1038/s42005-024-01727-1} {\bibfield  {journal} {\bibinfo  {journal} {Communications Physics}\ }\textbf {\bibinfo {volume} {7}},\ \bibinfo {pages} {246} (\bibinfo {year} {2024})},\ \bibinfo {note} {publisher: Nature Publishing Group}\BibitemShut {NoStop}%
\bibitem [{\citenamefont {Sridhar}\ \emph {et~al.}(2025)\citenamefont {Sridhar}, \citenamefont {Srikanth}, \citenamefont {Miller}, \citenamefont {McComb},\ and\ \citenamefont {Dutt}}]{sridhar_measuring_2025}%
  \BibitemOpen
  \bibfield  {author} {\bibinfo {author} {\bibfnamefont {S.~K.}\ \bibnamefont {Sridhar}}, \bibinfo {author} {\bibfnamefont {R.}~\bibnamefont {Srikanth}}, \bibinfo {author} {\bibfnamefont {A.~R.}\ \bibnamefont {Miller}}, \bibinfo {author} {\bibfnamefont {F.~J.}\ \bibnamefont {McComb}},\ and\ \bibinfo {author} {\bibfnamefont {A.}~\bibnamefont {Dutt}},\ }\href {https://doi.org/10.48550/arXiv.2505.04151} {\bibinfo {title} {Measuring \${\textbackslash}mathbb\{{Z}\}\_2\$ invariants in dimer models and cross-coupled ladders with a programmable photonic molecule}} (\bibinfo {year} {2025}),\ \bibinfo {note} {arXiv:2505.04151 [physics]}\BibitemShut {NoStop}%
\bibitem [{\citenamefont {Dinh}\ \emph {et~al.}(2024)\citenamefont {Dinh}, \citenamefont {Balčytis}, \citenamefont {Ozawa}, \citenamefont {Ota}, \citenamefont {Ren}, \citenamefont {Baba}, \citenamefont {Iwamoto}, \citenamefont {Mitchell},\ and\ \citenamefont {Nguyen}}]{dinh_reconfigurable_2024}%
  \BibitemOpen
  \bibfield  {author} {\bibinfo {author} {\bibfnamefont {H.~X.}\ \bibnamefont {Dinh}}, \bibinfo {author} {\bibfnamefont {A.}~\bibnamefont {Balčytis}}, \bibinfo {author} {\bibfnamefont {T.}~\bibnamefont {Ozawa}}, \bibinfo {author} {\bibfnamefont {Y.}~\bibnamefont {Ota}}, \bibinfo {author} {\bibfnamefont {G.}~\bibnamefont {Ren}}, \bibinfo {author} {\bibfnamefont {T.}~\bibnamefont {Baba}}, \bibinfo {author} {\bibfnamefont {S.}~\bibnamefont {Iwamoto}}, \bibinfo {author} {\bibfnamefont {A.}~\bibnamefont {Mitchell}},\ and\ \bibinfo {author} {\bibfnamefont {T.~G.}\ \bibnamefont {Nguyen}},\ }\bibfield  {title} {\bibinfo {title} {Reconfigurable synthetic dimension frequency lattices in an integrated lithium niobate ring cavity},\ }\href {https://doi.org/10.1038/s42005-024-01676-9} {\bibfield  {journal} {\bibinfo  {journal} {Communications Physics}\ }\textbf {\bibinfo {volume} {7}},\ \bibinfo {pages} {185} (\bibinfo {year} {2024})},\ \bibinfo {note} {publisher: Nature Publishing Group}\BibitemShut {NoStop}%
\bibitem [{\citenamefont {Leefmans}\ \emph {et~al.}(2024)\citenamefont {Leefmans}, \citenamefont {Parto}, \citenamefont {Williams}, \citenamefont {Li}, \citenamefont {Dutt}, \citenamefont {Nori},\ and\ \citenamefont {Marandi}}]{leefmans_topological_2024}%
  \BibitemOpen
  \bibfield  {author} {\bibinfo {author} {\bibfnamefont {C.~R.}\ \bibnamefont {Leefmans}}, \bibinfo {author} {\bibfnamefont {M.}~\bibnamefont {Parto}}, \bibinfo {author} {\bibfnamefont {J.}~\bibnamefont {Williams}}, \bibinfo {author} {\bibfnamefont {G.~H.~Y.}\ \bibnamefont {Li}}, \bibinfo {author} {\bibfnamefont {A.}~\bibnamefont {Dutt}}, \bibinfo {author} {\bibfnamefont {F.}~\bibnamefont {Nori}},\ and\ \bibinfo {author} {\bibfnamefont {A.}~\bibnamefont {Marandi}},\ }\bibfield  {title} {\bibinfo {title} {Topological temporally mode-locked laser},\ }\href {https://doi.org/10.1038/s41567-024-02420-4} {\bibfield  {journal} {\bibinfo  {journal} {Nature Physics}\ }\textbf {\bibinfo {volume} {20}},\ \bibinfo {pages} {852} (\bibinfo {year} {2024})},\ \bibinfo {note} {publisher: Nature Publishing Group}\BibitemShut {NoStop}%
\bibitem [{\citenamefont {Yang}\ \emph {et~al.}(2020)\citenamefont {Yang}, \citenamefont {Lustig}, \citenamefont {Harari}, \citenamefont {Plotnik}, \citenamefont {Lumer}, \citenamefont {Bandres},\ and\ \citenamefont {Segev}}]{yang_mode-locked_2020}%
  \BibitemOpen
  \bibfield  {author} {\bibinfo {author} {\bibfnamefont {Z.}~\bibnamefont {Yang}}, \bibinfo {author} {\bibfnamefont {E.}~\bibnamefont {Lustig}}, \bibinfo {author} {\bibfnamefont {G.}~\bibnamefont {Harari}}, \bibinfo {author} {\bibfnamefont {Y.}~\bibnamefont {Plotnik}}, \bibinfo {author} {\bibfnamefont {Y.}~\bibnamefont {Lumer}}, \bibinfo {author} {\bibfnamefont {M.~A.}\ \bibnamefont {Bandres}},\ and\ \bibinfo {author} {\bibfnamefont {M.}~\bibnamefont {Segev}},\ }\bibfield  {title} {\bibinfo {title} {Mode-{Locked} {Topological} {Insulator} {Laser} {Utilizing} {Synthetic} {Dimensions}},\ }\href {https://doi.org/10.1103/PhysRevX.10.011059} {\bibfield  {journal} {\bibinfo  {journal} {Physical Review X}\ }\textbf {\bibinfo {volume} {10}},\ \bibinfo {pages} {011059} (\bibinfo {year} {2020})},\ \bibinfo {note} {publisher: American Physical Society}\BibitemShut {NoStop}%
\bibitem [{\citenamefont {Mittal}\ \emph {et~al.}(2021)\citenamefont {Mittal}, \citenamefont {Moille}, \citenamefont {Srinivasan}, \citenamefont {Chembo},\ and\ \citenamefont {Hafezi}}]{mittal_topological_2021}%
  \BibitemOpen
  \bibfield  {author} {\bibinfo {author} {\bibfnamefont {S.}~\bibnamefont {Mittal}}, \bibinfo {author} {\bibfnamefont {G.}~\bibnamefont {Moille}}, \bibinfo {author} {\bibfnamefont {K.}~\bibnamefont {Srinivasan}}, \bibinfo {author} {\bibfnamefont {Y.~K.}\ \bibnamefont {Chembo}},\ and\ \bibinfo {author} {\bibfnamefont {M.}~\bibnamefont {Hafezi}},\ }\bibfield  {title} {\bibinfo {title} {Topological frequency combs and nested temporal solitons},\ }\href {https://doi.org/10.1038/s41567-021-01302-3} {\bibfield  {journal} {\bibinfo  {journal} {Nature Physics}\ }\textbf {\bibinfo {volume} {17}},\ \bibinfo {pages} {1169} (\bibinfo {year} {2021})},\ \bibinfo {note} {publisher: Nature Publishing Group}\BibitemShut {NoStop}%
\bibitem [{\citenamefont {Hu}\ \emph {et~al.}(2020)\citenamefont {Hu}, \citenamefont {Reimer}, \citenamefont {Shams-Ansari}, \citenamefont {Zhang},\ and\ \citenamefont {Loncar}}]{hu_realization_2020}%
  \BibitemOpen
  \bibfield  {author} {\bibinfo {author} {\bibfnamefont {Y.}~\bibnamefont {Hu}}, \bibinfo {author} {\bibfnamefont {C.}~\bibnamefont {Reimer}}, \bibinfo {author} {\bibfnamefont {A.}~\bibnamefont {Shams-Ansari}}, \bibinfo {author} {\bibfnamefont {M.}~\bibnamefont {Zhang}},\ and\ \bibinfo {author} {\bibfnamefont {M.}~\bibnamefont {Loncar}},\ }\bibfield  {title} {\bibinfo {title} {Realization of high-dimensional frequency crystals in electro-optic microcombs},\ }\href {https://doi.org/10.1364/OPTICA.395114} {\bibfield  {journal} {\bibinfo  {journal} {Optica}\ }\textbf {\bibinfo {volume} {7}},\ \bibinfo {pages} {1189} (\bibinfo {year} {2020})},\ \bibinfo {note} {publisher: Optica Publishing Group}\BibitemShut {NoStop}%
\bibitem [{\citenamefont {Zhang}\ \emph {et~al.}(2019)\citenamefont {Zhang}, \citenamefont {Buscaino}, \citenamefont {Wang}, \citenamefont {Shams-Ansari}, \citenamefont {Reimer}, \citenamefont {Zhu}, \citenamefont {Kahn},\ and\ \citenamefont {Lončar}}]{zhang_broadband_2019}%
  \BibitemOpen
  \bibfield  {author} {\bibinfo {author} {\bibfnamefont {M.}~\bibnamefont {Zhang}}, \bibinfo {author} {\bibfnamefont {B.}~\bibnamefont {Buscaino}}, \bibinfo {author} {\bibfnamefont {C.}~\bibnamefont {Wang}}, \bibinfo {author} {\bibfnamefont {A.}~\bibnamefont {Shams-Ansari}}, \bibinfo {author} {\bibfnamefont {C.}~\bibnamefont {Reimer}}, \bibinfo {author} {\bibfnamefont {R.}~\bibnamefont {Zhu}}, \bibinfo {author} {\bibfnamefont {J.~M.}\ \bibnamefont {Kahn}},\ and\ \bibinfo {author} {\bibfnamefont {M.}~\bibnamefont {Lončar}},\ }\bibfield  {title} {\bibinfo {title} {Broadband electro-optic frequency comb generation in a lithium niobate microring resonator},\ }\href {https://doi.org/10.1038/s41586-019-1008-7} {\bibfield  {journal} {\bibinfo  {journal} {Nature}\ }\textbf {\bibinfo {volume} {568}},\ \bibinfo {pages} {373} (\bibinfo {year} {2019})},\ \bibinfo {note} {publisher: Nature Publishing Group}\BibitemShut {NoStop}%
\bibitem [{\citenamefont {Buddhiraju}\ \emph {et~al.}(2021)\citenamefont {Buddhiraju}, \citenamefont {Dutt}, \citenamefont {Minkov}, \citenamefont {Williamson},\ and\ \citenamefont {Fan}}]{buddhiraju_arbitrary_2021}%
  \BibitemOpen
  \bibfield  {author} {\bibinfo {author} {\bibfnamefont {S.}~\bibnamefont {Buddhiraju}}, \bibinfo {author} {\bibfnamefont {A.}~\bibnamefont {Dutt}}, \bibinfo {author} {\bibfnamefont {M.}~\bibnamefont {Minkov}}, \bibinfo {author} {\bibfnamefont {I.~A.~D.}\ \bibnamefont {Williamson}},\ and\ \bibinfo {author} {\bibfnamefont {S.}~\bibnamefont {Fan}},\ }\bibfield  {title} {\bibinfo {title} {Arbitrary linear transformations for photons in the frequency synthetic dimension},\ }\href {https://doi.org/10.1038/s41467-021-22670-7} {\bibfield  {journal} {\bibinfo  {journal} {Nature Communications}\ }\textbf {\bibinfo {volume} {12}},\ \bibinfo {pages} {2401} (\bibinfo {year} {2021})},\ \bibinfo {note} {publisher: Nature Publishing Group}\BibitemShut {NoStop}%
\bibitem [{\citenamefont {Budich}\ and\ \citenamefont {Bergholtz}(2020)}]{budich_non-hermitian_2020}%
  \BibitemOpen
  \bibfield  {author} {\bibinfo {author} {\bibfnamefont {J.~C.}\ \bibnamefont {Budich}}\ and\ \bibinfo {author} {\bibfnamefont {E.~J.}\ \bibnamefont {Bergholtz}},\ }\bibfield  {title} {\bibinfo {title} {Non-{Hermitian} {Topological} {Sensors}},\ }\href {https://doi.org/10.1103/PhysRevLett.125.180403} {\bibfield  {journal} {\bibinfo  {journal} {Physical Review Letters}\ }\textbf {\bibinfo {volume} {125}},\ \bibinfo {pages} {180403} (\bibinfo {year} {2020})},\ \bibinfo {note} {publisher: American Physical Society}\BibitemShut {NoStop}%
\bibitem [{\citenamefont {McDonald}\ and\ \citenamefont {Clerk}(2020)}]{mcdonald_exponentially-enhanced_2020}%
  \BibitemOpen
  \bibfield  {author} {\bibinfo {author} {\bibfnamefont {A.}~\bibnamefont {McDonald}}\ and\ \bibinfo {author} {\bibfnamefont {A.~A.}\ \bibnamefont {Clerk}},\ }\bibfield  {title} {\bibinfo {title} {Exponentially-enhanced quantum sensing with non-{Hermitian} lattice dynamics},\ }\href {https://doi.org/10.1038/s41467-020-19090-4} {\bibfield  {journal} {\bibinfo  {journal} {Nature Communications}\ }\textbf {\bibinfo {volume} {11}},\ \bibinfo {pages} {5382} (\bibinfo {year} {2020})},\ \bibinfo {note} {publisher: Nature Publishing Group}\BibitemShut {NoStop}%
\bibitem [{\citenamefont {Lustig}\ \emph {et~al.}(2018)\citenamefont {Lustig}, \citenamefont {Sharabi},\ and\ \citenamefont {Segev}}]{lustig_topological_2018}%
  \BibitemOpen
  \bibfield  {author} {\bibinfo {author} {\bibfnamefont {E.}~\bibnamefont {Lustig}}, \bibinfo {author} {\bibfnamefont {Y.}~\bibnamefont {Sharabi}},\ and\ \bibinfo {author} {\bibfnamefont {M.}~\bibnamefont {Segev}},\ }\bibfield  {title} {\bibinfo {title} {Topological aspects of photonic time crystals},\ }\href {https://doi.org/10.1364/OPTICA.5.001390} {\bibfield  {journal} {\bibinfo  {journal} {Optica}\ }\textbf {\bibinfo {volume} {5}},\ \bibinfo {pages} {1390} (\bibinfo {year} {2018})},\ \bibinfo {note} {publisher: Optica Publishing Group}\BibitemShut {NoStop}%
\bibitem [{\citenamefont {Zheludev}(2024)}]{zheludev_time_2024}%
  \BibitemOpen
  \bibfield  {author} {\bibinfo {author} {\bibfnamefont {N.~I.}\ \bibnamefont {Zheludev}},\ }\bibfield  {title} {\bibinfo {title} {Time crystals for photonics and timetronics},\ }\href {https://doi.org/10.1038/s41566-024-01557-1} {\bibfield  {journal} {\bibinfo  {journal} {Nature Photonics}\ }\textbf {\bibinfo {volume} {18}},\ \bibinfo {pages} {1123} (\bibinfo {year} {2024})},\ \bibinfo {note} {publisher: Nature Publishing Group}\BibitemShut {NoStop}%
\bibitem [{\citenamefont {Fan}\ \emph {et~al.}(2023)\citenamefont {Fan}, \citenamefont {Wang}, \citenamefont {Wang}, \citenamefont {Dutt},\ and\ \citenamefont {Fan}}]{fan_experimental_2023}%
  \BibitemOpen
  \bibfield  {author} {\bibinfo {author} {\bibfnamefont {L.}~\bibnamefont {Fan}}, \bibinfo {author} {\bibfnamefont {K.}~\bibnamefont {Wang}}, \bibinfo {author} {\bibfnamefont {H.}~\bibnamefont {Wang}}, \bibinfo {author} {\bibfnamefont {A.}~\bibnamefont {Dutt}},\ and\ \bibinfo {author} {\bibfnamefont {S.}~\bibnamefont {Fan}},\ }\bibfield  {title} {\bibinfo {title} {Experimental realization of convolution processing in photonic synthetic frequency dimensions},\ }\href {https://doi.org/10.1126/sciadv.adi4956} {\bibfield  {journal} {\bibinfo  {journal} {Science Advances}\ }\textbf {\bibinfo {volume} {9}},\ \bibinfo {pages} {eadi4956} (\bibinfo {year} {2023})},\ \bibinfo {note} {publisher: American Association for the Advancement of Science}\BibitemShut {NoStop}%
\bibitem [{\citenamefont {Wanjura}\ and\ \citenamefont {Marquardt}(2024)}]{wanjura_fully_2024}%
  \BibitemOpen
  \bibfield  {author} {\bibinfo {author} {\bibfnamefont {C.~C.}\ \bibnamefont {Wanjura}}\ and\ \bibinfo {author} {\bibfnamefont {F.}~\bibnamefont {Marquardt}},\ }\bibfield  {title} {\bibinfo {title} {Fully nonlinear neuromorphic computing with linear wave scattering},\ }\href {https://doi.org/10.1038/s41567-024-02534-9} {\bibfield  {journal} {\bibinfo  {journal} {Nature Physics}\ }\textbf {\bibinfo {volume} {20}},\ \bibinfo {pages} {1434} (\bibinfo {year} {2024})},\ \bibinfo {note} {publisher: Nature Publishing Group}\BibitemShut {NoStop}%
\bibitem [{\citenamefont {Javid}\ \emph {et~al.}(2023)\citenamefont {Javid}, \citenamefont {Lopez-Rios}, \citenamefont {Ling}, \citenamefont {Graf}, \citenamefont {Staffa},\ and\ \citenamefont {Lin}}]{javid_chip-scale_2023}%
  \BibitemOpen
  \bibfield  {author} {\bibinfo {author} {\bibfnamefont {U.~A.}\ \bibnamefont {Javid}}, \bibinfo {author} {\bibfnamefont {R.}~\bibnamefont {Lopez-Rios}}, \bibinfo {author} {\bibfnamefont {J.}~\bibnamefont {Ling}}, \bibinfo {author} {\bibfnamefont {A.}~\bibnamefont {Graf}}, \bibinfo {author} {\bibfnamefont {J.}~\bibnamefont {Staffa}},\ and\ \bibinfo {author} {\bibfnamefont {Q.}~\bibnamefont {Lin}},\ }\bibfield  {title} {\bibinfo {title} {Chip-scale simulations in a quantum-correlated synthetic space},\ }\href {https://doi.org/10.1038/s41566-023-01236-7} {\bibfield  {journal} {\bibinfo  {journal} {Nature Photonics}\ }\textbf {\bibinfo {volume} {17}},\ \bibinfo {pages} {883} (\bibinfo {year} {2023})},\ \bibinfo {note} {publisher: Nature Publishing Group}\BibitemShut {NoStop}%
\bibitem [{\citenamefont {Bartlett}\ \emph {et~al.}(2024)\citenamefont {Bartlett}, \citenamefont {Long}, \citenamefont {Dutt},\ and\ \citenamefont {Fan}}]{bartlett_programmable_2024}%
  \BibitemOpen
  \bibfield  {author} {\bibinfo {author} {\bibfnamefont {B.}~\bibnamefont {Bartlett}}, \bibinfo {author} {\bibfnamefont {O.~Y.}\ \bibnamefont {Long}}, \bibinfo {author} {\bibfnamefont {A.}~\bibnamefont {Dutt}},\ and\ \bibinfo {author} {\bibfnamefont {S.}~\bibnamefont {Fan}},\ }\bibfield  {title} {\bibinfo {title} {Programmable photonic system for quantum simulation in arbitrary topologies},\ }\href {https://doi.org/10.1063/5.0181151} {\bibfield  {journal} {\bibinfo  {journal} {APL Quantum}\ }\textbf {\bibinfo {volume} {1}},\ \bibinfo {pages} {016102} (\bibinfo {year} {2024})}\BibitemShut {NoStop}%
\bibitem [{\citenamefont {Argüello-Luengo}\ \emph {et~al.}(2024)\citenamefont {Argüello-Luengo}, \citenamefont {Bhattacharya}, \citenamefont {Celi}, \citenamefont {Chhajlany}, \citenamefont {Grass}, \citenamefont {Płodzień}, \citenamefont {Rakshit}, \citenamefont {Salamon}, \citenamefont {Stornati}, \citenamefont {Tarruell},\ and\ \citenamefont {Lewenstein}}]{arguello-luengo_synthetic_2024}%
  \BibitemOpen
  \bibfield  {author} {\bibinfo {author} {\bibfnamefont {J.}~\bibnamefont {Argüello-Luengo}}, \bibinfo {author} {\bibfnamefont {U.}~\bibnamefont {Bhattacharya}}, \bibinfo {author} {\bibfnamefont {A.}~\bibnamefont {Celi}}, \bibinfo {author} {\bibfnamefont {R.~W.}\ \bibnamefont {Chhajlany}}, \bibinfo {author} {\bibfnamefont {T.}~\bibnamefont {Grass}}, \bibinfo {author} {\bibfnamefont {M.}~\bibnamefont {Płodzień}}, \bibinfo {author} {\bibfnamefont {D.}~\bibnamefont {Rakshit}}, \bibinfo {author} {\bibfnamefont {T.}~\bibnamefont {Salamon}}, \bibinfo {author} {\bibfnamefont {P.}~\bibnamefont {Stornati}}, \bibinfo {author} {\bibfnamefont {L.}~\bibnamefont {Tarruell}},\ and\ \bibinfo {author} {\bibfnamefont {M.}~\bibnamefont {Lewenstein}},\ }\bibfield  {title} {\bibinfo {title} {Synthetic dimensions for topological and quantum phases},\ }\href {https://doi.org/10.1038/s42005-024-01636-3} {\bibfield  {journal} {\bibinfo  {journal} {Communications Physics}\ }\textbf {\bibinfo {volume} {7}},\ \bibinfo {pages} {1}
  (\bibinfo {year} {2024})},\ \bibinfo {note} {publisher: Nature Publishing Group}\BibitemShut {NoStop}%
\bibitem{DA}
A. Chénier, B. d'Aligny, F. Pellerin, P.-É. Blanchard, T. Ozawa, I. Carusotto, P. St-Jean, Data for the Article "Quantized Hall drift in a frequency-encoded Chern insulator", \url{https://doi.org/10.17605/OSF.IO/ATRNV}.%

\end{thebibliography}
\end{document}